\documentclass[a4paper]{article}

\usepackage[hyphens]{url}
\usepackage[top=2cm,bottom=3cm,left=2.4cm,right=2.4cm]{geometry}
\usepackage{mathtools,amssymb}
\usepackage[usenames,dvipsnames,table]{xcolor}
\usepackage[titletoc]{appendix}
\usepackage{mathtools,amssymb}
\usepackage[makeroom]{cancel}
\usepackage{comment}
\usepackage{graphicx}
\usepackage{epstopdf}
\usepackage{mathrsfs}
\usepackage{enumitem}
\usepackage{stackrel}
\usepackage{float}
\usepackage{booktabs}
\usepackage[normalem]{ulem}  %%% to use strikeout  \sout{unwanted text}
\usepackage{cancel} %%% to use the user-defined cancel command in the mathmode
\usepackage{longtable}
\usepackage{physics}
\usepackage{empheq}
\usepackage{cite} %to generate compact citations
\usepackage[hyperfootnotes = false, colorlinks = true, linkcolor = blue, citecolor = blue]{hyperref}
\usepackage[final]{showkeys} %%% should be loaded after hyperref        %%% add final to avoid printing the labels
\usepackage{braket}
\usepackage{tikz}
\usepackage{subcaption}
\usepackage{xcolor}

\allowdisplaybreaks
\allowbreak

\renewcommand*\showkeyslabelformat[1]{%
  \fbox{\parbox[t]{0.8\marginparwidth}{\raggedright\normalfont\scriptsize\url{#1}}}}

\usepackage{etoolbox}
\makeatletter
\patchcmd{\hyper@makecurrent}{%
    \ifx\Hy@param\Hy@chapterstring
    \let\Hy@param\Hy@chapapp
    \fi
}{%
    \iftoggle{inappendix}{%true-branch
	\@checkappendixparam{chapter}%
	\@checkappendixparam{section}%
	\@checkappendixparam{subsection}%
	\@checkappendixparam{subsubsection}%
	\@checkappendixparam{paragraph}%
	\@checkappendixparam{subparagraph}%
    }{}%
}{}{ \errmessage{failed to patch}}

\newcommand*{\@checkappendixparam}[1]{%
	\def\@checkappendixparamtmp{#1}%
	\ifx\Hy@param\@checkappendixparamtmp
	\let\Hy@param\Hy@appendixstring
	\fi
}
\makeatletter

\newtoggle{inappendix}
\togglefalse{inappendix}

\apptocmd{\appendix}{\toggletrue{inappendix}}{}{\errmessage{failed to patch}}
\apptocmd{\subappendices}{\toggletrue{inappendix}}{}{\errmessage{failed to patch}}
\newcommand{\lsim}{\mathrel{\hbox{\rlap{\lower .55ex
\hbox{$\sim$}} \kern-.3em \raise.4ex \hbox{$<$}}}}
\newcommand{\gsim}{\mathrel{\hbox{\rlap{\lower.55ex
\hbox{$\sim$}} \kern-.3em \raise.4ex \hbox{$<$}}}}

\begin{document}

%%%%  Gautam's definitions    %%%%%%%

\newcommand{\partiald}[2]{\dfrac{\partial #1}{\partial #2}}
\newcommand{\be}{\[}
\newcommand{\ee}{\]}
\newcommand{\f}{\frac}
\newcommand{\s}{\sqrt}
\newcommand{\lm}{\mathcal{L}}
\newcommand{\wm}{\mathcal{W}}
\newcommand{\om}{\mathcal{O}_{n}}
\newcommand{\ep}{\epsilon}

\def\gap#1{\vspace{#1 ex}}
\def\del{\partial}
\def\eq#1{(\ref{#1})}
\def\fig#1{Fig \ref{#1}} 
\def\re#1{{\bf #1}}
\def\bull{$\bullet$}
\def\nn{\nonumber}
\def\ub{\underbar}
\def\nl{\hfill\break}
\def\ni{\noindent}
\def\bibi{\bibitem}
\def\vev#1{\langle #1 \rangle} 
\def\mattwo#1#2#3#4{\left(\begin{array}{cc}#1&#2\\#3&#4\end{array}\right)} 
\def\tgen#1{T^{#1}}
\def\half{\frac12}
\def\floor#1{{\lfloor #1 \rfloor}}
\def\ceil#1{{\lceil #1 \rceil}}

\def\Tr{{\rm Tr}}

\def\mysec#1{\gap1\ni{\bf #1}\gap1}
\def\mycap#1{\begin{quote}{\footnotesize #1}\end{quote}}

\def\Red#1{{\color{red}#1}}

\def\Om{\Omega}
\def\a{\alpha}
\def\b{\beta}
\def\l{\lambda}
\def\g{\gamma}
\def\e{\epsilon}
\def\Si{\Sigma}
\def\p{\phi}
\def\z{\zeta}

\def\lan{\langle}
\def\ran{\rangle}

\def\bit{\begin{item}}
\def\eit{\end{item}}
\def\benu{ \begin{enumerate} }
\def\eenu{ \end{enumerate} }

\def\tr{{\rm tr}}
\def\intk#1{{\int\kern-#1pt}}

%%%%%%%%%%%%%%  PRANJAL'S DEFS   %%%%%%%%%%%%%%%%%%%%%%%%%

\parindent=0pt
\parskip = 10pt

\def\al{\alpha}
\def\ga{\gamma}
\def\Ga{\Gamma}
\def\G{\Gamma}
\def\be{\beta}
\def\de{\delta}
\def\De{\Delta}
\def\ep{\epsilon}
\def\ro{\rho}
\def\la{\lambda}
\def\La{\Lambda}
\def\ka{\kappa}
\def\om{\omega}
\def\si{\sigma}
\def\th{\theta}
\def\ze{\zeta}
\def\ne{\eta}
\def\del{\partial}
\def\cdev{\nabla}

\def\gh{\hat{g}}
\def\Rh{\hat{R}}
\def\Boxh{\hat{\Box}}
\def\Kb{\mathcal{K}}
\def\phit{\tilde{\phi}}
\def\gt{\tilde{g}}
\newcommand{\h}{\hat}
\newcommand{\ti}{\tilde}
\newcommand{\sD}{{\mathcal{D}}}
\newcommand{\colored}[1]{ {\color{turquoise} #1 } }
\newcommand{\propBbd}{\mathcal{G}}
\newcommand{\propBB}{\mathbb{G}}
\newcommand{\christof}[3]{ {\Ga^{#1}}_{#2 #3}}
\def\ads{AdS$_{\text{2}}$~}
\def\GN{G$_{\text{N}}$~}
\def\zb{{\bar{z}}}
\def\fb{\bar{f}}
\def\delb{\bar{\del}}
\def\wb{\bar{w}}
\def\gb{\bar{g}}
\def\gp{g_+}
\def\gm{g_-}
\def\phit{\tilde{\phi}}
\def\mut{\tilde{\mu}}
\def\xb{\bar{x}}
\def\yb{\bar{y}}
\def\xp{x_+}
\def\xm{x_-}
\def\finv{\mathfrak{f}_i}
\def\fbinv{\bar{\mathfrak{f}}_i}
\def\gc{\mathfrak{g}}
\def\gcb{\bar{ \mathfrak{g}}}
\def\disc{\mathcal{D}}
\def\rhp{\mathbb{H}}
\def\picklemma{Schwarz-Pick lemma}
\def\mobius{M\"{o}bius~}
\def\ft{\tilde{f}}
\def\zet{\tilde{\ze}}
\def\taut{\tilde{\tau}}
\def\thet{\tilde{\theta}}
\def\slr{\ensuremath{\mathbb{SL}(2,\mathbb{R})}}
\def\slc{\ensuremath{\mathbb{SL}(2,\mathbb{C})}}
\def\nh{\hat{n}}
\def\cD{\mathcal{D}}
\def\Lfg{\mathfrak{f}}

%%%%%%%%%%%%%%  ADWAIT'S DEFS   %%%%%%%%%%%%%%%%%%%%%%%%%

\renewcommand{\real}{\mathbb{R}}
%%%%%% COMMAND FOR A BIGGER DOT

\newcommand*{\Cdot}[1][1.25]{%
  \mathpalette{\CdotAux{#1}}\cdot%
}
\newdimen\CdotAxis
\newcommand*{\CdotAux}[3]{%
    {%
	\settoheight\CdotAxis{$#2\vcenter{}$}%
	\sbox0{%
	    \raisebox\CdotAxis{%
		\scalebox{#1}{%
		    \raisebox{-\CdotAxis}{%
			$\mathsurround=0pt #2#3$%
		    }%
		}%
	    }%
	}%
    % Remove depth that arises from scaling.
	\dp0=0pt %
    % Decrease scaled height.
	\sbox_2{$#2\bullet$}%
	\ifdim\ht_2<\ht0 %
	\ht0=\ht_2 %
	\fi
    % Use the same width as the original \cdot.
	\sbox_2{$\mathsurround=0pt #2#3$}%
	\hbox to \wd2{\hss\usebox{0}\hss}%
    }%
}

%%%%%% COMMAND TO CANCEL INSIDE THE MATHMODE (REQUIRES PACKAGE CANCEL)
\newcommand\hcancel[2][black]{\setbox0=\hbox{$#2$}%
\rlap{\raisebox{.45\ht0}{\textcolor{#1}{\rule{\wd0}{1pt}}}}#2}

\renewcommand{\arraystretch}{2.5}%
\renewcommand{\floatpagefraction}{.8}%

%%%%%%%%%%  This paper %%%%%%%%%%%%%%%
\def\newthing{\marginpar{{\color{red}****}}}
\reversemarginpar
\def\bz{{\bar z}}
\def\by{{\bar y}}
\def\bw{{\bar w}}
\def\delb{{\bar \del}}
\def\bep{{\bar \epsilon}}

\hypersetup{pageanchor=false}
\begin{titlepage}
    \begin{flushright}
	 TIFR/TH/19-34
    \end{flushright}

    \vspace{.4cm}
    \begin{center}
	\noindent{\Large \bf{Quantum quench and thermalization to GGE in arbitrary dimensions and the odd-even effect}}\\
	\vspace{1cm}
        Parijat Banerjee$^b$\footnote{pbanerjee5@students.iiserpune.ac.in; work started during a research project in TIFR},
	Adwait Gaikwad$^a$\footnote{adwait@theory.tifr.res.in},
	Anurag Kaushal$^a$\footnote{anurag.kaushal@theory.tifr.res.in}
	and Gautam Mandal$^a$\footnote{mandal@theory.tifr.res.in} 

	\vspace{.5cm}
	\begin{center}
	    {\it a. Tata Institute of Fundamental Research, Mumbai 400005, 
	    India.}\\
    \vspace{.5cm} {\it b. Johns Hopkins University
    	3400 N. Charles Street
    	Baltimore, MD 21218, USA.}\\
	\end{center}

	\gap2

	\centerline{version \today}

    \end{center}

%%%%%%%%%%%%%%%%%%%%%%%%%%%%%%%
%%%%%%%%%%%%  ABSTRACT %%%%%%%%%%%%
%%%%%%%%%%%%%%%%%%%%%%%%%%%%%%%
    \begin{abstract}

In many quantum quench experiments involving cold atom systems the
post-quench system can be described by a quantum field theory of free
scalars or fermions, typically in a box or in an external potential. We work with free scalars in arbitrary dimensions generalizing the techniques employed in our earlier work \cite{Mandal:2015kxi} in 1+1 dimensions. In this paper, we generalize to $d$ spatial dimensions for arbitrary $d$. The system is considered in a box much larger than any other scale of interest. We start with the ground state, or a squeezed state, with a high mass and suddenly quench the system to zero mass (``critical quench'').  We explicitly compute
time-dependence of local correlators and show that at long times they
are described by a 
generalized Gibbs ensemble (GGE), which, in special cases, reduce to a thermal (Gibbs) ensemble. The equilibration of {\it local} correlators can be regarded as `subsystem thermalization' which we simply call 'thermalization' here (the notion of thermalization here also includes equlibration to GGE). The rate of approach to
equilibrium is exponential or power law depending on whether $d$ is odd or even
respectively.  As in 1+1
dimensions, details of the quench protocol affect the long time
behaviour; this underlines the importance of irrelevant operators at IR in
non-equilibrium situations. We also discuss quenches from a high mass
to a lower non-zero mass, and find that in this case the approach to equilibrium
is given by a power law in time, for all spatial dimensions $d$, even or odd.

\end{abstract}

\end{titlepage}

\pagenumbering{roman}
\tableofcontents
\enlargethispage{1000pt}
\pagebreak
\pagenumbering{arabic}
\setcounter{page}{1}

%%%%%%%%%%%%%%%%%%%%%%%%%%%%%%%
%%%%%%%%% INTRODUCTION %%%%%%%%%%
%%%%%%%%%%%%%%%%%%%%%%%%%%%%%%%
\section{Introduction and Summary}\label{sec:Intro}

Thermalization in integrable systems\footnote{We use the word
  ``thermalization'' here to mean asymptotic approach of a local
  correlator in a time-dependent pure state to its value in a
  generalized Gibbs ensemble (GGE). Sometimes the word
  ``thermalization'' is reserved only for cases where the asymptotia
  corresponds to a Gibbs ensemble; we will not make this
  distinction. Since we specifically refer to equilibration of {\it
    local correlators}, such equilibriation to GGE may be called
  ``subsystem equilibration'' or, in our generalized sense,
  ``subsystem thermalization''.} has grown to be an important area of
study in recent years \cite{Rigol:2007, Rigol:2007a, Barthel:2008GGE,
  Cramer:2008GGE, Mussardo:2009GGE, Iucci:2010GGE, Calabrese:2012GGE,
  Calabrese:2011GGE, Calabrese:2012GGE-II, Essler:2014GGE,
  Essler:2014qza, gogolin2015equilibration}. Integrable systems
possess an infinite number of conservation laws and are often
described by free (quasi)particles; thus, to understand thermalization
in integrable models, it is important to understand it in free field
theories. Besides the theoretical motivation, it turns out that in
many recent quantum quench experiments involving cold atom systems,
the post-quench phase can be described by a quantum field theory of
free scalars or fermions \cite{scully1997quantum, squeezed-2011}. In a
previous work \cite{Mandal:2015kxi} such systems were studied at
length where the spatial dimension $d$ of the system was $d=1$ and the
post-quench hamiltonian was taken to be critical. Both bosonic and
fermionic systems were studied with explicit quench protocols. Among
other things, it was found that (i) the post-quench state can be
represented by a generalized Calabrese-Cardy state, characterized by
an infinite number of $W_\infty$ charges, as postulated in earlier
works \cite{Mandal:2015jla, Cardy:2015xaa}, (ii) correlators of local
operators as well as the reduced density matrix of finite subsystems
asymptotically approached that of a generalized Gibbs ensemble (GGE),
(iii) the relaxation was exponential, characterized by a rate given by
conformal and $W_\infty$ algebraic properties of the operators in
question. Furthermore, it was found that (iv) it is possible to start
with specially tailored squeezed states so that the post-quench state
is a simple Calabrase-Cardy state, characterized by only one conserved
quantity--- namely the energy; in this case the equilibrium is
described by the familiar Gibbs ensemble and the relaxation rate is a
simple function of the energy and the conformal dimension of the
operator.

It is natural to ask, both from an experimental as well as a
theoretical point of view: (a) which of the above results, if any,
generalize to $d$ spatial dimensions for arbitrary $d$?, and, (b) in
particular, if there {\it is} an asymptotic equilibrium, what is the
rate of aproach to the equilibrium? There are several reasons why the
answer to these questions is not easy to guess. First of all, the
proof of thermalization, for 1+1 dimensional conformal field theories
in \cite{Mandal:2015jla}, relies on conformal maps which map a strip
of the complex plane to an upper half plane. The quantitative aspects
such as the relaxation rate, also arise from such conformal maps.  In
$d+1$ dimensions, there is no conformal map which maps the Euclidean
time circle (or interval) times $R^d$ to (a part of) $R^{d+1}$ for
$d>1$ \footnote{We thank Abhijit Gadde for a useful discussion in this
  regard.} Furthermore, \cite{Mandal:2015jla} extensively used
properties of $W_\infty$ algebra which are specific to 1+1 dimensions,
to derive thermalization to GGE. In addition, in holographic contexts,
the GGE in 1+1 dimensions finds an interpretation in terms of 3D
higher spin black holes which have an infinite dimensional algebra of
large diffeomorphisms (to be precise, $W_\infty$). There is no obvious
generalization of these to higher dimensions.

The explicit computations for free scalars and fermions,
studied in \cite{Mandal:2015kxi}, did not directly use the conformal
map or the $W_\infty$ algebra, but in terms of detail, used several
features specific to 1+1 dimensions.

In this paper we consider quantum quench of the mass parameter of free
scalar field theories in arbitrary dimensions (both to zero and
non-zero final mass) and explicitly compute time-dependence of local
correlators. For simplicity of the computation, we consider sudden
quench for most part of the paper, although our method is applicable
to more general quench protocols. A summary of our results is as
follows:

\begin{enumerate}

\item General results for mass quench ($m_{in} \to m_{out}$)

    \begin{enumerate}

  \item If we start with the ground state of the
    initial Hamiltonian, the post-quench state turns out to be
    equivalent to a generalized Calabrese-Cardy (gCC) state, which is
    a boundary conformal state with a cut-off applied to each of an
    infinite family of commuting conserved charges. If we start with
    specially tailored squeezed states, the post-quench state turns
    out to be an ordinary Calabrese-Cardy (CC) state, namely a
    conformal boundary state with a cut-off only on the total energy
    of the system. 

  \item We show that at long times (a) the correlators, starting from
    a CC state, are described by a thermal (Gibbs) ensemble, and (b)
    the correlators, starting from a gCC state are described by a
    generalized Gibbs ensemble (GGE).

  \item The GGE is characterized by chemical potentials which are
    determined by the infinite number of conserved charges of the
    initial gCC state. As found and remarked about in
    \cite{Mandal:2015kxi}, this implies not only a retention of
    detailed memory of the initial state, but also an apparent
    contradiction of naive intuition that higher dimensional conserved
    charges \eq{charges} should not affect long time behaviour. This
    naive Wilsonian intuition, which perfectly works at low energies,
    does not work in situations where the initial state is a high energy state.

\end{enumerate}

    %\begin{enumerate}

  \item Quench to zero mass: $m_{in}\to 0$ ({\it critical quench})

    \begin{enumerate}
      
  \item The rate of approach to equilibrium is exponential or power
    law depending on whether the number of spatial dimensions, $d$, is
    odd or even, respectively. This is one of the main results of the
    paper.

  \item
    In the case where the post-quench state is a CC state, there is a
    geometric understanding of thermalization in terms of the method
    of images, which also elucidates the odd-even effect.
    
  \item
    Unlike in 1+1 dimensions, the decay exponents are not uniquely
    defined by the conformal dimension of the operator. For example,
    in a given odd spatial dimension $d>1$, where $\vev{\phi \phi}$
    and $\vev{\partial_i\phi \partial_i\phi}$ both decay
    exponentially, with the same decay exponent.

  \item
    As in 1+1 dimensions, details of the quench protocol affect the
    long time behaviour. This underlines the importance of irrelevant
    operators at long times in non-equilibrium situations involving
    high energies.

\end{enumerate}

%    \begin{enumerate}
      
\item Quench to small non-zero mass: In case the system is quenched
  from a high mass to a low non-zero mass, exact time-dependence of
  correlators can again be computed. These generically approach a GGE
  with an oscillatory power law behaviour, with a period of
  oscillation given by the final mass. There is no odd-even effect unlike
  for the massless case mentioned above.

\end{enumerate}

We summarize the main points a bit more precisely in a tabular form below.

\begin{table}[H]
  \begin{center}
    Long time behaviour of equal time two-point functions $G(t)$
    \\
    ~\\
    \begin{tabular}{|c|c|c|}
    \hline & $m_{out}=0$ & $m_{out}\not= 0 \ll m_{in}$ \\ \hline $d=1$ &
    \color{red}{$e^{-\gamma t}$} & $\frac{\cos\left(2 m_{out} t+ \delta \right)}{t^{1/2}}$
    \\ \hline $d=2$ & \color{red}{$\frac1{t^2}$}& $\frac{\cos \left(2 m_{out} t+ \delta
    \right)}{t}$ \\ \hline odd $d>1$ & $
    e^{-\gamma t}$& $\frac{\cos\left(2 m_{out} t+ \delta \right)}{t^{d/2}}$ \\ \hline even $d>2$ & $\frac1{t^{d-2}}$ & ,,
    \\ \hline
  \end{tabular}\caption{\label{tab:overall} Here $d$ refers to the number of
      spatial dimensions. The table displays the structure of an equal
      time two-point function $G(t)$.  For $d>2$, $G(t) \equiv \lan
      \phi(x, t) \phi(y, t)\ran$; for {\color{red}{$d=1, 2,
          m_{out}=0$}} this quantity diverges at long times, hence we
      display $G(t)=\lan \del_t \phi(x, t) \del_t\phi(y, t)\ran$ (in
      all cases, we drop the spatial dependence). The decay
      coefficients $\gamma$ are given in the text (see e.g. Table \ref{tab:critical-phi} which lists various values of $\gamma$ for $d=3$: these values depend on the specific initial state); the exponential
      decay has a power law prefactor in case of ground state
      quench. The phase factors $\delta$ are given by $d\,\pi/4$.
      Note that the power laws in the massless case (for even
      dimensions) differ from those in the massive case.}
 \end{center}
\end{table}
   
The paper is organized as follows: 

In section \ref{sec:protocol} we discuss the main formalism of how to
describe the post-quench state as a Bogoliubov transform of the
out-vacuum, leading to its identification (with the help of Appendix
\ref{app:dirichlet}) as a generalized Calabrese-Cardy state
\eq{def-gcc}. We discuss starting with ground states as well as
squeezed states in the massive phase. Section \ref{sec:squeezed-2pt}
gives the formulae for the time-dependent two-point functions.

In section \ref{sec:time-dep-2pt} we explicitly calculate the
time-dependent part of the two-point function (additional material is
provided in Appendices \ref{app:recursion} and
\ref{app:crit-quench}). Tables \ref{tab:critical-phi},
\ref{tab:critical-del-phi} and \ref{tab:critical-del-phi-2} show the
decay of the time-dependent part for critical quench ($m_{out}=0$) for
$d=1,2,3,4$; the results for general dimensions $d$ are obtained by
using a recursion relation. For non-critical quench ($m_{out}\not=0$),
Tables \ref{tab:massive-phi-gcc}, \ref{tab:massive-phi} and
\ref{tab:massive-del-phi} present the time-depedent part of the
two-point functions for $d=1,2,3,4$; the formula for the general
dimension is given in \eq{eq:massive-2pt-general}. The overall
behaviour is as indicated in Table \ref{tab:overall} above.

The fact that the time-dependent parts of the two-point functions
vanish at large times, already implies that these correlators (and
hence all correlators, by the application of Wick's theorem),
asymptotically equilibrate.  In section
\ref{sec:thermalization-to-GGE}, using results from Appendix
\ref{app:gge}, we explicitly show that the time-independent parts of
the two-point functions, to which they equilibrate, are given by
two-point functions in a generalized Gibbs ensemble(GGE). This
constitutes a proof of thermalization of arbitrary local correlators
(see Section \ref{sec:eq-to-gge}).

In section \ref{sec:approach-thermalization} we look at the
geometrical interpretation of two-point functions in the CC state,
which allows up to have a better understanding of the odd-even effect.

In section \ref{sec:KK} we calculate the GGE correlator for purely
spatial separation for the simple case of the thermal ensemble and a
critical quench. The calculation has an interpretation in terms of a
Kaluza-Klein reduction along the thermal circle. For high enough
temperature, or equivalently large enough spatial sepation, only the
Kaluza-Klein zero mode contribution survies, which has a power law
behaviour. This result holds in any dimension, even or odd.

In section \ref{sec:discussions} we conclude with some comments and
discussion of future directions.

%%%%%%%%
\section{Quantum quench in free scalar theories}\label{sec:protocol}

\begin{figure}[H]
\centering
\begin{subfigure}{.5\textwidth}
  \centering
  \includegraphics[width=.6\linewidth]{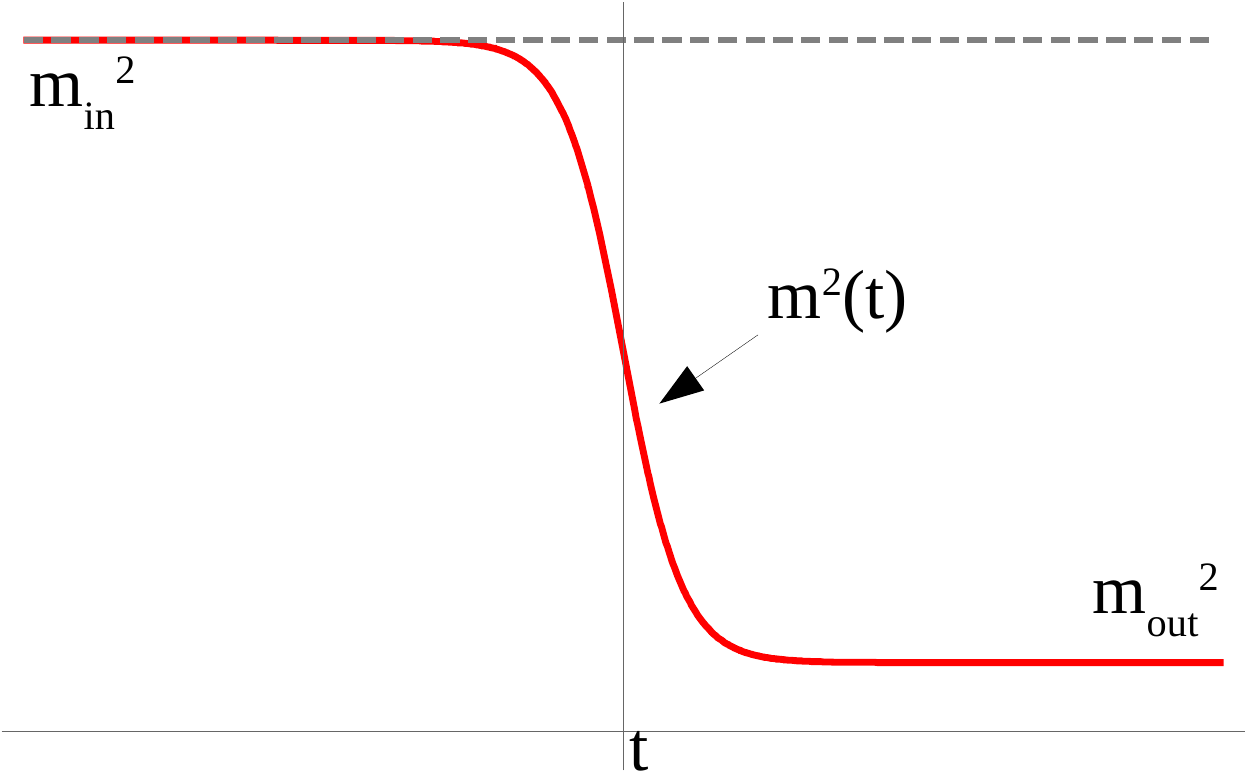}
  \caption{\label{fig:1a}Mass quench}
\end{subfigure}%
\begin{subfigure}{.5\textwidth}
  \centering
  \includegraphics[width=.6\linewidth]{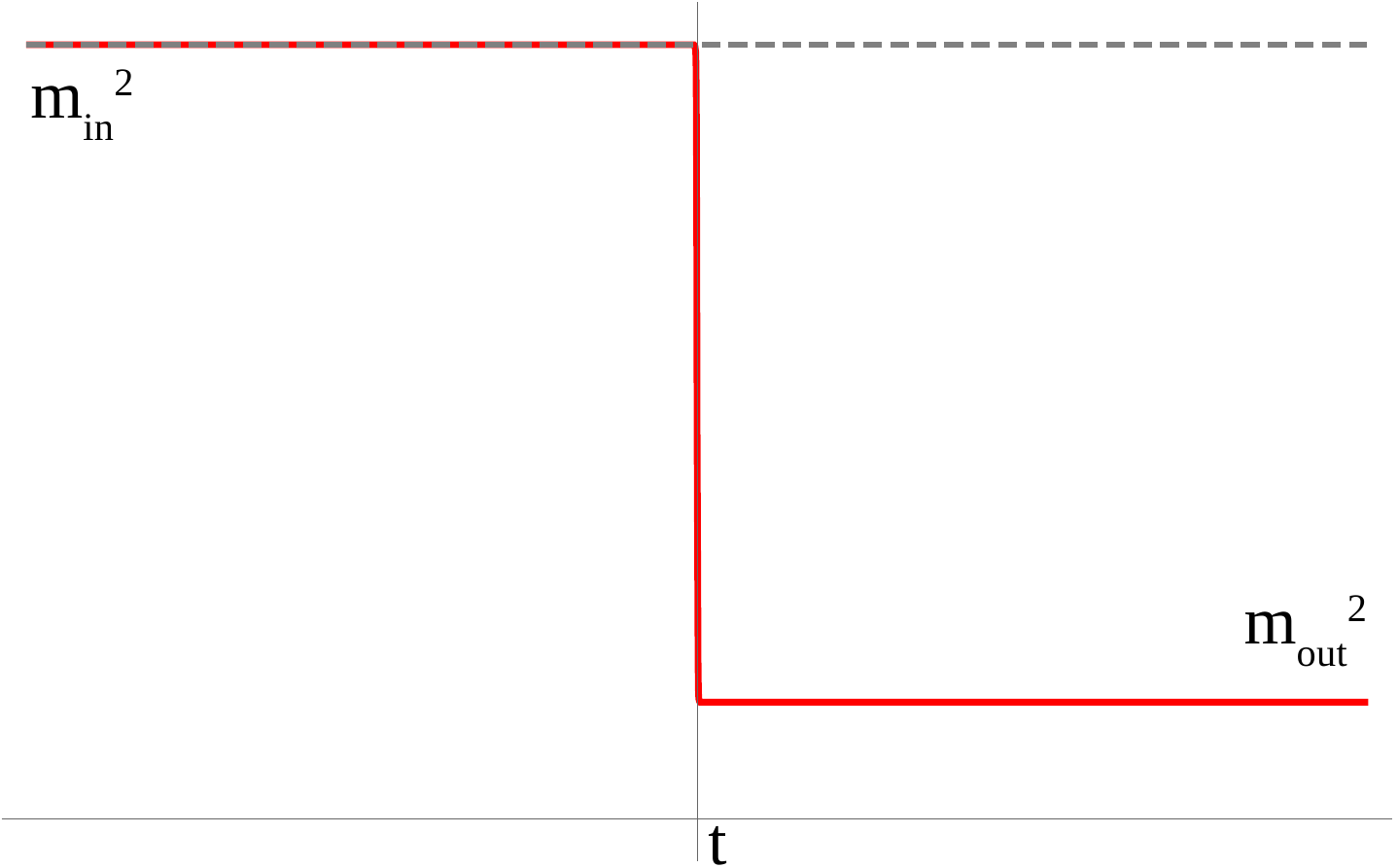}
  \caption{\label{fig:1b}Quench in sudden limit}
\end{subfigure}
\caption{Quantum quench of mass of a scalar field in $d+1$ dimensions with the protocol in eq. (\ref{eq:mass-profile}). This is shown in the left panel. For simplicity, we will consider a sudden quench for most of the paper, in which case, the mass profile looks like the figure on the right.}
\label{fig:quench}
\end{figure}
The basic set-up is as follows. Consider a relativistic scalar free field
theory in $d$ spatial dimensions, with a time dependent mass (see fig.
\ref{fig:quench})
\[
  {\cal L}=\f12\left( \del_t \phi \del_t \phi - \del_i \phi
  \del_i \phi - m^2(t) \phi \phi \right) ;\qquad i=1,\cdots,d
\]
In Fourier space,
\[
 \phi(\vec{x},t) = \int \frac{d^d k}{(2 \pi)^d} e^{i\vec{k}\cdot \vec{x}}
 \phi(\vec{k},t)
\]
the equation of motion for the Fourier mode $\phi(\vec{k},t)$
(similar equation for $\phi^*(\vec{k},t)$) is
\[
-\del_t^2 \phi(\vec{k},t)- m^2(t)\phi(\vec{k},t)= |\vec{k}|^2
\phi(\vec{k},t)
\]
For every $\vec{k}$, this can be identified with a Schrodinger
problem on a line (coordinatized by $y$, say), with the
identifications
\[
t \to y, m^2(t) \to -V(y), |\vec{k}|^2 \to E, \phi(\vec{k},t) \to  \psi_E(y)
\]
In the Schrodinger problem there are two equivalent bases of
solutions: one which corresponds to particles coming in from the left:
$\psi_E(y) =$ linear combination of $u_{in}(E,y), u_{in}^*(E,y)$,
where $u_{in}(E,y) \sim e^{i \om_{in} y}$ as $y\to -\infty$, and
another which corresponds to particles coming in from the right,
$u_{out}(y) \sim e^{-i \om_{out} y}$ as $y\to +\infty$. Taking cue
from this, we have two sets of normal mode expansions:
\begin{align}
&\phi(\vec{k},t)= a_{in}(\vec{k},t) u_{in}(\vec{k},t) + cc, \; u_{in}
\sim e^{-i\om_{in} t}, t\to -\infty, \nonumber\\
&\phi(\vec{k},t)= a_{out}(\vec{k},t) u_{out}(\vec{k},t) + cc, \; u_{out}
\sim e^{-i\om_{out} t},t\to \infty \nonumber\\
&\om_{in}= \sqrt{|\vec{k}|^2 + m_{in}^2}, \; \om_{out}= \sqrt{|\vec{k}|^2
+ m_{out}^2} \label{om-in-out}
\end{align}
Here. The two basis sets are of course linear combinations of each other
\[
u_{in}(\vec{k},t)=  \al(\vec{k}) u_{out}(\vec{k},t)+ \be(\vec{k})
u_{out}^*(-\vec{k},t)
\]
which implies
\begin{align}
 a_{in}=&\alpha^*(\vec{k})a_{out}(\vec{k})-\beta^*(\vec{k})
 a_{out}^\dagger(-\vec{k}) \nonumber \\
 a_{out}=&\alpha(\vec{k})a_{in}(\vec{k})+\beta^*(\vec{k})
 a_{in}^\dagger(-\vec{k}) \nonumber
\end{align}

Here $\al, \be$ are the Bogoliubov coefficients. We take the mass profile to be
(see Fig \ref{fig:1a})
\begin{equation}
    m^2(t)= \frac{1}{2}(m_{in}^2 +m_{out}^2) - \frac{1}{2}
    (m_{in}^2-m_{out}^2)\tanh(\rho t)\label{eq:mass-profile}
\end{equation}
Further we do all our calculations in the sudden limit $\rho\rightarrow\infty$\footnote{This limit has to be understood in the sense explained in detail in \cite{Mandal:2015kxi}}(see Fig \ref{fig:1b}) . 
For the quench protocol eq. (\ref{eq:mass-profile}), Bogoliubov coefficients are easy to compute explicitly \cite{birrell1984quantum}; in the sudden limit
they become \cite{Mandal:2015kxi}
\begin{align}
 \alpha(k)&=\frac{1}{2}\frac{\om_{out}+\om_{in}}{\sqrt{\om_{out}
 \om_{in}}}, \; \beta(k)=\frac{1}{2}\frac{\om_{out}-\om_{in}}
 {\sqrt{\om_{out}\om_{in}}}
 \label{ab-sudden}
\end{align}
In this limit the in- and out- waves also become especially simple
\begin{align}
u_{in}(k,t)= \frac{e^{-i\omega_{in}t}}{\sqrt{2 \omega_{in}}},\;
u_{out}(k,t) =\frac{e^{-i\omega_{out}t}}{\sqrt{2 \omega_{out}}}
\label{wave-sudden}
\end{align}
The final mass is arbitrary but taken to be less than the inital mass i.e. $m_{out}< m_{in}$. We also study the theory when the final mass is taken to be zero, when the final theory is critical. As we will see the more interesting results are obtained in the critical quench.

%%%%
\subsection{Quantum quench from the ground state}\label{sec:qq-gr}

A natural pre-quench initial state (just before $t=0$) is the ground
state $|0_{in}\rangle$ of the initial Hamiltonian $H_{in}$, which is
also the zero-particle state defined by the oscillators
$a_{in}|0\ran_{in}=0$. Just after $t=0$, the state remains $|\psi(0)
\rangle = |0\rangle_{in}$ (remember we are working in the sudden
quench limit, therefore the state has no time to change). We then
define $|\psi(t)\rangle = e^{-iHt}|\psi(0)\rangle$ where $H:=H_{out}$
is the final Hamiltonian.

The `in' ground state can be written in terms of the out ground state
through a Bogoliubov transformation \footnote{This is easily checked by
noting that the right hand side is annihilated by $a_{in}
=\alpha^*(\vec{k})a_{out}(\vec{k})-\beta^*(\vec{k})a_{out}^\dagger(
-\vec{k})$}
\[
 |\psi(0)\rangle= |0_{in}\rangle = \exp[\frac{1}{2} \sum_{\vec{k}}
  \gamma(k) a_{out}^\dagger (\vec{k}) a^\dagger_{out}(-\vec{k})]|0_{out}
  \rangle 
\]
where $\gamma(k)=\beta^*(k)/\alpha^*(k)$ depends on the Bogoliubov
coefficients and is only a function of $k:=|\vec{k}|$.\footnote{Note
that, because we focus on homogenous quench protocols in this paper,
we have rotational invariance in the $d$ spatial dimensions. Hence
the Bogoliubov coefficients are a function of $k=|\vec k|$.
\label{ftnt:rot-inv}} Further the `in' ground state is also related
to the Dirichlet boundary state \cite{Mandal:2015jla} (see appendix
\ref{app:dirichlet} for more details)
\[
 |D\rangle = \exp[-\frac{1}{2} \sum_{\vec{k}} a^\dagger_{out}(\vec{k})
 a^\dagger _{out}(-\vec{k})] |0_{out}\rangle
\]
through the relation
\[
 |0_{in}\rangle =\exp[\frac{1}{2}\sum_{\vec{k}}\kappa(k) a_{out}^\dagger
 (\vec{k}) a_{out}(\vec{k})]|D\rangle
 \label{eq:ground-as-gCC}
\]
where
\begin{equation}
 \kappa(k)=-\frac{1}{2}\log(-\gamma(k))\label{eq:kappa-gamma}
\end{equation}
In general one can expand $\kappa(k)$ for small $k$:
\[
\kappa(k)= \sum_{i=1}^\infty \kappa_i k^{i-1}= \kappa_1 + \kappa_2 k
+ \kappa_3 k^2 + ...
\]
With this, the expression for $|0_{in}\ran$ becomes
\begin{equation}
|\psi(0)\rangle= |0_{in}\rangle= \exp[-\sum_{i=1}^{\infty}\kappa_i Q_i]
|D\rangle \label{massive-gcc}
\end{equation}
where
\begin{align}
 Q_{i} = \sum_{\vec{k}} |\vec{k}|^{i-1}a^\dagger_{out}(\vec{k})a_{out}(\vec{k})
 \label{charges}
 \end{align}
are consereved charges (they obviously all commute with the `out'
Hamiltonian $H=\sum_{\vec{k}} $\\ $\sqrt{|\vec{k}|^2+m_{out}^2}
a^\dagger_{out}(\vec{k})a_{out}(\vec{k})$).\footnote{For free scalars,
  the number operators $N(k)=a^\dagger_{out}(\vec{k})a_{out}(\vec{k})$
  are themselves conserved, and provide an alternative basis of the
  algebra of conserved charges.}
It is easy to explicitly compute the
$\kappa_i$ coefficients by using the definition of $\kappa(k)$ in
terms of the Bogoliubov coefficients which are given in
\eq{ab-sudden}:
\begin{align}
  \kappa(k)= \f12 \log\left(\f{\sqrt{k^2+ m^2}+ \sqrt{k^2+ m_{out}^2}}
  {\sqrt{k^2+ m^2}- \sqrt{k^2+ m_{out}^2}}\right)
  \label{kappa-gr}
\end{align}
Note that henceforth we will call $m_{in}=: m$ for simplicity. It is assumed
that $m> m_{out}$. From the above equation, the small $k$ expansion
can be easily found.\\
For $m_{out}=0$, the expansion contains only odd powers
of $k$, thus the even $\kappa_i$'s are
non-zero, e.g. 
\begin{align}
  \kappa_2= \f1{m},\; \kappa_4= -\f1{6 m^3}, ...
  \label{cr-gr-kappas}
\end{align}
By contrast, for $m_{out}\not=0$,  only even powers of k survive,
leading to odd $\kappa_i$'s, e.g.
\begin{equation}
 \kappa_1= \frac{1}{2}\log\frac{m+m_{out}}{m-m_{out}},\; \kappa_3= \frac{1}{2 m m_{out}}, ...
 \label{ma-gr-kappas}
\end{equation}

\paragraph{Relation to a generalized Calabrese-Cardy (gCC) state}

In case of critical quench ($m_{out}=0$), the post-quench dynamics is
conformal. For critical quenches leading to a generic (non-integrable)
conformal field theory, Calabrese and Cardy [CC:2004-2005] postulated
the following form for the post-quench wavefunction\footnote{One
  assumes here that the quench takes a finite time to end, say
  $t=t_0$.  In this paper, we are interested in a sudden quench which
  takes place at $t=0$; thus $t_0=0$.\label{ftnt:finite-time}}
\begin{align}
|CC\rangle= \exp[-\kappa H]|Bd\rangle
\label{def-cc}
\end{align}
where, $\kappa$ is given by the inverse of the mass gap characterizing
the initial state (before the quench). The state $|Bd \ran$ is a
``boundary state'' representing a conformal state subject to an
appropriate boundary condition at imaginary time $t=-i\kappa$.

In case the post-quench conformal theory is integrable, characterized
by an infinite number of charges $Q_i$, a more appropriate ansatz for
the post-quench state is the gCC (generalized Cardy-Calabrese) state
\cite{Mandal:2015jla, Mandal:2015kxi}
\begin{align}
|gCC\rangle= \exp[-\sum_{i=1}^\infty \kappa_i Q_i]|Bd\rangle
\label{def-gcc}
\end{align}
The state \eq{massive-gcc} is clearly a special case of such a state,
which was first found in \cite{Mandal:2015kxi} in 1+1 dimensions. Note
that the Dirichlet state $|D\ran$ is a particular example of a
boundary state. As mentioned above \eq{cr-gr-kappas}, here only the
even charges $Q_{2n}, n>0$ are non-zero; note that $Q_2= H$, as can be
seen from the definition \eq{charges}.

In case of a massive quench, the state \eq{massive-gcc} is of the form
\eq{def-gcc}, where the conformal boundary state $|Bd\ran$ is replaced
by the Dirichlet boundary state $|D\ran$. Although the post-quench
theory is not conformal in this case, we will continue to use the name
gCC for the state \eq{massive-gcc}. In fact we will continue to use
this nomenclature also for states like \eq{massive-gcc-squeezed}
obtained by noncritical quench from squeezed states.

\subsubsection{Two-point functions}\label{sec:gr-2pt}
We will be interested in computing quantities like
\begin{equation}
\lan \psi(t_1)| O(x_1) O(x_2)| \psi(t_2) \ran
= \braket{\psi(0)|O(\vec{x_1},t_1)O(\vec{x_2},t_2)|\psi(0)}
\end{equation}
The operators appearing on the RHS $O(x,t)= e^{i H_{out} t} O(x) e^{-i H_{out} t}$ are in the Heisenberg picture defined with the Hamiltonian $H_{out}$. This is applicable for $t\ge 0$. For time evolution to $t<0$ we must use the Hamiltonian $H_{in}$. With this understanding it is clear that the above definition comes with the prescription $t_1,t_2\ge 0$. The relation to time-ordered correlator is noted below \eq{eq:ground-quench}. This definition includes in particular, the equal-time correlator (ETC)
\begin{equation}
 \braket{0_{in}|O(\vec{x_1},t)O(\vec{x_2},t)|0_{in}}
\end{equation}
In our theory these are related to the ground-state two-point function
\begin{align}
  &G(\vec{x_1},t_1;\vec{x_2},t_2) \equiv
  \braket{0_{in}|\phi(\vec{x_1},t_1)\phi(\vec{x_2},t_2)|0_{in}} = \int
  \frac{d^d k}{(2 \pi)^d} u_{in}(\vec{k},t_1)u_{in}^*(\vec{k},t_2)e^{i\vec{k}.(\vec{x_1}-\vec{x_2})} \qquad
  \nonumber \\
  &= \int \frac{d^dk}{(2 \pi)^k} \left[|\alpha(k)|^2 u_{out}(k,t_1) u^*_{out}(k,t_2) + \alpha(k) \beta^*(k) u_{out}(k,t_1) u_{out}(-k,t_2)  \qquad \qquad \qquad  \right. \nonumber \\ 
 &~~~\left. + \alpha^*(k) \beta(k) u^*_{out}(-k,t_1) u^*_{out}(k,t_2) + |\beta(k)|^2 u^*_{out}(-k,t_1) u_{out}(-k,t_2)\right]e^{ik.(x_1-x_2)}
 \label{2-pt}
\end{align}
Using the expressions (\ref{ab-sudden}) and (\ref{wave-sudden}), the ground-state two-point function becomes
\begin{align}
  &\braket{0_{in}|\phi(\vec{x_1},t_1)\phi(\vec{x_2},t_2)|0_{in}} = \int \frac{d^d k}{(2\pi)^d} \frac{e^{i \vec{k}\cdot (\vec{x_1}-\vec{x_2})}}{4 \sqrt{\left(k^2+m^2\right)}\left(k^2+m_{out}^2\right)}\Bigg[\left(2 k^2+m^2+m_{out}^2\right) \cos \left(\sqrt{k^2+m_{out}^2} (t_1-t_2)\right) \nonumber\\
  &+ \left(m_{out}^2-m^2\right) \cos \left(\sqrt{k^2+m_{out}^2} (t_1+t_2)\right) -2i\sqrt{\left(k^2+m^2\right)\left(k^2+m_{out}^2\right)} \sin (\sqrt{k^2+m_{out}^2} (t_1-t_2))\Bigg]
  \label{eq:ground-quench}
\end{align}
As mentioned above this correlator is not time-ordered. The time-ordered correlator can be obtained from this by replacing $t_1-t_2$ by $|t_1-t_2|$ but leaving $t_1+t_2$ unaltered. As a consequence, the ETC's considered here are already time-ordered. The expression above simplifies for $t_1=t_2$ (ETC):
\begin{align}
  \braket{0_{in}|\phi(\vec{x_1},t)\phi(\vec{x_2},t)|0_{in}} = & \int \frac{d^d k}{(2\pi)^d} \frac{e^{i \vec{k}\cdot (\vec{x_1}-\vec{x_2})}}{4\sqrt{k^2+m^2} \left(k^2+m_{out}^2\right)} \nonumber \\&\Bigg[\left(m_{out}^2-m^2\right) \cos(2t \sqrt{k^2+m_{out}^2})+2 k^2+m^2+m_{out}^2 \Bigg]
  \label{eq:ground-quench-etc}
\end{align}
We also look at the 2-point function of the $\partial_t \phi$
operator.  Note that this correlator is directly obtainable from the
unequal time two-point function $\lan \phi \phi \ran$
\eq{eq:ground-quench} by applying the operator
$\del_{t_1}\del_{t_2}|_{t_1=t_2=t}$. Since the unequal time correlator
is of the form $f(t_1-t_2) + g(t_1 + t_2)$, it follows that $\lan
0_{in}|\partial_{t_1}\phi(\vec{x_1},t_1)$
$\partial_{t_2}\phi(\vec{x_2},t_2)|0_{in}\ran$ is of the form $-
f''(t_1-t_2) + g''(t_1+t_2)$. Hence the equal-time two-point function
of $\del_t\phi$ is of the form $-f''(0) + g''(2t)$; in particular, the
time-dependent part of this correlator is $\del_t^2$
applied to the time-dependent part of \eq{eq:ground-quench-etc}.
By explicit calculation, one finds the expression
\begin{align}
  \lan 0_{in}|\partial_{t}\phi(\vec{x_1},t)\partial_{t} \phi(\vec{x_2},t)&|0_{in}\ran = \int \frac{d^d k}{(2\pi)^d}
  \frac{e^{i \vec{k}\cdot (\vec{x_1}-\vec{x_2})}}{4\sqrt{k^2+m^2}}
  \nonumber \\
  & \Bigg[-\left(m_{out}^2-m^2\right) \cos \left(2 t \sqrt{k^2+m_{out}^2}
  \right)+2 k^2+m^2+m_{out}^2 \Bigg]
  \label{eq:ground-quench-etc-delt}
\end{align}
which indeed follows the general form $-f''(0) + g''(2t)$ mentioned
above.

A primary motivation for considering two-point functions of the
derivative operators $\lan\del_t \phi \del_t \phi\ran$ is as follows.
In low dimensions the time-dependent part $g(2t)$ of the equal time
correlator $\lan\del_t \phi \del_t \phi\ran$ \eq{eq:ground-quench-etc}
grows in time which masks the transients that signal thermalization
(it grows linearly in time in $d=1$, and logarithmically in
$d=2$). The extra time derivatives, leading to $g''(2t)$, get rid of
these divergences.

For the same reason, we may also consider the correlators of the
spatial derivatives $\vec{\partial}\phi$,
\[
\lan 0_{in}| \del_i \phi(\vec x_1, t_1) \del_j \phi(\vec x_2, t_2)
|0_{in}\ran = \del_{x_{1i}}\del_{x_{2j}} \lan \phi(\vec x_1, t_1) 
\phi(\vec x_2, t_2) \ran,
\]
The extra derivatives ensure decay with increasing $|\vec x|$ (in
$d=1,2$ $\lan \phi \phi \ran$ grows, respectively, linearly and
logarithmically with $|\vec x|$). It is enough for this purpose to
focus on
\begin{align}
  \lan  0_{in}|\del_i \phi(\vec x_1, t) \del_i \phi(\vec x_2, t)|
   0_{in}\ran
= - \del_i\del_i \lan \phi(\vec x, t_1) \phi(\vec 0, t_2) \ran,
\;\; \vec x= \vec x_1 - \vec x_2
\label{deli-deli}
\end{align}
We will discuss more details of two-point functions of these
derivative operators and their relations in Section \ref{sec:squeezed-2pt}
below. 
 
%%%%

\subsection{More general quantum quench: from squeezed states}\label{sec:squeeze}

As we saw above, the post quench state (\ref{massive-gcc}) is built out of infinite number of chemical potentials acting on the Dirichlet boundary state. However certain special gCC states can be obtained if the initial state are chosen to be specific squeezed states of the pre-quench Hamiltonian
\[
|\psi(0)\rangle = |f\rangle_{in} \equiv \exp \left[\frac{1}{2}\sum_k f(\vec{k}) a^\dagger_{in}(\vec{k}) a^\dagger_{in}(-\vec{k})\right]|0_{in}\rangle
\]
This is just the Bogoliubov transformation of $|0_{in}\rangle$. As the $|0_{in}\rangle$ state can itself be written as a Bogoliubov transform of $|0_{out}\rangle$, the post quench state $|f_{in}\ran$ is a composite
Bogoliubov transform of $|0_{out}\rangle$:
\begin{align}
  |f\rangle_{in} &= \exp\left[\frac{1}{2}\sum_{\vec{k}} \gamma_{eff} (\vec{k}) a^\dagger_{in}(\vec{k}) a^\dagger_{in}(-\vec{k})\right]|0_{out}\rangle
  \nonumber\\
  &= \exp[\frac{1}{2}\sum_{k}\kappa_{eff}(k) a_{out}^\dagger (\vec{k}) a_{out}(\vec{k})]|D\rangle
  \label{massive-gcc-squeezed}
\end{align}
where in the first line $\gamma_{eff}$ denotes the composite $\gamma$:
\begin{align}
 \gamma_{eff}(k)=\frac{\beta^*(\vec{k})+f(\vec{k})\alpha(\vec{k})}{\alpha^*(\vec{k})+f(\vec{k})\beta(\vec{k})}
 \label{gamma-eff}
 \end{align}
and in the second line we have the relation
$\kappa_{eff}=-\frac{1}{2}\log(-\gamma_{eff})$, following similar
arguments as for the ground state. Here $\alpha, \beta$ are given
by \eq{ab-sudden}. The state \eq{massive-gcc-squeezed} is of the form
of a gCC state. Can we get {\it any} gCC (with a {\it given} $\ka(k)$)
starting from a suitably chosen squeezing function $f$?

The answer is clearly yes. By solving \eq{gamma-eff} for $f$, we can
clearly find an $f$ for any $\gamma_{eff}$. If we wish to generate a
{\it given} $\ka_{eff}(k) \equiv \ka(k)$ in \eq{massive-gcc-squeezed},
we must choose $\gamma_{eff}(k)= -\exp[-2\ka(k)]$. This gives us
\begin{align}
f(\vec{k}) = 1- \frac{2\omega_{out}}{\omega_{in}
  \tanh(\kappa(k)) +\omega_{out}}
\label{eq:squeezing-function}
\end{align}
This proves that we can prepare any gCC state, with a given $\kappa(k)$, from squeezed states:
\begin{align}
  |f\rangle_{in} = \exp[\frac{1}{2}\sum_{k}\kappa(k)
    a_{out}^\dagger (\vec{k}) a_{out}(\vec{k})]|D\rangle=: |gCC\ran
  \label{gen-gCC}
  \end{align}

\paragraph{Critical quench from specific squeezed states}
Let us consider the case of critical quench $m_{out}=0$. We look at
the following special states.  From \eq{eq:squeezing-function} it is
clear that the choice $f(\vec{k})= f_4(k) = 1-
\frac{2k}{\sqrt[]{k^2 +m^2}
  \tanh(\kappa_2k+\kappa_4k^3) +k} $ leads to
a state with only two non-zero tunable parameters $\kappa_2$ and
$\kappa_4$
\begin{equation}
|\psi(0)\rangle = |f_4\rangle = \exp[-\kappa_2 H-\kappa_4
  Q_4]|D\rangle \equiv |gCC_4\rangle \label{eq:f4-state}
\end{equation}
we call this the $gCC_4$ state because it is characterized by only two
charges $Q_2$ and $Q_4$; we will find that the equilibrium state
describing asymptotic correlators in $|gCC_4\ran$ is a grand canonical
ensemble chracterized by a temperature and one chemical
potential. Further if $\kappa_4$ is also zero, i.e. $f(\vec{k})=
f_2(k) = 1- \frac{2k}{\sqrt[]{k^2 +m^2}
  \tanh(\kappa_2 k) +k}$ then we end up with the CC
state
\begin{equation}
    |\psi(0)\rangle = |f_4\rangle = \exp[-\kappa_2 H ]|D\rangle \equiv
    |CC\rangle \label{eq:f2-state}
\end{equation}
Note that $\kappa_2$ here is not related to the mass parameter $m$ unlike
in \eq{cr-gr-kappas} where $\kappa_2=1/m$.

\subsubsection{The squeezed state 2-point function}\label{sec:squeezed-2pt}

The 2-point function in the squeezed state is the same as in the
ground state with $\al$ and $\be$ replaced by $\al_{eff}$ and
$\be_{eff}$ above. The 2-point function in the general squeezed state
\eq{gen-gCC} is given by
\begin{align}
\braket{gCC|\phi(\vec x_1,t_1)\phi(\vec x_2,t_2)|gCC} = \int \frac{d^d k}{(2\pi)^d}& \frac{e^{i \vec{k}\cdot (\vec{x_1}-\vec{x_2})}}{2\sqrt{k^2+m_{out}^2}} \csch(2\kappa(k)) \Bigg[\cos \left(\sqrt{k^2+m_{out}^2}(t_1-t_2)+ 2i\kappa(k)\right)-\nonumber \\
&\cos\left(\sqrt{k^2+m_{out}^2}(t_1+t_2)\right) \Bigg] \label{2pt-gcc}
\end{align}
When $t_1=t_2=t$ this is
\begin{equation}
\braket{gCC|\phi(\vec x_1,t)\phi(\vec x_2,t)|gCC} = \int \frac{d^d k}{(2\pi)^d} \frac{e^{i \vec{k}\cdot (\vec{x_1}-\vec{x_2})}}{2 \sqrt{k^2+m_{out}^2}} \left[\coth \left(2\kappa(k)\right)-\cos \left(2t \sqrt{k^2+m_{out}^2}\right) \csch\left(2 \kappa(k)\right) \right] \label{2pt-gcc-etc}
\end{equation}
For the equal-time two-point function of $\del_t\phi$, we apply to
\eq{2pt-gcc} the rule $-f''(0)+ g''(2t)$ mentioned above
\eq{eq:ground-quench-etc-delt}, which gives us
\begin{align}
  \braket{gCC|\del_t \phi(\vec x_1,t)\del_t\phi(\vec x_2,t)|gCC} =
  \int \frac{d^d k}{(2\pi)^d} \frac{e^{i \vec{k}\cdot
      (\vec{x_1}-\vec{x_2})}}{2} \sqrt{k^2+m_{out}^2} \left[\coth
    \left(2\kappa(k)\right)+ \cos \left(2t \sqrt{k^2+m_{out}^2}\right)
    \csch\left(2 \kappa(k)\right) \right]
  \label{2pt-gcc-etc-delt}
  \end{align}
For the two-point function of spatial derivtives, these
derives act only on the exponential term, leading to
\begin{align}
  \braket{gCC|\del_i \phi(\vec x_1,t)\del_i\phi(\vec x_2,t)|gCC} =
  \int \frac{d^d k}{(2\pi)^d} \frac{k^2 e^{i \vec{k}\cdot
      (\vec{x_1}-\vec{x_2})}}{2 \sqrt{k^2+m_{out}^2}} \left[\coth
    \left(2\kappa(k)\right)- \cos \left(2t \sqrt{k^2+m_{out}^2}\right)
    \csch\left(2 \kappa(k)\right) \right]
  \label{2pt-gcc-etc-deli}
  \end{align}
Note that the time-dependent parts of the above correlators, calling
them $\lan ... \ran^{(td)}_{gCC}$, satisfy the relation
\begin{align}
  \lan \del_i \phi(\vec x_1, t) \del_i \phi(\vec x_2, t)
  \ran^{(td)}_{gCC} + \lan \del_t \phi(\vec x_1, t) \del_t \phi(\vec
  x_2, t) \ran^{(td)}_{gCC} + m_{out}^2 \lan \phi(\vec x_1, t)
  \phi(\vec x_2, t) \ran^{(td)}_{gCC} =0
  \label{derivative-relation}
  \end{align}

\subsection{gCC correlators encompass all cases}\label{sec:gCC-all}

As is clear from the foregoing discussion, all the particular
post-quench states $|\psi(0)\ran$ mentioned above, specifically
$|CC\ran$, $|gCC_4\ran$, and $|0_{in}\ran$, are examples of $|gCC\ran$
states, with specific $\kappa(k)$ as in Table \ref{tab:gCC}:
\begin{table}[H]
 \begin{center}
  \begin{tabular}{|c|c|c|c|c|}
         \hline post-quench state & $|gCC\ran$ & $| CC\ran \big|_{m_{out}=0}$ & $| gCC_4 \ran \big|_{m_{out}=0}$ &
         $|0_{in}\ran$\\ \hline
         form of $\kappa(k)$ & $\kappa(k)$ & $\kappa_2 k$ & $\kappa_2
         k+ \kappa_4 k^2$ & $\f12 \log\left(\f{\sqrt{k^2+ m^2}+
           \sqrt{k^2+ m_{out}^2}}{\sqrt{k^2+ m^2}- \sqrt{k^2+
             m_{out}^2}}\right) $ \\ \hline
  \end{tabular}\caption{\label{tab:gCC} Ground states and specific
  squeezed states as examples of gCC states. The $\kappa(k)$ for the
  ground state (last column) is reproduced from \eq{kappa-gr}. This table
  paves way for a uniform discussion all post-quench states discussed
  in this paper.}
 \end{center}
\end{table}
Consequently, the above formula \eq{2pt-gcc-etc}, and the other
related two-point functions, give the corresponding two-point
functions in all these different cases.  In particular, using the form
of $\kappa(k)$ \eq{kappa-gr} for ground state quench, one can recover
all results of Section \ref{sec:gr-2pt} from those of Section
\ref{sec:squeezed-2pt}, and in particular recover
\eq{eq:ground-quench-etc} from \eq{2pt-gcc-etc} (after all, the ground
state is a special squeezed state!). This implies, in principle, that
if we derive some results on thermalization and relaxation for the
general gCC states, using \eq{2pt-gcc-etc}, we need not consider the
above particular cases separately. However because each of these cases
is significant on its own, we find it useful to often state the
results separately.

%%%%%%%%
\section{Time-dependence  of two-point Functions}\label{sec:time-dep-2pt}

Since we are dealing with free field theories, all correlators are
related to the basic two point function $\lan \psi(0)| \phi(x_1, t_1)
\phi(x_2,t_2)| \psi(0)\ran$. The primary goal of this paper is to
study the long time behaviour of this quantity. We will find that it
asymptotes (``thermalizes'') to the corresponding observable in a GGE;
we will find the characterization of the GGE and find the rate of
approach to equilibrium. It is easy to generalize these results to
multi-point functions, by Wick's theorem. It is also straightforward
to compute correlators of composite operators from the above two-point
function by appropriate regularization procedures. 

In this section we will analytically evaluate the various 2-point
functions \eq{eq:ground-quench-etc},\eq{eq:ground-quench-etc-delt},
\eq{2pt-gcc-etc} and \eq{2pt-gcc-etc-delt} (note that the first two
are a special case of the last two). We will indicate the main steps
and list the results, leaving some details to the various appendices.

\subsection{General remarks}\label{sec:gen-remarks}

Note that the various two-point functions mentioned in the previous
paragraph are all of the form
\begin{align}
\int \frac{d^d k}{(2\pi)^d} e^{i\vec k.\vec r} \left[ F(k) + G(k)
  \cos(2t \omega_k) \right], \quad \omega_k= \sqrt{k^2 + m_{out}^2}
\label{form-2-pt}
\end{align}
where $\vec r= \vec x_1 - \vec x_2$. For correlators of the $\braket{\phi \phi}$ type,
\begin{equation}%\label{key}
F(k)=\frac{1}{2\omega_k}\coth(2\kappa(k))\, ,\quad G(k)=-\frac{1}{2\omega_k}\csch(2\kappa(k))
\end{equation}
To proceed further, we write the
vector $\vec k$ in polar coordinates $(k, \theta, \varphi_1, ...,
\varphi_{d-2})$, oriented such that $\vec r$ points to the `north
pole' which ensures $\vec k.\vec r= kr \cos\theta$; the $\varphi_i$
parameterize a unit sphere $S^{d-2}$ (we assume $d>1$ here). With this,
the above integral becomes
\begin{align}
 &\int_0^\infty dk \left[F(k) + G(k) \cos(2t \sqrt{k^2 + m_{out}^2}) 
  \right] A_d(k,r) \label{radial-integral}\\
 & A_d(k,r) := k^{d-1}\frac{\Omega_{d-2}}{(2\pi)^{d}}
  \int_0^{\pi} d\theta e^{ikr \cos\theta} (\sin\theta)^{d-2}
\label{angular-part}
\end{align}
where $\Omega_{d-2}= \f{2\pi^{(d-1)/2}}{\Gamma((d-1)/2)}$ is the
volume of $S^{d-2}$.
\paragraph{Angular integrals}
The angular integration in \eq{angular-part} can be exactly done,
which gives
\begin{align}
A_d(k,r) =k^{d-1} 2^{1-d} \pi^{-\frac{d}{2}}\,
_0\tilde{F}_1\left(\frac{d}{2};-\frac{1}{4} k^2
r^2\right)\label{eq:angular-integral}
\end{align}
The regularized Hypergeometric function $_0\tilde F_1$ is some
combination of trigonometric functions $\sin(kr)$ and $\cos(kr)$ in
odd $d$ while it is some Bessel function in even $d$. The specific
forms of $A_d(k,r)$ for various dimensions $d$ are
tabulated below in Table \ref{tab:angular-function}:

\begin{table}[H]
 \begin{center}
  \begin{tabular}{|c|c|}
    \hline
    number of spatial dimensions $d$ & $ A_d(k,r)$ \\
    \hline
    1~\protect\footnotemark & $\frac{1}{\pi}\cos (kr)$
      \\
    \hline
    2 & $\frac{k J_0(k r)}{2\pi}$\\
    \hline
    3 & $\frac{k \sin (k r)}{2\pi^2 r}$\\
    \hline
    4 & $\frac{k^2 J_1(k r)}{4\pi^2 r}$\\
    \hline
    5 & $\frac{k (\sin (k r)-k r \cos(k r))}{4\pi^3 r^3}$\\
    \hline
  \end{tabular}\caption{\label{tab:angular-function} The angular integral
    $A_d(k,r)$, (see eq.(\ref{eq:angular-integral})) in the first few
    dimensions.}
 \end{center}
\end{table}
\footnotetext{We added $d=1$ here for uniformity, using
$\int_{-\infty}^\infty \f{dk}{2\pi}e^{ikx} f(|k|)$ =$ \int_0^\infty
d|k| \f{1}{\pi}\cos(|k||x|) f(|k|)$.\label{ftnt:d=1}}

%%%%
\paragraph{Recursion relations for 2-point functions}\label{sec:recursion}
The structure of the integrals \eq{form-2-pt} occuring in the various
correlators defined above makes it possible to connect them across
dimensions through recursion relations(see Appendix \ref{app:recursion}
for details) \footnote{Similar recursion relations apppear in a
somewhat different context in \cite{Bhattacharyya:2016nhn}}. From a
correlator in $d+1$ spacetime dimensions, one can obtain the correlator
in $(d+2)+1$ dimensions by acting with a particular differential
operator. This relation holds for both even and odd $d>1$ (here $t_
-=t_1-t_2$ and $t_+=t_1+t_2$).
\begin{equation}
 \langle \phi(x_1,t_1)\phi(x_2,t_2)\rangle^{(d+2)}= \frac{\Omega_{d-2}}
 {4\pi^2 \Omega_{d-4}} \left(-\partial^2 _{t_{-}} -\partial^2 _{t_{+}}
 +m_{out}^2 +\partial_r^2 \right) \langle \phi(x_1,t_1)\phi(x_2,t_2)
 \rangle^{(d)},\;\;d>1 \label{eq:recursion}
\end{equation}
where $\lan ... \ran^{(d)}$ denotes an expectation value, in a $d+1$
dimensional theory, in a general post-quench state. 

For the special case of equal time correlators which is our main focus, the recursion relation becomes
\begin{equation}
  \langle \phi(x_1,t_1)\phi(x_2,t_2)\rangle^{(d+2)}= \frac{\Omega_{d-2}}
  {4\pi^2 \Omega_{d-4}} \left(-\partial^2 _{t_{+}} +m_{out}^2 +
  \partial_r^2 \right) \langle \phi(x_1,t_1)\phi(x_2,t_2)\rangle^{(d)},\;
  \;d>1
  \label{eq:recursion-etc}
\end{equation}
A similar relation holds for the equilibrium correlators. In an arbitrary GGE, we have the following recursion relation ($t$ here is same as $t_-$)
\begin{equation}
  \langle \phi(\vec{r},t) \phi(0,0)\rangle^{(d)}_{\mu(k)}=
  \frac{\Omega_{d-2}}{4\pi^2 \Omega_{d-4}}\left(-\partial_t^2 +m_{out}^2
  +\partial_r^2 \right) \langle \phi(\vec{r},t) \phi(0,0) \rangle^{(d-2)}
  _{\mu(k)}
  \label{eq:recursion-gge}
\end{equation}
The 1+1 dimensional 2-pt function is not connected to its 3+1
dimensional counterpart in the above fashion, since in one dimensional
space there is no angular coordinate which is crucial for this
recursion relation to work (see Appendix \ref{app:recursion}).
Nevertheless, there is a different, specific relation connecting
the two:
\begin{align}
 \langle \phi(x_1,t_1)\phi(x_2,t_2)\rangle^{(3+1)} = -\frac{1}{2\pi r}
 \partial_r \langle \phi(x_1,t_1)\phi(x_2,t_2)\rangle^{(1+1)}
 \label{recursion-1-3}
 \end{align}
which follows straightforwardly, by inspection.

\paragraph{Time-dependence}

As remarked before, there is a time-dependent part and a
time-independent part of the generic equal time two-point function
\eq{form-2-pt}. We will come back to an analysis of the
time-independent part in Section \ref{sec:thermalization-to-GGE}
where we show that this part exactly matches the corresponding
two-point function in a GGE (generalized Gibbs ensemble).

In the remainder of the section, we will therefore consider only the
time-dependent, non-equilibrium, part. We will show that this part
decays to zero at large times, which amounts to a proof of
thermalization, and we will derive the form of the decay.  It will
also be conveninent, for this purpose, to differentiate between
massive quench $m_{out}\not=0$ and critical quench $m_{out}=0$. This is
because in the latter case, the post-quench theory is conformal and we
expect, and indeed verify, that this theory will have special
proprties as compared to the massive case.  We will comment later (see
Section \ref{sec:crit-massive}) why we cannot obtain results for the
critical quench by simply taking the $m_{out} \to 0$ limit of the
power series expansion for $m_{out} \not= 0$.

%%%%
\subsection{Critical Quench Correlators}

As mentioned above, we will mainly focus on the time-dependent parts
of the 2-point functions, while the time-independent parts will be
considered in Section \ref{sec:thermalization-to-GGE}.

\subsubsection{Time-dependent part of the
  $\lan \phi \phi \ran$ correlator}\label{sec:crit-phi-phi}

Let us put $m_{out}=0$. To keep the discussion general, let us
consider the general gCC correlator \eq{2pt-gcc-etc} (as emphasized in
Section \ref{sec:gCC-all}, we can infer about the specific cases of
ground states and the squeezed states from here). In the notation
of \eq{radial-integral}, we now have
\begin{align}
F(k)= \f1{2k} \coth(2\kappa(k)), \;
G(k) = -\f1{2k} \csch(2\kappa(k)),\;
\cos(2t\sqrt{k^2+m_{out^2}})=\cos(2kt)
\label{f-g-crit}
\end{align}
The functions $\ka(k)$ are tabulated for the special cases
of interest in Table \ref{tab:gCC}, where for the ground state,
we now have 
\begin{align}
\ka(k)= \f12 \log\left(\f{\sqrt{k^2+ m^2}+k}{\sqrt{k^2+ m^2}-k}\right)
\label{kappa-gr-cr}
\end{align}
which has an expansion in odd powers of $k$, as mentioned above
\eq{cr-gr-kappas}. Following this, {\it we will focus, in the case of
critical quench, only on gCC states with $\ka(k)$ which have an
expansion in odd powers of $k$}.  The $\ka(k)$ for the CC and gCC$_4$
states, of course, satisfy this by definition.

With this assumption, the functions $F(k), G(k)$ are even in $k$.

\subsubsection{Large time behaviour for odd $\boldsymbol{d}$\label{critical-odd}}

Note from the Table \ref{tab:angular-function}, $A_d{k,r}$ that for
odd $d$ are even functions of $k$. Using the evenness of $F(k), G(k)$,
we find that the integrand in \eq{radial-integral} is even, which
allows us to extend the integration contour to the entire real line
(for even $d$, this step is not allowed, hence we will do something
else). If one can close the contour on the upper or the lower half
plane the integral can be evaluated by the method of residues. To see
how it works, let us take the example of $d=3$ (the procedure
described below is similar to the case of $d=1$ \cite{Mandal:2015kxi};
and in special cases, where exact results are known in $d=1$, the
3-dimensional results can be derived by using \eq{recursion-1-3}).
For more details see Appendix \ref{app:calc-details-3+1}.

For the critical gCC correlator at $d=3$, we get
\begin{align}
 & \braket{gCC|\phi(x_1,t)\phi(x_2,t)|gCC} = \frac{1}{4\pi^2 r} \int_0^\infty dk \sin(k r) \left[\coth(2\kappa(k)) - \csch(2\kappa(k)) \cos(2 k t)\right] \nonumber \\
 & = \frac{1}{8 \pi^2 i r} \int_{-\infty}^\infty dk e^{i k r} \left[\coth(2\kappa(k)) - \csch(2\kappa(k)) \cos(2 k t)\right] \nonumber \\
  & = \frac{1}{8 \pi^2 i r} \int_{-\infty}^\infty dk \left[\underbrace{\frac{\cosh(2\ka(k)) e^{i k r}}{\sinh(2\ka(k))}}_{\text{A}} - \underbrace{\frac{e^{i k (r+ 2t)}}{2\sinh(2\kappa(k))}}_{\text{B}}- \underbrace{\frac{e^{i k (r- 2t)}}{2\sinh(2\ka(k))}}_{\text{C}}\right]
  \label{gCC-contour}
\end{align}
Let us focus on the time-dependent terms B and C (the time-independent
term A is described later on, and in Appendix
\ref{app:calc-details-3+1}). We will also assume large $t$ so that
$2t>r$. Then from the asymptotic behaviour of functions $\ka(k)$ in
the Table \ref{tab:gCC}, one can see that the functions $\kappa(k)$
grow as $k \to \infty$, in a way that ensures that the integrals are
convergent, and for large $t$ (so that $2t > r$) it is possible to
close the contour for the integral B on the UHP and for C on the
LHP. At large times $t$, the slowest transient is given by the pole of
the integrand singularity nearest to the origin (in the UHP for B and
LHP for C; the pole at the origin itself cancels between the terms B
and C; see Appendix \ref{app:calc-details-3+1}, especially Figure
\ref{fig:f4s}). It is easy to see that for a gCC with a finite number
of chemical potentials, the nearest such singularity is a pole. This
is given by the relevant zero of $\sinh(2\ka(k))$, or, eqivalently by
the corresponding root of
\begin{align}
2\ka(k)= \pm i \pi
\label{roots}
\end{align}
The root discussed above turns out to be of the form as $k= k_1 \pm i
k_0, k_0>0$; here we choose the + and $-$ signs respectively for the
B and C terms; the value of $k_1$ may be zero, positive or negative.
By applying residue calculus, we find that the slowest decay at large
$t$ of the $\phi\phi$ correlator is given by
\begin{align}
  b \exp[-k_0(t+2r)] + c \exp[-k_0(t-2r)]
  \label{exp-3-dim}
\end{align}
For non-zero $k_1$ this comes multiplied by an oscillatory term, of
time period $2\pi/k_1$.  In case of the CC state, the exact
time-dependence can be calculated by summing over the poles (see
Appendix \ref{app:calc-details-3+1} for details). The exact two-point
function for the CC state as well as for the ground state can also be
derived from the their 1+1 dimensional counterparts found in
\cite{Mandal:2015kxi} using the recursion relation \eq{recursion-1-3}
(see Appendix \ref{app:calc-details-3+1}). The net result of all this
is that at large times the time-dependent part of the critical
$\lan\phi\phi \ran$ fall off exponentially, as in \eq{exp-3-dim};
these are tabulated in Table \ref{tab:critical-phi}.

In case of a gCC with an infinite number of chemical potentials, such
as when the gCC is the ground state, the poles can merge into a branch
cut (see also a similar discussion in \cite{Mandal:2015kxi}). The
structure \eq{exp-3-dim} is retained even here, with additional
power-law prefactors. Details of the above procedure for the CC, gCC$_4$
and the ground state, are worked out in Appendix \ref{app:calc-details-3+1}.
Note that although for specificity we had taken the example of $d=3$
here, the arguments go through for $d=5$ or any higher odd space
dimensions.

\begin{table}[h!]
\begin{center}
    \begin{tabular}{ |p{1cm}|p{8.6cm}|p{5.8cm}| }
        \hline
        \multicolumn{3}{|c|}{$\braket{\psi(0)|\phi(\vec{x_1},t)
        \phi(\vec{x_2},t)|\psi(0)}$} \\
        \hline
        $| \psi(0) \ran $& $ \mathbf{d=3}$ & $ \mathbf{d=4}$ \\
        \hline
        Ground state & $\frac{-m^2}{16\pi^{5/2}\sqrt{mt}} e^{-2m t}
        \left(1+ \mathcal{O}(m r)^2 \right) \left(1+ \mathcal{O}(m t)^{-1}
        \right)$ & $\frac{m}{128\pi^2} \frac{1}{t^2}+ \mathcal{O}(\frac{1}
        {t^4})$\\
        \hline
        CC state & $\frac{-1}{16\kappa_2^2}e^{-\pi t/\kappa_2} (1+
        \mathcal{O}(\frac{r}{\kappa_2})^2)+ \mathcal{O}\left(e^{-2\pi t/
        \kappa_2}\right)$ & $\frac{1}{128\pi^2\kappa_2} \frac{1}{t^2}+
        \mathcal{O} (\frac{1}{t^4})$\\
        \hline
        gCC$_4$ state & $\frac{-(1+ \pi^2 \bar{\kappa_4} + \cdots)}
        {16\kappa_2^2} e^{-\frac{\pi t}{\kappa_2}(1 + \frac{\pi^2}{4}
        \bar{\kappa_4}+ \cdots)} (1+\mathcal{O}(\frac{r}{\kappa_2})^2)+ 
        \mathcal{O}\left(e^{-2\pi t/\kappa_2}\right)$ & $\frac{1}{128
        \pi^2\kappa_2} \frac{1}{t^2} + \frac{3r^2 +16\kappa_2^2 +
        24\kappa_4/\kappa_2}{2048\pi^2\kappa_2} \frac{1}{t^4} +\mathcal{O}
        (\frac{1}{t^6})$\\
        \hline
    \end{tabular}\caption{\label{tab:critical-phi}Time-dependent part of 2-point function $\braket{\psi(0)|\phi(\vec{x_1},t)\phi(\vec{x_2},t)|\psi(0)}$ in critical quench for $d=3$ (left column) and $d=4$ (right column). Here $r=|\vec{x_1}-\vec{x_2}|$. Note that for even $d$, e.g. $d=4$, the leading large $t$ dependence is the same in all three types of post-quench state $|\psi(0)\ran$, with the identification $m=1/\kappa_2$ (see \eq{cr-gr-kappas}). Further the gCC answer tells us that since the higher $\kappa_i$'s for $i>2$ show up in subleading terms, the higher conserved charges do not affect the leading transient towards thermalization. However, as we will see later, the time-independent part of these two-point functions reflects an equilibrium given by a GGE which is characterized by {\it all} the $\kappa_i$ parameters.}
\end{center}
\end{table}
Since we are interested in the leading approach to equilibrium
for simplicity, in the tables we only display the correlators in the limit
$mt\gg 1\gg mr$ or equivalently $\frac{t}{\kappa_2}\gg 1 \gg \frac{r}
{\kappa_2}$. The ellipsis in the gCC$_4$ correlator represents
$\mathcal{O}(\bar{\kappa_4}^2)$ terms.

\paragraph{General odd $\boldsymbol{d}$ result:} From the recursion relation
\eq{eq:recursion} it follows that for all odd $d\ge 3$, the slowest
decay at large time of the time-dependent part of the $\phi \phi$
correlator is given by replacing \eq{exp-3-dim} with 
\begin{equation}\label{eq-general-odd}
b_d \exp[-k_0(t+2r)] + c_d \exp[-k_0(t-2r)]
\end{equation}
where $k_0$ has the same value discussed above and is independent of
dimension $d$. The constants $b_d, c_d$ depend on the dimension $d$. The expression above may come with a possible oscillatory part as explained below \eq{exp-3-dim}.

\subsubsection{Large time behaviour for even $\boldsymbol{d}$ \label{critical-even}}
Here the angular function, $A_d(k,r)$ for even $d$ are odd functions
of $k$. Therefore, unlike in odd $d$, we cannot extend the integration
contour to the entire real line and apply the method of residues.
Fortunately there does exist a different approach in this case.
Let us take the example of $d=4$ to illustrate the method. As in
the case of odd $d$, we allow the most general 
$\kappa(k)=\sum_{i=1}^\infty \kappa_{2i} k^{2i-1}$ which grows at
infinity such that the integral is convergent. The critical gCC
correlator at $d=4$ is
\begin{align}
 & \braket{gCC|\phi(x_1,t)\phi(x_2,t)|gCC} = \frac{1}{8\pi^2 r}
 \int_0^\infty dk k J_1(kr)\left[\coth(2\kappa(k)) - \csch(2\kappa(k))
 \cos(2 k t)\right] \nonumber
\end{align}
In the notation of equation \ref{radial-integral},
$G(k)=-\csch(2\kappa(k))/(2k)$. As before we are interested in large t.
Consider the following scaling
\begin{equation}
 k\rightarrow p=2kt \label{eq:scaling}
\end{equation}
and rewrite the time-dependent part in terms of this new dimensionless
momenta $p$
\begin{align}
 &-\frac{1}{8\pi^2 r} \int _0^{\infty} dk\,k J_1(k r) \csch(2\kappa(k))
 \cos(2kt) =-\frac{1}{32\pi^2 r} \frac{1}{t^2} \int_0^{\infty} dp\,p
 J_1\left(\frac{pr}{2t}\right) \csch(2\kappa \left(\frac{p}{2t}\right))
 \cos(p) \nonumber \\
 &=-\frac{1}{64\pi^2 r} \frac{1}{t^2} \left\{ \int_0^{\infty(1+i\epsilon)}
 dp\,p J_1\left(\frac{pr}{2t}\right) \csch(2\kappa \left(\frac{p}{2t}
 \right)) e^{ip} + \int_0^{\infty(1-i\epsilon)} dp\,p J_1\left(\frac{pr}
 {2t}\right) \csch(2\kappa \left(\frac{p}{2t}\right)) e^{-ip} \right\}
 \nonumber \\
 & =-\frac{1}{256\pi^2 \kappa_2} \left[\frac{I(1)}{t^2}-\frac{\left(16
 \kappa_2^3+24 \kappa_4+3 \kappa_2 r^2\right)}{96 \kappa_2 t^4} I(3) 
 +\mathcal{O}\left(t^{-6}\right)\right] \label{eq:gcc-expansion}
\end{align}
In the second line we have rotated the contour anti-clockwise for
$e^{ip}$ and clockwise for $e^{-ip}$. We can do this because the
integrand $J_1(kr) \csch(2\kappa_2 k)$ vanishes on the arc at
infinity due to the exponential damping in the $\csch(2\kappa(k))$
as $k\rightarrow\infty$. In the last line we have Taylor expanded
the integrand at large $t$. We have also introduced the integrals
$I(n)$ defined by
\begin{align}
 I(n) &= \lim_{\epsilon \to 0}\int_0^{\infty(1+i\epsilon)}dp e^{ip} p^n
 + \lim_{\epsilon \to 0}\int_0^{\infty(1-i\epsilon)} dp e^{-ip}p^n
 \nonumber \\
 &= \lim_{\epsilon \to 0}\int_0^{\infty}dp_+ e^{ip_+} p_+^n +
 \lim_{\epsilon \to 0} \int_0^{\infty}dp_- e^{-ip_-}p_-^n\nonumber \\
 &= \lim_{\epsilon \to 0} \int_0^{\infty} dp\left(e^{p(i-\epsilon)}+ 
 e^{-p(i+\epsilon)}\right)p^n + \mathcal{O}(\epsilon) \nonumber \\
 &= \lim_{\epsilon \to 0} (\partial_{-\epsilon})^n \int _0^{\infty}
 dp\left( e^{p(i-\epsilon)}+ e^{-p(i+\epsilon)}\right) = \lim_{\epsilon
 \to 0} (\partial_{-\epsilon})^n \left(\frac{2\epsilon}{1+\epsilon^2}
 \right) \nonumber \\
 I(n) &= \begin{cases}
    (-1)^{(n+1)/2} 2\times n! & \text{n is odd}\\
    0 & \text{n is even}
\end{cases}\label{eq:I(n)}
\end{align}
where in the second line we have introduced $p_\pm= p(1\pm i\epsilon)$.
Note that in spite of the growth of the $I(n)$ with $n$, because
of the  even more strongly convergent power series expansion
for the product of the Bessel and cosech functions, the equation
\eq{eq:gcc-expansion} has a convergent expansion in $1/t^2$ for
large enough $t$.

Notice that $\kappa_4$ shows up only from the $O(t^{-4})$ onwards.
Similarly $\kappa_6$ starts appearing only from the 
$\mathcal{O}\left(t^{-6}\right)$ term; explicitly that term is
\[
 \frac{64 \left(14 \kappa_2^6-30 \kappa_2^3 \kappa_4-45 \kappa_2
 \kappa_6 +45 \kappa_4^2\right)+15 \kappa_2^2 r^4+120 r^2 \left(2
 \kappa_2^4+3 \kappa_2 \kappa_4\right)}{46080 \kappa_2^2 t^6} I(5)
\]
Having calculated $I(n)$'s, we find that the time-dependent part
of the gCC 2-point function goes as
\begin{align}
 -\frac{1}{8\pi^2 r} \int _0^{\infty} dk\,k J_1(k r) \csch(2\kappa(k))
 \cos(2kt)= \frac{1}{128 \pi^2 \kappa_2} \frac{1}{t^2} + \frac{3r^2
 +16\kappa_2^2 + 24\kappa_4/\kappa_2}{2048\pi^2\kappa_2} \frac{1}{t^4} 
 +\mathcal{O}(t^{-6})
\label{power-law-subleading}
\end{align}
The effect of $\kappa_4$ is suppressed by $\mathcal{O}(t^{-2})$
relative to the leading term and therefore it makes no difference
to the leading power law fall off in the large $t$ limit. The same
is true for higher $\kappa_n$'s, as they are even more suppressed.

\begin{figure}[H]
  \begin{center}
    \includegraphics[scale=0.28]{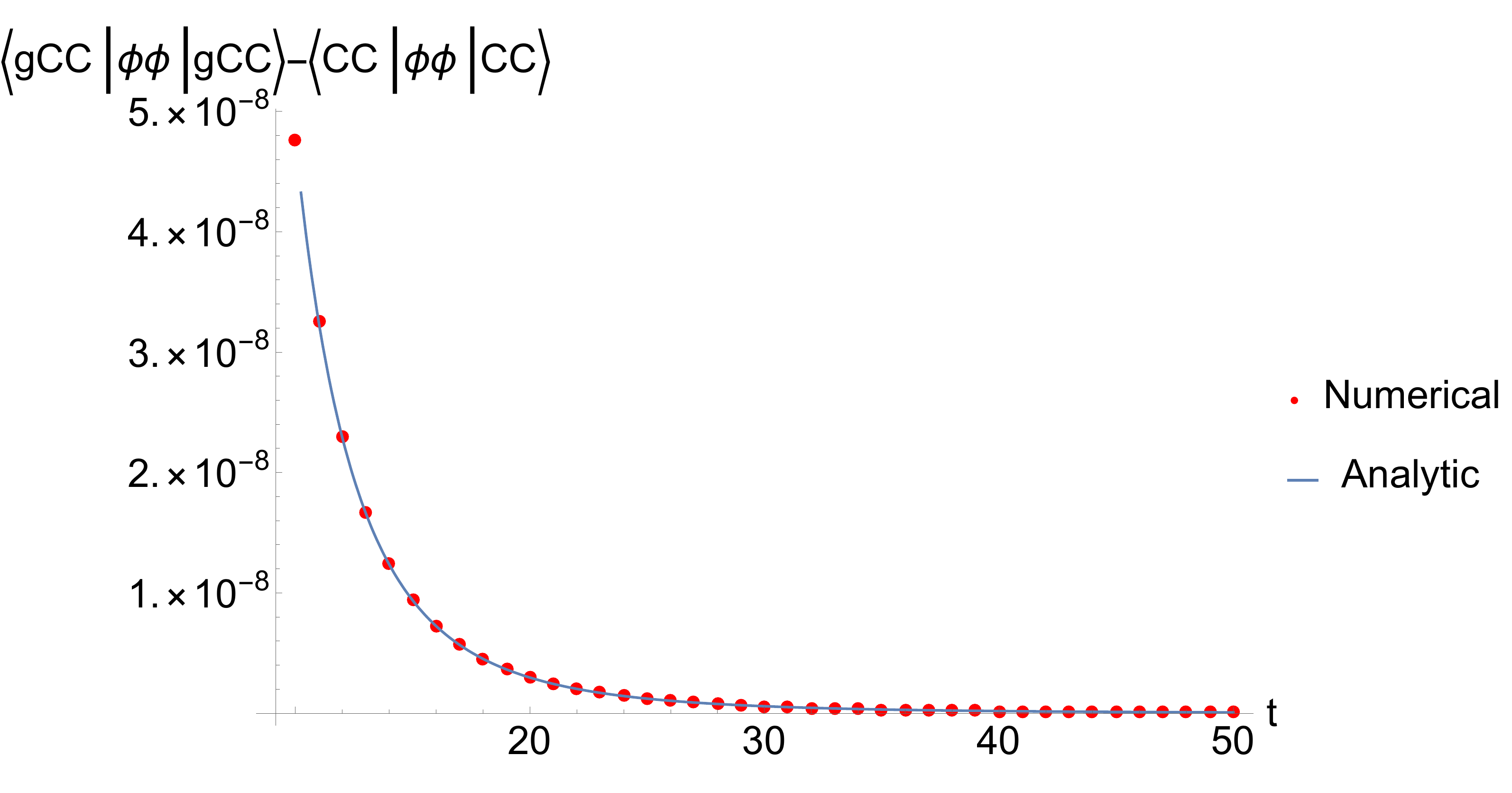}
    \caption{We plot the gCC$_4$ equal-time correlator after
      subtracting off the CC part. This gives us the contribution
      depending solely on $\kappa_4$. There is near perfect match
      between the analytic and numerical results. (Numerics done for
      values $r=0.1$,$\kappa_2 =0.25$ and
      $\kappa_4=0.025$)}\label{fig:gce-cc}
    \end{center}
\end{figure}

The equation \eq{power-law-subleading} can be verified against
explicit numerical integration of the momentum integral. In
practical terms, the subleading term, of order $1/t^4$, is possible
to read off by first also calculate the time-dependent part in the
gCC$_4$ state numerically and see exact agreement in subleading
$\mathcal{O}(t^{-4})$ behaviour after subtracting off the $\kappa_2$
-dependent part (fig. \ref{fig:gce-cc}).
\paragraph{General even $\boldsymbol{d}$ result:} From the recursion relation \eq{eq:recursion} it follows that for all $d\ge 4$, the time-dependent part of the $\phi \phi$ correlator decays as (up to time-independent factors)
\begin{equation}\label{eq-general-even}
	\frac{1}{t^{d-2}}
\end{equation}
The $d=2$ case is discussed separately below
(see table \ref{tab:critical-del-phi-2}) for a summary and also appendix \ref{app:calc-details-2+1} for details).

One might wonder if the same reasoning would have given us the answer
in odd $d$, then we would not have to do the contour integrals there.
In $d=3$ for example, the time-dependent part of the 2-point function
has even powers of $p$ (eq. \ref{eq:scaling})
\begin{align}
 &-\frac{1}{4 \pi^2 r} \int_0^{\infty} dk \sin(kr) \csch(2 \kappa_2 k)
 \cos(2kt) \nonumber \\
 &=-\frac{1}{16\pi^2 \kappa_2} \int_0^{\infty}dp \left[\frac{1}{t}
 -\frac{\left(4 \kappa_2^3+6 \kappa_4 +\kappa_2 r^2\right)p^2}{24
 \kappa_2 t^3} +\mathcal{O}(t^{-5})\right] \cos(p) \nonumber \\
 & =-\frac{1}{32\pi^2 \kappa_2} \left[\frac{I(0)}{t} -\frac{\left(4
 \kappa_2^3+6 \kappa_4 +\kappa_2 r^2\right)I(2)}{24 \kappa_2 t^3}
 +\mathcal{O}(t^{-5})\right] \nonumber \\
 & = 0
\end{align}
What this means is that either the answer is zero or something non-perturbative, i.e. a function with no Taylor expansion at $t=\infty$.
Since we already know the answer is an exponentially decaying function
$e^{-\pi t/\kappa_2}$, we know it is the latter. The above argument
elucidates the odd-even difference in the approach to thermalization.
We give another geometric understanding in the next section
\ref{sec:approach-thermalization}.

%%%%
\subsubsection{$\partial_i \phi \partial_i \phi$ and $\partial_t \phi \partial_t \phi$ Correlator}

\begin{table}[h!]
\begin{center}
    \begin{tabular}{ |p{1cm}|p{9.1cm}|p{5.6cm}| }
        \hline
        \multicolumn{3}{|c|}{$\braket{\psi(0)|\partial_t\phi(\vec{x_1},t)
        \partial_t\phi(\vec{x_2},t)|\psi(0)}$} \\
        \hline
        $|\psi(0)\ran$ & $ \mathbf{d=3}$ & $\mathbf{d=4}$ \\
        \hline
        Ground state & $\frac{-m^4}{16\pi^{3/2}\sqrt{mt}} e^{-2m t}
        \left(1+ \mathcal{O}(m r)^2 \right) \left(1+ \mathcal{O}(m t)^{-1}
        \right)$ & $\frac{3m}{256\pi^2}\frac{1}{t^4}+ \mathcal{O}
        (\frac{1}{t^6})$ \\
        \hline
        CC state & $\frac{-\pi^2}{64\kappa_2^4}e^{-\pi t/\kappa_2}(1+
        \mathcal{O}(\frac{r}{\kappa_2})^2)+ \mathcal{O}\left(e^{-2\pi
        t/\kappa_2}\right)$ & $\frac{3}{256\pi^2 \kappa_2}\frac{1}{t^4}
        + \mathcal{O}(\frac{1}{t^6})$\\
        \hline
        gCC$_4$ state & $\frac{-\pi^2(1+\frac{3\pi^2 \bar{\kappa_4}}{2}+
        \cdots)}{64\kappa_2^4} e^{-\frac{\pi t}{\kappa_2}(1+ \frac{\pi^2}
        {4}\bar{\kappa_4}+ \cdots)} (1+\mathcal{O}(\frac{r}{\kappa_2})^2)
        + \mathcal{O}\left(e^{-2\pi t/\kappa_2}\right)$ & $\frac{3}{256
        \pi^2 \kappa_2}\frac{1}{t^4}-\frac{15r^2+80\kappa_2^2+120\kappa_4}
        {2048\pi^2\kappa_2t^6} + \mathcal{O}(\frac{1}{t^8})$\\
        \hline
    \end{tabular}\caption{\label{tab:critical-del-phi}Time-dependent
    part of 2-point function $\braket{\psi(0)|\partial_t\phi(\vec{x_1},t)
    \partial_t\phi(\vec{x_2},t)|\psi(0)}$ in critical quench in $d=3$ in
    the left column and $d=4$ in the right column. Here $r=|\vec{x_1}
    -\vec{x_2}|$. $|\psi(0)\ran$ denotes the type of post-quench state.}
\end{center}
\end{table}

As argued in section \ref{sec:gr-2pt},
the 2-point function of $\partial \phi$ is not independent of the
2-point function of $\phi$. As explained in that section, in $d=1$
and 2, our primary interest is in $\lan \del\phi \del\phi\ran$. For $d=1$ \cite{Mandal:2015kxi} this quantity for the general gCC
state can be obtained by using the equations \eq{2pt-gcc-etc-delt}, \eq{radial-integral} and \eq{angular-part}:
\begin{align}
 &\braket{gCC|\partial_t\phi(x_1,t) \partial_t\phi(x_2,t)|gCC} =
 \frac{1}{4\pi} \int_{-\infty}^\infty dk e^{i k r} k
 \left[\coth(2\kappa(k)) + \csch(2\kappa(k)) \cos(2kt)\right] \nonumber
\end{align}
This expression  can be evaluated  in the same manner as described in
Section \ref{critical-odd}. Similarly for $d=2$, the time dependent part of the gCC 2-point function can be evaluated by again using equations \eq{2pt-gcc-etc-delt}, \eq{radial-integral} and \eq{angular-part}:
\begin{align}
 \braket{\partial_t\phi(\vec{x_1,t}) \partial_t\phi(\vec{x_2},t)}_{td}
 &= \frac{1}{4\pi}\int_{0}^{\infty} dk e^{ikr} k^2 J_0(kr) \cos(2kt)
 \csch(2\kappa(k)) \nonumber
\end{align}
This expression  can be evaluated  in the same manner as described in
Section \ref{critical-even}. We tabulate the results below \ref{tab:critical-del-phi-2}.
\begin{table}[h!]
\begin{center}
    \begin{tabular}{ |p{1cm}|p{8.9cm}|p{5.5cm}| }
        \hline
        \multicolumn{3}{|c|}{$\braket{\psi(0)|\partial_t\phi(\vec{x_1},t)\partial_t\phi(\vec{x_2},t)|\psi(0)}$} \\
        \hline
        $|\psi(0)\ran$ & $\mathbf{d=1}$ & $\mathbf{d=2}$ \\
        \hline
        Ground state & $\frac{m^2}{8\pi^{1/2}\sqrt{mt}} e^{-2m t} \left(1+ \mathcal{O}(m r)^2 \right) \left(1+ \mathcal{O}(m t)^{-1} \right)$ & $-\frac{m}{32\pi  t^2}+ \mathcal{O}(\frac{1}{t^4})$ \\
        \hline
        CC state & $\frac{\pi}{8\kappa_2^2}e^{-\pi t/\kappa_2}(1+\mathcal{O}(\frac{r}{\kappa_2})^2)+ \mathcal{O}\left(e^{-2\pi t/\kappa_2}\right)$ & $-\frac{1}{32\pi \kappa_2 t^2}+ \mathcal{O}(\frac{1}{t^4})$\\
        \hline
        gCC$_4$ state & $\frac{\pi(1+\pi \bar{\kappa_4}+ \cdots)}{8\kappa_2^2}e^{-2 t \left(\frac{\pi }{2 \kappa _2}+\frac{\pi^3 \bar{\kappa_4}+\cdots}{8\kappa_2}\right)}(1+\mathcal{O}(\frac{r}{\kappa_2})^2)+ \mathcal{O}\left(e^{-2\pi t/\kappa_2}\right)$ & $-\frac{1}{32\pi \kappa_2 t^2}- \frac{8 \kappa_2^3+12 \kappa_4+3 \kappa_2 r^2}{128 \pi \kappa_2^2 t^4} + \mathcal{O}(\frac{1}{t^6})$\\
        \hline
    \end{tabular}\caption{\label{tab:critical-del-phi-2}Time-dependent part of 2-point function $\braket{\psi(0)|\partial_t\phi(\vec{x_1},t)\partial_t\phi(\vec{x_2},t)|\psi(0)}$ in critical quench in $d=1$ in the left column and $d=2$ in the right column. Here $|\psi(0)\ran$ denotes the type of post-quench state. $r=|\vec{x_1}-\vec{x_2}|$).}
\end{center}
\end{table}

%%%%
\subsection{Massive Quench Correlators}

In the massive or non-critical quench the final mass $m_{out}$ is
non-zero. Here we incorporate stationary phase approximation to
perform the momentum integrals.

\subsubsection{$\phi \phi$ Correlator}

In the notation of (\ref{form-2-pt}), the time dependent part of
2-point function in the gCC state (\ref{2pt-gcc-etc}) can be
written as
\begin{align}
 & \int_0^{\infty} dk\, G(k) \cos\left(2t \sqrt{k^2+m_{out}^2}
 \right)A_d(k,r) \nonumber \\
 & =\frac{1}{2}\int_0^{\infty} dk\, G(k) \left(e^{i 2t\sqrt{k^2+m_{out}
 ^2}}+e^{-i 2t\sqrt{k^2+m_{out}^2}} \right) A_d(k,r) \nonumber \\
 & =\int_0^{\infty} dk\, f_d(k,r) e^{i t w(k)} + \int_0^{\infty} dk\,
 f_d(k,r) e^{-i t w(k)} \nonumber \\
 & =I_1+I_2
\end{align}
$G(k)= -\frac{\csch(2\kappa(k))}{2\sqrt{k^2+m_{out}^2}}$ for an arbitrary
gCC state, while $G(k)= \frac{(m_{out}^2-m^2)}{4 \sqrt{k^2+m^2}
(k^2+m_{out}^2)}$ for the ground state. $A_d(k,r)=k^{d-1} 2^{1-d}
\pi^{-\frac{d}{2}}\,_0\tilde{F}_1 \left(\frac{d}{2};-\frac{1}{4} k^2
r^2\right)$ is the angular fucntion defined before (\ref{angular-part})
and $f_d(k,r)= \frac{1}{2} G(k)A_d(k,r)$. We consider the gCC states
defined by
\begin{equation}
 \kappa(k)= \sum_{i=1}^\infty \kappa_{2i-1} k^{2(i-1)} =\kappa_1+
 \kappa_3 k^2+ \kappa_5 k^4 + \cdots
\end{equation}
The reason for choosing only even powers of $k$ is motivated by only
even powers appearing in eq.(\ref{ma-gr-kappas}). Let $k=k_0$ be a
stationary point of the integral $I_1$ (for which $w''(k_0) > 0$) having
the general form
\begin{align}
 I_1(t)=& m_{out}^{d-1}\int dk f(k) e^{i t w(k)} \xrightarrow[t\rightarrow
 it]{\text{Wick rotate}} m_{out}^{d-1}\int dk f(k) e^{-t w(k)} \nonumber \\
 =& m_{out}^{d-1} \frac{e^{-t w_0}}{\sqrt{t}} \int dy \left(f_0 + f'_0
 \frac{y}{\sqrt{t}} + \frac{1}{2!} f''_0 \frac{y^2}{t} + \frac{1}{3!}
 f'''_0 \frac{y^3}{t^{3/2}} + \cdots \right) \times \nonumber \\
 & \exp -\left(\frac{w''_0}{2!}y^2 + \frac{w'''_0}{3!}\frac{y^3}{\sqrt{t}}
 + \frac{w''''_0}{4!}\frac{y^4}{t}+ \cdots \right) \nonumber \\
 =& m_{out}^{d-1} \frac{e^{-2t}}{\sqrt{t}} \int dy \left(f_0 + f'_0
 \frac{y}{\sqrt{t}} + \frac{1}{2!} f''_0 \frac{y^2}{t} + \frac{1}{3!}
 f'''_0 \frac{y^3}{t^{3/2}} + \cdots \right) \exp -\left(y^2 -\frac{1}{4}
 \frac{y^4}{t}+ \cdots \right) \nonumber \\
 =& m_{out}^{d-1} \frac{e^{-2t}}{\sqrt{t}} \int dy e^{-y^2} \left(f_0 +
 f'_0 \frac{y}{\sqrt{t}} + \frac{1}{2!} f''_0 \frac{y^2}{t} + \frac{1}{3!}
 f'''_0 \frac{y^3}{t^{3/2}} + \cdots \right) \left(1 +\frac{1}{4}
 \frac{y^4}{t}+ \cdots \right) \label{eq:massive-saddle}
\end{align}
where we Wick rotate $t\rightarrow it$ in the first line (when
$w''(k_0)<0$ which is the case in $I_2$, we rotate in the opposite
direction i.e, $t\rightarrow -it$). We expand both $f$ and $w$ around
$k=k_0$ and introduce the variable $y=\sqrt{t}(k-k_0)$ in the second line.
We are using compact notation to write $f_0=f(k_0)$, $f'_0=f'(k=k_0)$,
etc. and similarly for $w$. In the third line plug in $w(k)=\sqrt{k^2+1}$
and finally in the last line we have brought down all terms in the
exponent except for $y^2$. The factor $m_{out}^{d-1}$ is pulled out to
make rest of the integral dimensionless. So every quantity inside the
integral is multiplied by approriate power of $m_{out}$ to make it
dimensionless.
\[
t\rightarrow m_{out}t; \qquad r\rightarrow m_{out}r;
\qquad k\rightarrow k/m_{out}; \qquad m\rightarrow m/m_{out}.
\]
$G(k)$ is an even function of $k$, while $A_d(k,r)$ is an even function
only in odd dimensions but an odd function in even dimensions. This means
for even $d$, the integration is neccesarily to be done for $k$ belonging
to $[0,\infty)$. This has important consequences for us because the
stationary point which is the solution of the equation $w'(k_0)=0$ is
exactly $k_0=0$. So for even $d$, the odd powers of $y$ will also
contribute. Since the most contribution comes near $k=0$, we look at the
behaviour of $f(k)$ near the origin
\begin{equation}
 f(k)=-\frac{\pi^{-\frac{d}{2}}}{2^{d+1} \Gamma \left(\frac{d}{2}\right)}
 \csch(2\kappa_1) k^{d-1} + \frac{\csch(2\kappa_1) \left(d+r^2 +4d\,
 \kappa_3 \coth(2\kappa_1)\right)}{2^{d+3} \pi^{\frac{d}{2}}\Gamma
 \left(\frac{d}{2}+1\right)}k^{d+1} + \mathcal{O}(k^{d+3})
\end{equation}
Therefore the leading contibution comes from the $(d-1)$th derivative
i.e. $f^{(d-1)}_0 \frac{y^{d-1}}{t^{(d-1)/2}}$ which together with the
$e^{-2t}/\sqrt{t}$ goes as $t^{-d/2}$. More explicitly, keeping track
upto the sub-leading term in $I_1$
\begin{align}
 \frac{I_1(t)}{m_{out}^{d-1}} \approx& -\frac{\csch(2\kappa_1)}{2^{d+1}
 \pi^{\frac{d}{2}} \Gamma \left(\frac{d}{2}\right)} \frac{e^{-2t}}
 {t^{d/2}} \int_0^{\infty} dy e^{-y^2}y^{d-1} + \frac{\csch(2\kappa_1)
 \left(d+r^2 +4d\, \kappa_3 \coth(2\kappa_1)\right)}{2^{d+3}
 \pi^{\frac{d}{2}}\Gamma \left(\frac{d}{2}+1\right)} \frac{e^{-2t}}
 {t^{1+d/2}} \int_0^{\infty} dy e^{-y^2} y^{d+1} \nonumber \\
 & -\frac{\csch(2\kappa_1)}{2^{d+1} \pi^{\frac{d}{2}} \Gamma
 \left(\frac{d}{2}\right)} \frac{e^{-2t}}{t^{d/2}} \left(\frac{1}{4t}
 \right) \int_0^{\infty} dy e^{-y^2} y^{d+3} \nonumber \\
 =& -\frac{\csch(2\kappa_1)}{2^{d+2} \pi^{\frac{d}{2}}} \frac{e^{-2t}}
 {t^{d/2}} + \frac{\csch(2\kappa_1) \left(d+r^2 +4d\, \kappa_3
 \coth(2\kappa_1)\right)}{2^{d+4} \pi^{\frac{d}{2}}} \frac{e^{-2t}}
 {t^{1+d/2}} - \frac{d(d+2)\csch(2\kappa_1)}{2^{d+6} \pi^{\frac{d}{2}}}
 \frac{e^{-2t}}{t^{1+d/2}} \nonumber \\
 =& -\frac{\csch(2\kappa_1)}{2^{d+2} \pi^{\frac{d}{2}}} \frac{e^{i2t+i\,
 d\pi/4}}{t^{d/2}} + \frac{\csch(2\kappa_1)(r^2 +4d\, \kappa_3
 \coth(2\kappa_1) - d(d-2)/4)}{2^{d+4} \pi^{\frac{d}{2}}} \frac{e^{i2t+i\,
 (d+2)\pi/4}}{t^{1+d/2}}
\label{tmp1}
\end{align}

\begin{table}[h!]
\begin{center}
    \begin{tabular}{ |l|r| }
        \hline
        \multicolumn{2}{|c|}{$\braket{gCC|\phi(\vec{x_1},t)\phi(\vec{x_2},t)|gCC}$} \\
        \hline
        $d=1$ & $-\frac{\csch(2m_{out} \kappa_1)}{4\sqrt{\pi}}
        \cos(2m_{out}t +\frac{\pi}{4})\frac{1}{\sqrt{m_{out} t}} +
        \mathcal{O} \left(\frac{1}{(m_{out}t)^{3/2}}\right)$\\
        \hline
        $d=2$ & $-\frac{m_{out} \csch(2 m_{out}\kappa_1)}{8\pi} \cos(2
        m_{out} t+ \frac{\pi}{2})\frac{1}{m_{out}t} +\mathcal{O}
        \left(\frac{1}{(m_{out}t)^2}\right)$ \\
        \hline
        $d=3$ & $-\frac{m_{out}^2 \csch(2 m_{out}\kappa_1)}{16\pi^{3/2}}
        \cos(2 m_{out} t+ 3\frac{\pi}{4}) \frac{1}{(m_{out} t)^{3/2}}
        +\mathcal{O} \left(\frac{1}{(m_{out}t)^{5/2}}\right)$\\
        \hline
        $d=4$ & $-\frac{m_{out}^3 \csch(2 m_{out}\kappa_1)}{32\pi^2}
        \cos(2 m_{out} t + \pi)\frac{1}{(m_{out}t)^2} +\mathcal{O}
        \left(\frac{1}{(m_{out}t)^3}\right)$ \\
        \hline
    \end{tabular}\caption{\label{tab:massive-phi-gcc}Time-dependent
    part of 2-point function $\braket{gCC|\phi(\vec{x_1},t)
      \phi(\vec{x_2},t)|gCC}$ in the gCC state defined by $\kappa(k)
    =\kappa_1+ \kappa_3 k^2+ \kappa_5 k^4+\cdots$. Here $m_{out}$ is
    the final mass and $r=|\vec{x_1} -\vec{x_2}|$.}
\end{center}
\end{table}

\begin{table}[h!]
\begin{center}
    \begin{tabular}{ |l|r| }
        \hline
        \multicolumn{2}{|c|}{$\braket{0_{in}|\phi(\vec{x_1},t)\phi(\vec{x_2},t)|0_{in}}$} \\
        \hline
        $d=1$ & $\frac{\left(m_{out}^2-m^2\right)}{8\sqrt{\pi} m m_{out}}
        \cos(2 m_{out} t+\frac{\pi}{4}) \frac{1}{\sqrt{m_{out} t}} +\mathcal{O} \left(\frac{1}{(m_{out}t)^{3/2}}\right)$ \\
        \hline
        $d=2$ & $\frac{\left(m_{out}^2-m^2\right)}{16\pi m} \cos(2 m_{out} t+\frac{\pi}{2}) \frac{1}{m_{out}t} +\mathcal{O}\left(\frac{1}{(m_{out}t)^2}\right)$ \\
        \hline
        $d=3$ & $\frac{\left(m_{out}^2-m^2\right)m_{out}}{32\pi^{3/2} m}  \cos(2 m_{out} t+\frac{3\pi}{4}) \frac{1}{(m_{out}t)^{3/2}}+\mathcal{O} \left(\frac{1}{(m_{out}t)^{5/2}}\right)$\\
        \hline
        $d=4$ & $\frac{\left(m_{out}^2-m^2\right) m_{out}^2}{64\pi^2m} \cos(2 m_{out}t +\pi) \frac{1}{(m_{out}t)^2} +\mathcal{O} \left(\frac{1}{(m_{out}t)^3}\right)$ \\
        \hline
    \end{tabular}\caption{\label{tab:massive-phi}Time-dependent part
    of 2-point function $\braket{0_{in}|\phi(\vec{x_1},t)
    \phi(\vec{x_2},t)|0_{in}}$ in massive quench. $r=|\vec{x_1}
    -\vec{x_2}|$.}
\end{center}
\end{table}

The third term in the first line of \eq{tmp1} comes from the product of the leading $k^{d-1}$ term in expansion of $f$ and $w^{(4)}y^4/t$ brought down from the exponential. In the second line we have performed the integral using the general formula $\int_0^\infty dy\, y^{d} e^{-y^2}= \frac{1}{2} \Gamma(\frac{d+1}{2})$ and finally we Wick rotate back to  $t\rightarrow -it$. $I_2$ is calculated in a similar manner with the only difference being that in the final answer replace $i\rightarrow -i$ (due to Wick rotation in opposite direction). $I_1$ and $I_2$ thus combine nicely to give a cosine. Reintroducing the dimensions, the time-dependent part of 2-point function in general is given by
\begin{align}
 \braket{\phi(\vec{x_1},t) \phi(\vec{x_2},t)}^{(td)}=& -\frac{m_{out}^{d-1} \csch(2 \kappa_1)}{2^{d+1} \pi^{d/2}} \frac{\cos \left(2 m_{out}
 t+ d\frac{\pi}{4}\right)}{(m_{out}t)^{d/2}} + \nonumber \\
 & \kern-60pt\frac{\csch(2\kappa_1)(r^2 +4d\, \kappa_3 \coth(2\kappa_1) - d(d-2)/4)}
 {2^{d+4} \pi^{\frac{d}{2}}} \frac{\cos \left(2 m_{out} t+ (d+2)\frac{\pi}
 {4}\right)}{(m_{out}t)^{1+d/2}} + \mathcal{O}\left(\frac{1}{(m_{out}
 t)^{d/2 + 2}}\right) \label{eq:massive-2pt-general}
\end{align}
The above expression tells us that the leading transient only depends
on $\kappa_1$. It neither cares about the the higher conserved charges
nor the separation $r$ which only show at subleading order. We
tabulate the leading approach to equilibrium of (\ref{2pt-gcc-etc}) in
Table \ref{tab:massive-phi-gcc}. We also tabulate the 2-point function
in the ground state state (\ref{eq:ground-quench-etc}) in Table
\ref{tab:massive-phi}.

% There is another scaling regime which might be of interest. Here we scale $r=vt$ with $v\le 1$, the speed of massless excitations. In particular we would be look at $s\ll 1$.
%%%%
\subsubsection{$\partial_t \phi \partial_t \phi$ and $\partial_i \phi \partial_i \phi$ Correlators}
We also calculate the time-dependent piece of the correlator $\braket{\partial_{t_1}\phi(\vec{x_1},t_1) \partial_{t_2}\phi(\vec{x_2},t_2)}\rvert_{t_1=t_2=t}$ eq. (\ref{eq:ground-quench-etc-delt}).

\begin{table}[h]
%\begin{flushleft}
\hskip-1.5cm   \begin{tabular}{ |p{0.83cm}|p{8.58cm}|p{8.45cm}| }
        \hline
        &$\braket{\partial_t \phi \partial_t \phi}_{td}$ & $\braket{\partial_i \phi \partial_i \phi}_{td}$ \\
        \hline
        $d=1$ & $-\frac{m_{out} \left(m_{out}^2-m^2\right) \cos(2m_{out} t+\frac{\pi}{4})}{8\sqrt{\pi} m} \frac{1}{\sqrt{m_{out} t}} +\mathcal{O} \left(\frac{1}{(m_{out}t)^{3/2}}\right)$ & $\frac{m_{out} \left(m_{out}^2-m^2\right)\cos(2m_{out}t+\frac{3\pi}{4})}{16\sqrt{\pi} m} \frac{1}{(m_{out}t)^{3/2}}+ \mathcal{O}\left(\frac{1}{(m_{out}t)^{5/2}}\right)$\\
        \hline
        $d=2$ & $-\frac{m_{out}^2 \left(m_{out}^2-m^2\right) \cos(2m_{out} t+\frac{\pi}{2})}{16\pi m} \frac{1}{m_{out}t} +\mathcal{O} \left(\frac{1}{(m_{out}t)^2}\right)$ & $\frac{m_{out}^2 \left(m_{out}^2-m^2\right)\cos(2m_{out}t+\pi)}{16\pi m} \frac{1}{(m_{out}t)^2}+ \mathcal{O}\left(\frac{1}{(m_{out}t)^3}\right)$\\
        \hline
        $d=3$ & $-\frac{m_{out}^3 \left(m_{out}^2-m^2\right) \cos(2m_{out} t+\frac{3\pi}{4})}{32\pi^{3/2} m} \frac{1}{(m_{out}t)^{3/2}} + \mathcal{O} \left(\frac{1}{(m_{out}t)^{5/2}}\right)$ & $\frac{3 m_{out}^3 \left(m_{out}^2-m^2\right)\cos(2m_{out}t+ \frac{5\pi}{4})}{32\pi^{3/2} m} \frac{1}{(m_{out}t)^{5/2}}+ \mathcal{O}\left(\frac{1}{(m_{out}t)^{7/2}}\right)$\\
        \hline
        $d=4$ & $-\frac{m_{out}^4 \left(m_{out}^2-m^2\right) \cos(2m_{out}t+ \pi)}{64\pi^2 m} \frac{1}{(m_{out}t)^2} +\mathcal{O} \left(\frac{1}{(m_{out}t)^3}\right)$ & $\frac{m_{out}^4 \left(m_{out}^2-m^2\right)\cos(2m_{out}t+ \frac{3\pi}{2})}{16\pi^2 m} \frac{1}{(m_{out}t)^3}+ \mathcal{O}\left(\frac{1}{(m_{out}t)^4}\right)$\\
        \hline
    \end{tabular}\caption{\label{tab:massive-del-phi}Time-dependent
    part of 2-point function
    $\braket{0_{in}|\partial_t\phi(\vec{x_1},t)
      \partial_t\phi(\vec{x_2},t)|0_{in}}$ on the left and
    $\braket{0_{in}|\partial_i\phi(\vec{x_1},t)\partial_i
      \phi(\vec{x_2},t)|0_{in}}$ on the right, both for massive
    quench. Here $r= |\vec{x_1}-\vec{x_2}|$.}
%\end{flushleft}
\end{table}
It is a bit surprising that the leading power of $t$ here is also same as before, but it is easy to understand why. For a moment let us rewrite $2t=t_1+t_2$. Then take derivatives on $\braket{\phi \phi}$  correlator (in 1+1 dim. for example) wrt $t_1$ and  $t_2$, and then restore $t_1=t_2=t$.
\[
 \partial_{t_1}\partial_{t_2}\left[\cos \left( m_{out} (t_1+t_2)
 +\frac{\pi}{4}\right) \frac{2^{1/2}}{(t_1+t_2)^{1/2}} \right]\Bigg|
 _{t_1=t_2=t} =-m_{out}^2 \cos(2m_{out}t+\frac{\pi}{4}) \left(\frac{1}{t}
 \right)^{1/2} +\mathcal{O} \left(\frac{1}{t^{3/2}}\right)
\]
Clearly the leading power is unchanged when both the derivatives act on the cosine. However the $\partial_i \phi$ 2-point function in any gCC state, following the same argument as given after eq.(\ref{eq:massive-saddle}), behaves as
\begin{equation}
 \braket{\partial_i \phi(\vec{x_1},t) \partial_i \phi(\vec{x_2},t)}_{td}
 =-\frac{d\, m_{out}^{d-1} \csch(2 m_{out} \kappa_1)}{2^{d+2} \pi^{d/2}}
 \frac{\cos \left(2 m_{out} t+ (d+2)\frac{\pi}{4}\right)}{(m_{out}t)^{d
 /2}} +\mathcal{O}\left(\frac{1}{(m_{out}t)^{d/2+ 1}}\right)
\end{equation}
Explicit expressions in various dimensions for these two-point functions for the ground state are tabulated in Table \ref{tab:massive-del-phi}.

%%%%
\subsection{Some comments on 2-point functions in critical vs. massive quench}\label{sec:crit-massive}

From the results of the previous subsections it seems that the
critical correlators behave in a qualitatively different fashion
from the those for the massive case, especially in odd spatial
dimensions. This seems to indicate an apparent discontinuity as
$m_{out}\to 0$. This is surprising since the master formula
(\ref{f-g-crit}) for the critical correlator was obtained from that
of the massive case (\ref{eq:ground-quench-etc}) by putting
$m_{out}=0$!

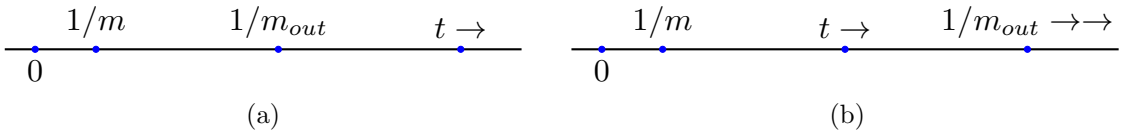
\begin{figure}[h!]
  \begin{minipage}{.45\textwidth}
 \centering
 \begin{tikzpicture}[scale = .4, every node/.style={scale=1.2}]
 \draw[black,thick] (-8,0)--(9,0);
 \node [below] at (-7,0) {$0$};
% \node [below] at (1,0) {$r$};
 \node [above] at (1,0) {$1/m_{out}$};
 \node [above] at (7,0) {$t \to $};
 \node [above] at (-5,0) {$1/m$};
% \draw [fill,blue] (1,0) circle [radius=0.1];
 \draw [fill,blue] (1,0) circle [radius=0.1];
 \draw [fill,blue] (7,0) circle [radius=0.1];
 \draw [fill,blue] (-7,0) circle [radius=0.1];
 \draw [fill,blue] (-5,0) circle [radius=0.1];
 %\draw[gray,dashed] (3,-1)--(3,1);
 %\node [below] at (3,-1) {B: $m_{out}t$};
 %\draw[gray,dashed] (-3,-1)--(-3,1);
 %\node [below] at (-3,-1) {A: $m_{out}t$};
 \end{tikzpicture}\\
 (a)\\
  \end{minipage}~~
\begin{minipage}{.45\textwidth}
 \centering
 \begin{tikzpicture}[scale = .4, every node/.style={scale=1.2}]
 \draw[black,thick] (-8,0)--(10,0);
 \node [below] at (-7,0) {$0$};
% \node [below] at (1,0) {$r$};
 \node [above] at (1,0) {$t \to $};
 \node [above] at (7,0) {$1/m_{out} \to\to$};
 \node [above] at (-5,0) {$1/m$};
% \draw [fill,blue] (1,0) circle [radius=0.1];
 \draw [fill,blue] (1,0) circle [radius=0.1];
 \draw [fill,blue] (7,0) circle [radius=0.1];
 \draw [fill,blue] (-7,0) circle [radius=0.1];
 \draw [fill,blue] (-5,0) circle [radius=0.1];
 %\draw[gray,dashed] (3,-1)--(3,1);
 %\node [below] at (3,-1) {B: $m_{out}t$};
 %\draw[gray,dashed] (-3,-1)--(-3,1);
 %\node [below] at (-3,-1) {A: $m_{out}t$};
 \end{tikzpicture}\\
(b)\\
  \end{minipage}  
\caption{\label{fig:scales} Relevant scales in the theory for the
  present consideration. Here $r/t$ is considered fixed as the large
  $t$ limit is taken.  In (a) $m_{out}$ is a finite non-zero
  quantity. $t \to \infty$ limit means $m_{out} t \to \infty$. This is the case we discuss in the analysis of large time behaviour for quench to non-zero mass. In (b) the limit $m_{out}\to 0$ limit is taken first, so that $m_{out}
  t \to 0$; the large $t$ limit in this case can be defined as $m\, t
  \to \infty$. We do not consider this limit in the paper except when $m_{out}=0$ (this limit is discussed a bit more at the end of the subsection).}
\end{figure}

Actually there is no true discontinuity at $m_{out}=0$ in the exact
time-dependent correlators. The apparent discontinuity emerges
because the limits $m_{out} \to 0$ and $t\to \infty$ do not
commute. This can be understood by considering the scales
involved in the theory. For any given  mass $m_{out}$, however small,
at very large times we will have $t \gg 1/m_{out}$  (see 
Fig. \ref{fig:scales} (a)). In other words, the dimensionless
quantity $\tilde t = m_{out} t \gg 1$.

However when $m_{out}=0$, $1/m_{out}=\infty$. No $t$, however large,
can exceed $1/m_{out}$. Indeed, the limit $m_{out} \to 0$
automatically implies $\tilde t= m_{out} t \to 0$ (see
Fig. \ref{fig:scales} (b)).

More mathematically speaking, the two-point functions involve scaling
functions of the form $F(m t, m_{out} t)$ (omitting the dimensionless
variable $r/t$). For the massive two-point functions, late times imply
that both scaled variables are taken to infinity; in particular
$m_{out} t \to \infty$. Indeed, the late time massive correlators in
the previous section are given in terms of an expansion in $1/(m_{out}
t)$; as explained above, this expansion cannot capture the limit
$m_{out} t\to 0$. Of course, if an exact evaluation of the relevant integral is possible, this limit can be taken.\footnote{Note that
in $d=1$ and 2, there are infrared divergences associated with this limit for $\braket{\phi \phi}$ correlators but these problems are not there in correlators of the type  $\braket{\partial \phi \partial \phi}$.} Such a
function would then interpolate between the critical and non-critical
behaviour. The apparent discontinuity at $m_{out}=0$ is only an
artifact of the large time asymptotic expansion.

%%%%%%%%
\section{The Generalized Gibbs Ensemble (GGE)}\label{sec:thermalization-to-GGE}

We have seen above that the time-dependent part of the two-point
functions (the part containing the cosine term in \eq{form-2-pt})
vanishes at large times. The remaining term, therefore, represents
the asymptotic value of the two-point function. We will now show 
that the asymptotic value corresponds to that of the two-point function
in a generalized Gibbs ensemble (GGE). Before that, we will first
recall some relevant facts about a GGE.

\subsection{GGE}\label{sec:GGE}

The generalized Gibbs ensemble (GGE) can be regarded as a certain
grand canonical ensemble with an infinite number of chemical
potentials \cite{Rigol:2007, Rigol:2007a, Calabrese:2012GGE} which are
appropriate for the description of equilibrium for an integrable
model. An integrable model has an infinite number of
commutative conserved charges $\hat Q_i$. A GGE is described by
a density matrix
\[
\rho_{GGE}= \f1{Z_{GGE}}\exp[-\sum_i \mu_i \hat Q_i],\;
Z_{GGE}= \Tr \exp[-\sum_i \mu_i \hat Q_i]
\]
Let us consider a free scalar field $\phi(x,t)$, described by
the standard mode expansion
\begin{align}
\phi(\vec{k},t)= a(\vec{k},t) u(\vec{k},t) + cc,\;
u(k,t) =\frac{e^{-i\omega(k) t}}{\sqrt{2 \omega(k)}},\;\omega(k) =
\sqrt{k^2 + m_0^2}
\label{free-scalar}
\end{align} 
and a Hamiltonian $H= \sum_k \omega(k) \hat N(k)$, where $\hat N(k)
= a^\dagger(k) a(k)$ are the occupation numbers.

This system is clearly integrable;
a standard basis of the commuting conserved charges is ({\it cf.}
\eq{charges})
$\hat Q_i= \sum_k |k|^{i-1} \hat N(k)$, $i=1,2,...$.
Another simple basis of charges are the set of occupation numbers
$\hat N(k)$ themselves. With this choice the GGE is described by
the density matrix 
\begin{equation}
  \rho_{GGE} = \frac{1}{Z_{GGE}}\exp(-\sum_{\vec{k}}\mu(k)\hat{N}(k)),\;
  Z_{GGE}=\Tr \exp(-\sum_{\vec{k}}\mu(k)\hat{N}(k))
  \label{rho-gge}
\end{equation}
The GGE partition is easy to evaluate (most calculations in
this subsection are presented in greater detail in Appendix
\ref{app:gge}):
\begin{equation}
  Z_{GGE}=\sum_{\{N(k)\}, N(k)=0,1,2,..}\, \exp(-\sum_{\vec{k}}\mu(k){N}(k))
  = \prod_k (1- \exp[-\mu(k)])^{-1}
  \label{z-gge}
\end{equation}
It is also easy to evaluate the average value of the number operator
\begin{align}
  \lan \hat N(k) \ran_{GGE}= \Tr(\rho_{GGE} \hat N(k))=
  -\frac\del{\del \mu(k)} \ln Z_{GGE}= 
  \f1{e^{\mu(k)}-1}
    \label{nk-gge}
    \end{align}
With the above ingredients, we can now compute the equal time
two-point function in the GGE (for more details and also the
case of unequal time correlator, see Appendix \ref{app:gge}):
\begin{align}
  & \lan \phi(\vec{x_1},t) \phi(\vec{x_2},t) \ran_{GGE}
  = \Tr [\rho_{GGE} \phi(\vec{x_1},t) \phi(\vec{x_2},t)]
 \nonumber \\  
  =& \int \frac{d^d k}{(2\pi)^d} |u(\vec{k},t)|^2 e^{i\vec{k}\cdot(\vec{x_1}-\vec{x_2})}[2 \lan N(k)\ran_{GGE}+1]
% \nonumber \\
 = \int \frac{d^d k}{(2\pi)^d}\f1{2 \sqrt{k^2+m_0^2}} e^{i\vec{k}\cdot(\vec{x_1}-\vec{x_2})} \coth(\mu(k)/2)
\label{2pt-gge}
\end{align}

Note that for massless scalars, $H=\sum_k |k| a^\dagger(k) a(k)= Q_2$.
Also the total occupation number is always $N= \sum_k a^\dagger(k)
a(k)= Q_1$.  Hence, we can write \eq{rho-gge} as 
\begin{align}
&\rho_{GGE}= \f1{Z_{GGE}}\exp\left[-\left(\mu N + \beta H + \sum_{i=3}^\infty
   \mu_k \hat Q_i\right)\right],\; Z_{GGE}= \Tr
\exp\left[-\left(\mu N + \beta H + \sum_{i=3}^\infty \mu_k \hat
  Q_i\right)\right]
\nonumber\\
&\beta := \mu_2, \mu:= \mu_1 
\label{GGE-GE}
\end{align}
Thus, a standard thermal (Gibbs) ensemble is a GGE with all $\mu_{i\not=2}=0$.

\subsection{Equilibration to GGE}\label{sec:eq-to-gge}

Note that the two-point function in the most general gCC state
\eq{2pt-gcc-etc} is of the form
\begin{equation}
\braket{gCC|\phi(\vec x_1,t)\phi(\vec x_2,t)|gCC} = \int \frac{d^d k}{(2\pi)^d} \frac{e^{i \vec{k}\cdot (\vec{x_1}-\vec{x_2})}}{2 \sqrt{k^2+m_{out}^2}} \left[\coth \left(2\kappa(k)\right) + \hbox{time dependent} \right] 
\label{2pt-gcc-again}
\end{equation}
In the preceding sections we have shown that the time-dependent part
decays at late times exponentially or by a power law. Hence the gCC
two-point function \eq{2pt-gcc-again} asymptotically approaches the
GGE two-point function \eq{2pt-gge} provided we identify
\begin{align}
  m_0 = m_{out}, \; \mu(k)= 4 \kappa(k),\; \hbox{equivalently}\;
  \mu_i=4\kappa_i
  \label{mu-kappa}
\end{align}
Indeed, these relations ensure
\begin{align}
\lan N(k) \ran_{GGE}= \lan gCC| N(k) | gCC \ran
  \label{n-n}
  \end{align}
To prove this, note that by identifying $|gCC\ran$ with $|f_{in}\ran$
and using the top line of \eq{massive-gcc-squeezed}, we get that the
right hand side of the above equation is given by
$ (1/\g_{eff}^2 - 1)^{-1}$. By using $\g_{eff}(k)= \exp[-2\kappa(k)]$
(see immediately above \eq{eq:squeezing-function}), the RHS
becomes $(e^{2\kappa(k)}-1)^{-1}$ which agrees with equation (\eq{nk-gge}) with the identification \eq{mu-kappa}. Q.E.D.

Thus, we have established the following statement of equilibration:
\begin{align}
  \braket{gCC|\phi(\vec x_1,t)\phi(\vec x_2,t)|gCC}
  = \lan \phi(\vec{x_1},t) \phi(\vec{x_2},t) \ran_{GGE} +
   \left(\sim e^{-\gamma t} ~\hbox{or}~ \sim 1/t^p\right)
\label{stmt-thermalization}
\end{align}
where the GGE is defined in terms of the gCC by the basic relation
\eq{n-n}. Since for a free field theory, all correlation functions
can be essentially built from the two-point function, the above
statement of equilibration is true for all correlators. This
constitutes a proof of the GGE version of quantum ergodic hypothesis
in the quench models considered in this paper:
\[
\lan \psi(t) | O(x_1) O(x_2) O(x_3) ... | \psi(t)\ran
\xrightarrow{t\to\infty}
\lan O(x_1) O(x_2) O(x_3)\ran_{GGE}
\]

Eq. \eq{mu-kappa} gives us the important relation between the
parameters of the equilibrium ensemble (GGE) with those of the initial
state. In the massless case, in terms of the notation of \eq{GGE-GE},
we can rewrite \eq{mu-kappa} as
\begin{align}
  \beta=4 \kappa_2;\;\; \mu_i = 4 \kappa_i, i\not=2
  \label{mu-kappa-m=0}
\end{align}
It is clear that the CC state \eq{eq:f2-state} equilibrates to a standard
Gibbs ensemble with $\mu_{i\not=2}=0$. 

%%%%%%%%
\section{Geometrical interpretation of the correlators in the CC state}\label{sec:approach-thermalization}

\begin{figure}[h]
\centering
\begin{tikzpicture}[scale = .6, every node/.style={scale=1.0}]

\draw[-,black, thick] (-12,3.0) -- (-4,3.0);

\node[above] at (-3.5,3.0) {$\tau=\frac{L}{2}$};
\node [above] at (-8.0,0.8) {$\phi(\tau_2)$};
\node [above] at (-8.0,2.0) {$\phi(\tau_1)$};

\draw [fill,black] (-8.0,0.8) circle [radius=0.08];
\draw [fill,red] (-8.0,2.0) circle [radius=0.08];

\draw [<->,black,thick] (-11,-3.0)--(-11,3.0);
\node [above] at (-11.5,0){$\frac{\beta}{2}$};

\draw [<->,black,thick] (-10,-3.0)--(-10,0.0);
\node [above] at (-9.6,-1.6){$\kappa$};
\draw[-,black, dotted] (-12,0.0) -- (-4,0.0);
\node [above] at (-3.5,-0.2) {$\tau=0$};

\draw[-,black, thick] (-12,-3.0) -- (-4,-3.0);
\node[below] at (-3.5,-3.0) {$\tau=-\frac{L}{2}$};

\draw[-,black, dotted] (0,3.0) -- (8,3.0);
\draw[-,black, dotted] (0,1.5) -- (8,1.5);
\draw[-,black, dotted] (0,-1.5) -- (8,-1.5);
\draw[-,black, dotted] (0,-3.0) -- (8,-3.0);
\draw[-,black, dotted] (0,0) -- (8,0);

\draw [<->,black,thick] (1,0.0)--(1,1.5);
\node [right] at (1.4,0.5){$\frac{L}{2}$};
\draw [fill,red] (4.0,0.5) circle [radius=0.08];
\node [above] at (4.0,0.5) {$+$};
\node [right] at (4.0,0.5) {$\phi(\tau_1)$};
\draw [fill,green] (4.0,-0.5) circle [radius=0.08];
\node [above] at (4.0,-0.6) {$-$};
\node [right] at (4.0,-0.6) {$(-\tau_1)$};
\draw [fill,red] (4.0,-2.5) circle [radius=0.08];
\node [above] at (4.0,-2.5) {$+$};
\node [right] at (4.0,-2.5) {$(-L+\tau_1)$};
\draw [fill,green] (4.0,-3.5) circle [radius=0.08];
\node [above] at (4.0,-3.6) {$-$};
\node [right] at (4.0,-3.6) {$(-L-\tau_1)$};
\draw [fill,green] (4.0,2.5) circle [radius=0.08];
\node [above] at (4.0,2.4) {$-$};
\node [right] at (4.0,2.4) {$(L-\tau_1)$};
\draw [fill,red] (4.0,3.5) circle [radius=0.08];
\node [above] at (4.0,3.5) {$+$};
\node [right] at (4.0,3.5) {$(L+\tau_1)$};
\end{tikzpicture}
\caption{\label{fig:slab} The two-point function between $\phi(\tau_1)$ and $\phi(\tau_2)$ can be regarded as the electrostatic potential between two positive charges of unit magnitude\cite{Sotiriadis:2010si}. This follows from the fact that the massless Klein-Gordon equation in Euclidean space is simply the Laplace equation. Here we consider $\phi(\tau_1)$ as a source charge whose electrostatic potential in the slab geometry can be found out by the method of images by regarding the slab as two parallel mirrors. In the right panel we depict by green(red) circles the images obtained by an odd(even) number of reflections of the source. The electrostatic potential in the slab can now be regarded as a sum of that between the probe charge placed at $\tau_2$ and each of the images.}
\end{figure}
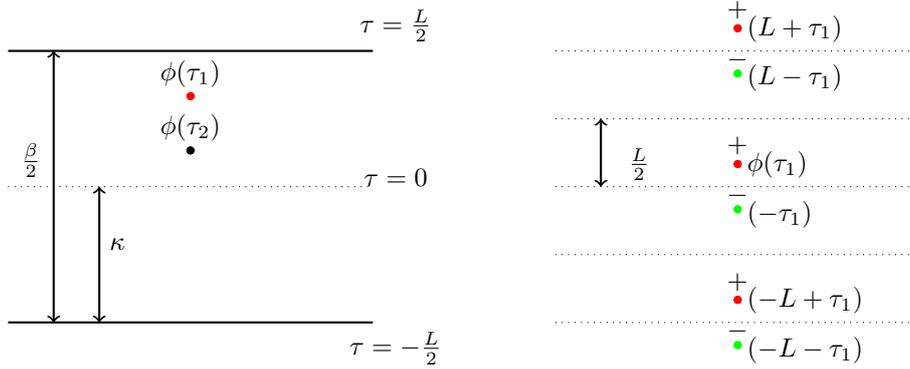

Here we follow \cite{Sotiriadis:2010si} to show how the two-point
function in a CC state can be identified with an image sum in a slab
geometry $R^d \times \hbox{interval}$. This would help us better
understand the origin of the difference in thermalization between odd
and even dimensions (in short, the {\it odd-even effect}). The slab
propagator is defined as
\begin{align}
G_{slab}(\tau_1,\vec{x_1};\tau_2,\vec{x_2}) &= \langle \phi(\tau_1,\vec{x_1}) \phi(\tau_2,\vec{x_2}) \rangle_{slab} \nonumber \\
&= \int \frac{d^d k}{(2 \pi)^d} e^{i \vec{k}\cdot (\vec{x_1}-\vec{x_2})} G_{slab}(\tau_1,\tau_2;\vec{k}) \nonumber \\
&= \int \frac{d^d k}{(2 \pi)^d}\frac{e^{i \vec{k}\cdot (\vec{x_1}-\vec{x_2})}}{2|\vec{k}|} \left(S_{th}(\tau_1-\tau_2) + S_{\text{diff}}(\tau_1+\tau_2)\right)
\label{g-slab}
\end{align}
where
\begin{align}
S_{\text{th}}(\tau)=& \sum_{n=0}^{\infty} e^{-|\vec{k}|(2nL + \tau)} + \sum_{n=1}^{\infty} e^{-|\vec{k}|(2nL- \tau)}\nonumber \\
S_{\text{diff}}(\tau')=& -\sum_{n=0}^{\infty} e^{-|\vec{k}|((2n+1)L + \tau')} - \sum_{n=1}^{\infty} e^{-|\vec{k}|((2n-1)L- \tau')} 
\end{align}
The expression \eq{g-slab} is obtained by the method of images as
shown in the diagram below (see figure \ref{fig:slab}). The geometry
reflects the fact that a CC state of the form $e^{-\kappa_2 H}|Bd\ran$
can be regarded as a Euclidean time evolution $\Delta \tau= \kappa_2$,
here denoted by $L/2$. The suffices $_{th}$ and $_{diff}$ will be
explained below.

It is easy to verify that \eq{g-slab} reproduces the two-point function
in a CC state. All the four terms in \eq{g-slab} can be easily summed to
obtain
\[
S_{\text{th}}(\tau_1-\tau_2) + S_{\text{diff}}(\tau_1+\tau_2) = \frac{1}{(1- e^{-2 |\vec{k}| L})} \left[e^{-|\vec{k}| \left| \tau_1- \tau_2 \right|} + e^{-|\vec{k}| (2L - \left| \tau_1- \tau_2 \right|)} -2 e^{-|\vec{k}| L} \cosh(|\vec{k}| (\tau_1 + \tau_2)) \right]
\]
Now analytically continuing to real time by $\tau \rightarrow it$ and identifying $\kappa_2=L/2$ we find the slab progator is the same as (\ref{2pt-gcc}) with $\kappa_2=L/2$ and all other $\kappa_i=0$
\begin{align}
G_{slab}(t_1,\vec{x_1};t_2,\vec{x_2}) &= \int \frac{d^d k}{(2\pi)^d} \frac{e^{i\vec{k}\cdot (\vec{x_1}-\vec{x_2})}}{2|\vec{k}|} \csch(2 \kappa_2|\vec{k}|) \left[ \cos(|\vec{k}| (2i\kappa_2 + t_1-t_2)) -\cos(|\vec{k}|(t_1 +t_2))\right]
\end{align}
This completes the explicit check that the slab propagator \eq{g-slab}
indeed represents the two-point function in the CC state.

%%%%
\subsection{Comments on approach to thermalization and the odd-even effect}

\begin{comment}
From the tables, one can see that for the critical quench, i.e. when the final theory becomes conformal,

(1) For odd $d>1$
\[
 \langle \phi(r,t)\phi(0,t)\rangle \rightarrow \frac{1}{Z} \text{tr} \left(\rho\,\phi(r)\phi(0) \right) + e^{-\gamma t}
\]
with
\[
 \gamma = \frac{4\pi}{\beta}
\]
(2) For even $d>2$
\[
 \langle \phi(r,t)\phi(0,t)\rangle \rightarrow \frac{1}{Z} \text{tr}\left(\rho\,\phi(r)\phi(0)\right) + \frac{1}{t^{d-2}}
\]
This is true for all the initial states, ground state, CC state even for any arbitrary gCC state.
\end{comment}

%

The understanding of thermalization from the above geometrical picture
goes as follows. First, note that the terms in $S_{\text{th}}$
correspond to Euclidean $R^{d+1}$ 2-point function between the test
charge with the `+' images above and below it. These are precisely the
terms which would appear in a cylindrical geometry, whch represents a
thermal 2-point function with Euclidean time period $\beta= 2L_2= 4
\kappa_2$; hence the suffix $_{th}$.  The terms in $S_{\text{diff}}$,
on the other hand, correspond to $R^{d+1}$ 2-point functions between
the test charge with the `-' images above and below it; below we will
find that these terms vanish at long times, proving that the CC
two-point function asymptotically approaches the thermal 2-point
function $S_{\text{th}}$, $\beta= 4 \kappa_2$, which verifies the
relation \eq{mu-kappa-m=0}.

It is clear from the above that $S_{\text{diff}}$ represents the
difference between CC correlator and the thermal correlator. For
thermalization of the auto-correlator $(\vec x=\vec y= \vec r, t_1
\not= 0)$, we need to consider the following difference
\begin{align}
&G_{CC}(\tau_1,\vec{r};\tau_2,\vec{r})-G_{th}(\tau_1,\vec{r};\tau_2,\vec{r})_\beta %\nonumber \\ &
= \frac{\Omega_{d-1}}{2 (2 \pi)^d}\int_0^{\infty} dk\, k^{d-2} S_{\text{diff}}(\tau_1+\tau_2) \nonumber \\
&= -\frac{\Omega_{d-1}}{2 (2 \pi)^d}\int_0^{\infty} dk\, k^{d-2} \left( \sum_{n=0}^{\infty} e^{-k((2n+1)L + \tau)} + \sum_{n=1}^{\infty} e^{-k((2n-1)L- \tau)} \right) \nonumber \\
&= -\frac{\Omega_{d-1} (d-2)!}{2 (2 \pi)^d} \left[\left(\frac{1}{L+\tau}\right)^{d-1}+ \sum_{n=1}^{\infty} \left(\frac{1}{(2n+1)L +\tau}\right)^{d-1} + \sum_{n=1}^{\infty} \left(\frac{1}{(2n-1)L-\tau}\right)^{d-1} \right]
\end{align}
Where $\tau=\tau_1+\tau_2$. In the last line we have performed
the $k$ integral. Note that since $\tau_1$ and $\tau_2$ lie in the interval $[-L/2,L/2]$, allowed values of $\tau$ are in the interval $[-L,L]$.

To see the difference between odd and even $d$, let
us choose particular values of $d$.

\paragraph{$\boldsymbol{d=4:}$}

We have
\begin{align}
 &G_{CC}(\tau_1,\vec{r};\tau_2,\vec{r})-G_{th}(\tau_1,\vec{r};\tau_2,\vec{r})_\beta \nonumber \\&= -\frac{1}{8\pi^2} \left[\left(\frac{1}{L+\tau}\right)^3+ \sum_{n=1}^{\infty} \left(\frac{1}{(2n+1)L +\tau}\right)^3 + \sum_{n=1}^{\infty} \left(\frac{1}{(2n-1)L-\tau}\right)^3 \right]
\end{align}
Now we use the Euler-Maclaurin sum formula for an infintely differentiable function
\begin{equation}
 \sum_{n=1}^{\infty} f(n)= \int_0^{\infty} dy\,f(y) +\frac{f(\infty)-f(0)}{2} + \sum_{k=1}^{\infty} \frac{B_{2k}}{(2k)!} \left(f^{(2k-1)}(\infty)-f^{(2k-1)}(0) \right)
 \label{Euler-Maclaurin}
\end{equation}
and apply it to the 2 sums above to get
\begin{align}
 & S_+= \sum_{n=1}^{\infty} \left(\frac{1}{(2n+1)L +\tau}\right)^3 = \frac{1}{4 L(L+\tau)^2}-\frac{1}{2(L+\tau)^3}+\frac{L}{2(L+\tau)^4}+ \mathcal{O}((L+\tau)^{-6}) \nonumber \\
 & S_-= \sum_{n=1}^{\infty} \left(\frac{1}{(2n-1)L -\tau}\right)^3 =\frac{1}{4 L(L+\tau)^2}+\frac{1}{2(L+\tau)^3} +\frac{L}{2(L+\tau)^4} + \mathcal{O}((L+\tau)^{-6})
\end{align}
and therefore
\begin{align}
 G_{CC}(\tau_1,\vec{r};\tau_2,\vec{r})-G_{th}(\tau_1,\vec{r};\tau_2,\vec{r})_\beta = -\frac{1}{8\pi^2} \left(\frac{1}{2 L(L+\tau)^2}+\frac{1}{(L+\tau)^3}+\frac{L}{(L+\tau)^4} \right) + \mathcal{O}((L+\tau)^{-6})
\end{align}
Analytically continuing to real time ($\tau_1 \rightarrow it$ and $\tau_2 \rightarrow it$), for $t\rightarrow \infty$\footnote{Note that although $|\tau| \le L$, there is no such restriction on $t$ which can be taken to infinity.} we find (after putting $L=2\kappa_2$)
\begin{align}
 G_{CC}(t,\vec{r};t,\vec{r})-G_{th}(t,\vec{r};t,\vec{r})_\beta = \frac{1}{128 \pi^2 \kappa_2 t^2} +\frac{\kappa_2}{128 \pi^2 t^4} + \mathcal{O}(t^{-6})
\end{align}
This gives us the power law decay in even dimensions. Also this is precisely the expression for the 2-point function of $\phi$ in the CC state ($\kappa_{2n}=0,n>1$) for $r=0$. We match the Euler-Maclaurin result with the method used before to calculate eq. (\ref{power-law-subleading}) to $\mathcal{O}(t^{-20})$ and find exact agreement. Notice that there is a coincident singularity because of $t_1$=$t_2$, but that would show up in `thermal' part $G_{th}$, so we do not need to worry about it here. Now lets see what happens in odd dimensions.

\paragraph{$\boldsymbol{d=3:}$} 
Here we have
\begin{align}
 &G_{CC}(\tau_1,\vec{r};\tau_2,\vec{r})-G_{th}(\tau_1,\vec{r};\tau_2,\vec{r})_\beta \nonumber \\&= -\frac{1}{8\pi^2} \left[\left(\frac{1}{L+\tau}\right)^2+ \sum_{n=1}^{\infty} \left(\frac{1}{(2n+1)L +\tau}\right)^2 + \sum_{n=1}^{\infty} \left(\frac{1}{(2n-1)L-\tau}\right)^2 \right]
\end{align}
The Euler-Maclaurin sum formula applied here results in
\begin{align}
 & S_+= \sum_{n=1}^{\infty} \left(\frac{1}{(2n+1)L +\tau}\right)^2 =\frac{1}{2 L(L+\tau)}-\frac{1}{2(L+\tau)^2} +\frac{L}{3 (L+\tau)^3}+ \mathcal{O}((L+\tau)^{-5}) \nonumber \\
 & S_-= \sum_{n=1}^{\infty} \left(\frac{1}{(2n-1)L -\tau}\right)^2 =-\frac{1}{2L (L+\tau)}-\frac{1}{2 (L+\tau)^2}-\frac{L}{3 (L+\tau)^3} + \mathcal{O}((L+\tau)^{-5})
\end{align}
Together with the $n=0$ term
\begin{align}
 G_{CC}(\tau_1,\vec{r};\tau_2,\vec{r})-G_{th}(\tau_1,\vec{r};\tau_2,\vec{r})_\beta = -\frac{1}{8\pi^2} \left(\frac{1}{(L+\tau)^2} + S_+ + S_-\right) = 0 
\end{align}
We expect the vanishing result found above to continue to all orders in the $1/t$ expansion (we have explicitly checked this fact till $\mathcal{O}(t^{-20}$). Indeed this had to be the case because we know the answer goes as $e^{-\pi t/\kappa_2}$ (see \eq{cc3d}) which does not have a Taylor expansion in $1/t$ around $0$.

\begin{figure}[ht!]
    \centering
    \begin{tikzpicture}
        %\draw[step=1cm,gray,very thin] (0,0) grid (5,6);
        \draw[dotted] (-3,0) -- (1,0);
        \draw[dotted] (-3,1) -- (1,1);
        \draw[dotted] (-3,2) -- (1,2);
        \draw[dotted] (-3,3) -- (1,3);
        \draw[dotted] (-3,4) -- (1,4);
        \draw[dotted] (-3,5) -- (1,5);
        
        \draw[fill,black] (-1,2.5) circle [radius=0.05]; 
        \node[right] at (-1,2.5) {$\phi(\tau_2)$};
        \draw[fill,red] (-1,0.8) circle [radius=0.05];
        \node[right] at (-1,0.8) {$(-L+\tau_1)$};
        \draw[fill,green] (-1,1.2) circle [radius=0.05];
        \node[right] at (-1,1.2) {$(-\tau_1)$};
        \draw[fill,red] (-1,2.8) circle [radius=0.05];
        \node[right] at (-1,2.8) {$\phi(\tau_1)$};
        \draw[fill,green] (-1,3.2) circle [radius=0.05];
        \node[right] at (-1,3.2) {$(L-\tau_1)$};
        \draw[fill,red] (-1,4.8) circle [radius=0.05];
        \node[right] at (-1,4.8) {$(L+\tau_1)$};
        \draw[fill,green] (-1,5.2) circle [radius=0.05];
        \node[right] at (-1,5.2) {$(2L-\tau_1)$};
        
        \draw[->] (1.5,3) -- (3.5,3);
        \node[above] at (2.5,3) {analytically};
        \node[below] at (2.5,3) {continue};
        
        \draw[dotted] (5,0) -- (9,0);
        \draw[dotted] (5,1) -- (9,1);
        \draw[dotted] (5,2) -- (9,2);
        \draw[dotted] (5,3) -- (9,3);
        \draw[dotted] (5,4) -- (9,4);
        \draw[dotted] (5,5) -- (9,5);
        
        \draw[] (7,2.5) circle [radius=0.05];
        \draw[] (7,2.5) -- (9,3.16);
        \draw[fill] (9,3.16) circle [radius=0.05];
        \node[right] at (9,3.16) {$\phi(-L+\tau_2+ i t_2)$};
        \draw[] (7,0.8) circle [radius=0.05];
        \draw[] (7,0.8) -- (10,1.8);
        \draw[fill,red] (10,1.8) circle [radius=0.05];
        \node[right] at (10,1.8) {$(\tau_1+ i t_1)$};
        \draw[] (7,1.2) circle [radius=0.05];
        \draw[] (7,1.2) -- (4,0.2);
        \draw[fill,green] (4,0.2) circle [radius=0.05];
        \node[right] at (4,0.2) {$(-\tau_1- i t_1)$};
        \draw[] (7,2.8) circle [radius=0.05];
        \draw[] (7,2.8) -- (10,3.8);
        \draw[fill,red] (10,3.8) circle [radius=0.05];
        \node[right] at (10,3.8) {$\phi(\tau_1+ i t_1)$};
        \draw[] (7,3.2) circle [radius=0.05];
        \draw[] (7,3.2) -- (4,2.2);
        \draw[fill,green] (4,2.2) circle [radius=0.05];
        \node[right] at (4,2.2) {$(L-\tau_1- i t_1)$};
        \draw[] (7,4.8) circle [radius=0.05];
        \draw[] (7,4.8) -- (10,5.8);
        \draw[fill,red] (10,5.8) circle [radius=0.05];
        \node[right] at (10,5.8) {$(L+\tau_1+ it_1)$};
        \draw[] (7,5.2) circle [radius=0.05];
        \draw[] (7,5.2) -- (4,4.2);
        \draw[fill,green] (4,4.2) circle [radius=0.05];
        \node[right] at (4,4.2) {$(2L-\tau_1- it_1)$};
    \end{tikzpicture}
    \caption{In the above diagram we consider $\phi(\tau_1)$ as the source and $\phi(\tau_2)$ as a probe. As we analytically continue $\tau_j \rightarrow \tau_j + i t_j$, and $t_1$ and $t_2$ are increased by the same amount, the red images (even reflections of the source) move in the same direction as the probe, while the green images (odd reflections) move in the opposite direction. For large $t_1$ and $t_2$ with $t_1-t_2$ held fixed, the probe $\phi(\tau_2 + i t_2)$ sees the green images at progressively farther distances and hence their correlators decay. Thus in this limit the slab correlator approaches the thermal correlator.}
    \label{fig:my_label}
\end{figure}
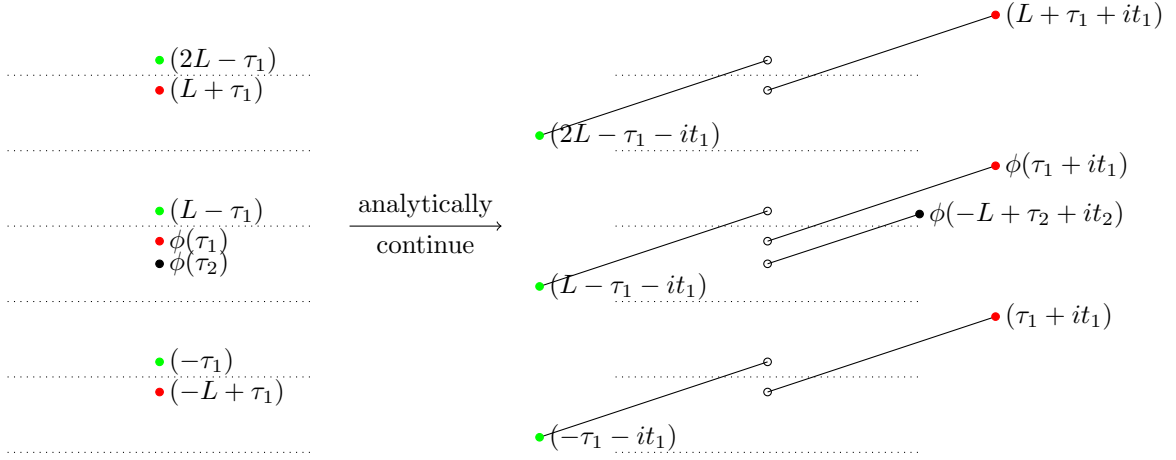

It is not difficult to generalize the above results to arbitrary dimensions $d>2$ for example by using the recursion relations (see Appendix \ref{app:recursion}). The summary is that
\begin{align}
G_{CC}(t_1,\vec{r};t_2,\vec{r})-G_{th}(t_1,\vec{r};t_2,\vec{r})_\beta\thicksim \begin{cases}
t^{-(d-2)} & \text{d is even}\\
e^{-\pi t/ \kappa_2} & \text{d is odd}
\end{cases}
\end{align}
There is a pictorial way to understand the decay of the time-dependent part of the CC correlator. In the slab diagram imagine the analytic continuation of time ($\tau_j \rightarrow \tau_j + i t_j$) with real time axis perpendicular to the plane of the paper (see figure \ref{fig:my_label}). The spatial dimensions are all in the horizontal direction. Note that upon analytic continuation to real time, the `even' and `odd' images move in opposite directions. Also notice that $S_{th}$ involves the correlator between the probe(black) and the even images of the source (marked red); the way to see this is to note that such correlators all involve the time difference $t_1-t_2$ as expected from \eq{g-slab}. Similarly while $S_{diff}$ involves the correlator between the probe(black) and the odd images of the source (marked green); the way to see this is to note that such correlators all involve the time difference $t_1+t_2$ again as expected from \eq{g-slab}. When $t_1$ and $t_2$ are increased by the same amount, the Euclidean separations involved in $S_{th}$ remain the same whereas they go to infinity in case of $S_{diff}$ . This is the reason why $S_{diff}$ decays in time in this limit and the slab correlator approaches the thermal value.\footnote{We thank Abhijit Gadde for suggesting this.}

In the above we have considered separation between the source and the probe only in time but the entire argument goes through for fixed, finite spatial separation $r$ as we take $t\to \infty$.

\subsection{The thermal auto-correlator}

In the previous subsection we showed how the dfference between the CC
correlator and the thermal auto-correlator vanishes as $t_1, t_2 \to
\infty$. In this section we will compute the thermal auto-correlator
itself. We will find that this too shows an odd-even difference between
dimensions as one takes the limit of large $t\equiv t_1 - t_2$.

The quantity to compute is
\begin{align}
G_{th}(\tau_1-\tau_2=\tau;r=0)_\beta &= \frac{\Omega_{d-1}}{2 (2 \pi)^d} \int_0^{\infty} dk\, k^{d-2} \left[ \sum_{n=0}^{\infty} e^{-k(n\beta +\tau)} + \sum_{n=1}^{\infty} e^{-k(n\beta-\tau)} \right] \nonumber \\
&= \frac{\Omega_{d-1} (d-2)!}{2 (2\pi)^d} \left[ \sum_{n=0}^{\infty} \left(\frac{1}{n\beta +\tau}\right)^{d-1} + \sum_{n=1}^{\infty} \left(\frac{1}{n\beta-\tau}\right)^{d-1} \right]
\end{align}
Note that here $\tau=\tau_1+\tau_2$. As in the previous subsection lets analyze this in $d=4$ and $d=3$ separately.

\paragraph{$\boldsymbol{d=4:}$}
Here
\begin{equation}
 G_{th}(\tau_1-\tau_2=\tau;r=0)_\beta= \frac{1}{8\pi^2} \left[\frac{1}{\tau^3} + \sum_{n=1}^{\infty} \left(\frac{1}{n\beta +\tau}\right)^3 + \sum_{n=1}^{\infty} \left(\frac{1}{n\beta-\tau}\right)^3 \right]
\end{equation}
Applying the Euler-Maclaurin formula
\begin{align}
 & S_+= \sum_{n=1}^{\infty} \left(\frac{1}{n\beta +\tau}\right)^3 = \frac{1}{2\beta \tau^2}-\frac{1}{2\tau^3} +\frac{\beta}{4\tau^4} + \mathcal{O}(\tau^{-6}) \nonumber \\
 & S_-= \sum_{n=1}^{\infty} \left(\frac{1}{n\beta -\tau}\right)^3 =\frac{1}{2\beta \tau ^2}+\frac{1}{2\tau^3} +\frac{\beta}{4\tau^4} + \mathcal{O}(\tau^{-6})
\end{align}
After analytically continuing $\tau=\tau_1 + \tau_2 \rightarrow it_1+ it_2=it$, together with the $n=0$ term, for large $t$
\begin{equation}
 G_{th}(\tau_1-\tau_2=\tau;r=0)=\frac{i}{8\pi^2 t^3} -\frac{1}{8\pi^2 \beta t^2}+\frac{\beta}{16\pi^2 t^4}+ \mathcal{O}(t^{-6})
\end{equation}
This matches the precisely the answer (\ref{thermal-auto}), obtained from
the integral formula \eq{2pt-gge-general}.

\paragraph{$\boldsymbol{d=3:}$}
Here
\begin{equation}
 G_{th}(\tau_1-\tau_2=\tau;r=0)_\beta= \frac{1}{8\pi^2} \left[\frac{1}{\tau^2} + \sum_{n=1}^{\infty} \left(\frac{1}{n\beta +\tau}\right)^2 + \sum_{n=1}^{\infty} \left(\frac{1}{n\beta-\tau}\right)^2 \right]
\end{equation}
Applying the Euler-Maclaurin formula
\begin{align}
 & S_+= \sum_{n=1}^{\infty} \left(\frac{1}{n\beta +\tau}\right)^2 = \frac{1}{\beta  \tau }-\frac{1}{2 \tau ^2} + \frac{\beta }{6 \tau ^3}+ \mathcal{O}(\tau^{-5}) \nonumber \\
 & S_-= \sum_{n=1}^{\infty} \left(\frac{1}{n\beta -\tau}\right)^2 =-\frac{1}{\beta \tau}-\frac{1}{2\tau^2} -\frac{\beta}{6\tau^3} +\mathcal{O}(\tau^{-5})
\end{align}
After analytically continuing $\tau=\tau_1 + \tau_2 \rightarrow it_1+ it_2=it$, for large $t$, together with the $n=0$ term
\begin{align}
 G_{th}(\tau_1-\tau_2=\tau;r=0)_\beta = \frac{1}{8\pi^2} \left( \frac{1}{\tau^2} + S_+ + S_-\right) = 0~ \hbox{up to}~ \mathcal{O}(t^{-5})
\end{align}
In fact, it turns out that the Taylor coefficients vanish to all
orders. This is consistent with the behaviour at $d=3$,
$G_{th}\sim e^{-\gamma t}$ which is independently obtained  in Appendix
\ref{app:calc-details-3+1} see equation (\ref{thermal-auto-3d}). Note
that $e^{-\gamma t}$ has an essential singularity at $1/t=0$ and hence
does not have a Taylor expansion in terms of $1/t$.

%%%%%%%%
\section{Kaluza-Klein Interpretation of the thermal correlators}\label{sec:KK}

In the previous section we considered the time-dependence of thermal
correlators, that is, thermal two-point functions with no spatial
separation.  In this section, we will look at thermal 2-point function
at space-like separation. A convenient way to look at this is the
Kaluza-Klein reduction. Demanding periodicity in Euclidean time, the
normal mode expansion of a scalar field in $R^d \times S^1$ is given
by
\[
 \phi(\tau,\vec{r}) = \sum_n \phi_n(\vec{r}) e^{i 2 \pi n \tau /\beta}
\]
In case the post-quench theory is critical, $\phi$ obeys massless
Klein-Gordon equation in $d+1$ dimensions $\partial^\mu \partial_\mu
\phi(\tau,\vec{r})=0$. In terms of the modes one gets
\[
 (\partial^i \partial_i - \frac{4\pi^2 n^2}{\beta^2})\phi_n(\vec{r})=0
\]
which is again a Klein-Gordon equation in one less dimension but now
the fields $\phi_n(\vec{r})$ aquire a (thermal) mass $m_n=\frac{4\pi^2
  n^2}{\beta^2}$. The equal-time thermal 2-point function in terms of
the modes is
\begin{align}
 \langle \phi(0,\vec{r}) \phi(0,\vec{0}) \rangle_\beta &= \frac{1}{\beta}\sum_n \langle \phi_n(\vec{r}) \phi_n(\vec{0})\rangle _\beta \nonumber \\
 &= \frac{1}{\beta}\sum_n \int \frac{d^dk}{(2\pi)^d} \frac{e^{i k r \cos\theta}}{k^2+m_n^2}
\end{align}
This can be evaluated exactly
\[
 \langle \phi(0,\vec{r}) \phi(0,\vec{0}) \rangle_\beta = \frac{a_d}{r^{d-2}} + 2a_d \sum_{n=1} ^\infty \frac{e^{-2\pi n r/\beta}}{r^{d-2}}
\]
where $a_d$ is a dimension-dependent constant. In $d=2$, $r^{2-d}$ is replaced by $\log r$. In all dimensions we get a Yukawa like exponential decay for the
non-zero modes. But in the $r \rightarrow \infty$ limit or
equivalently the $\beta \rightarrow 0$ limit, only the power law term,
corresponding to the $n=0$ survives. This happens in all dimensions
even or odd. This is expected because in the strict $\beta \rightarrow
0$, the circumference of the cylinder shrinks so that the insertions
of the operators do not see the cylinder. Note that the surviving
power law term corresponds to the Green's function for the dimensionally
reduced Euclidean goemetry $R^d$.

Note the appearance of a power law behaviour at all dimensions in
contrast with the thermal auto-correlator in the previous section, which
has an exponential/power law behaviour depending on whether the spatial
dimension $d$ is odd/even.

%%%%%%%%
\section{Discussion \label{sec:discussions}}

In this paper, we presented a detailed investigation of thermalization
of local correlators in free scalar field theories. The main results,
including a general proof of thermalization to GGE, and the difference
in approach to thermalization depending on odd or even dimensions, are
already summarised in the introduction. Here we will briefly mention
some further questions and puzzles:

\begin{itemize}

\item In this paper we have considered mass quench of free scalar
  field theories. It should be straightforward to generalize to other
  quenches involving free scalars, such as a quench in an external
  confining potential, or in an external electromagnetic field. One
  expects that the post-quench state will be again given by a
  Bogoliubov transform of the out-vacuum, and hence will correspond to
  a generalized Calabrese-Cardy state. It would be interesting to see
  whether thermalization to GGE happens in all these cases, and
  whether the approach to thermalization follows similar patterns as
  in this paper.  It would also be interesting to study theories of
  free fermions; once again the post-quench state should be
  describable in terms of a Bogoliubov transformation and the
  techniques of the present paper should be applicable.

\item Along similar lines, it would be interesting to see if our
  results are true in more general integrable models. For transverse
  field Ising models, which can be reduced to free fermions (possibly
  with mass), our results are known to be true (see, e.g.
  \cite{Calabrese:2012GGE}). It would be interesting to see if our
  results are true for integrable models without the explicit use of
  free field techniques.
  
\item It is important to investigate whether the results obtained
  above generalize to interacting theories, away from integrability.
  As shown in \cite{Sotiriadis:2010si}, quantum quench in large $N$
  $O(N)$ models can be reduced to a mass quench at long times in terms
  of an effective mass parameter. Using this result and following the
  techniques of the present paper, one can try to study thermalization
  in these models. Work along these lines is in progress
  \cite{AK-GM:2019}.

\item One of our results, that the leading transient towards
  thermalization is exponential or power law depending on,
  respectively, odd or even dimensionality of space, is strongly
  reminiscent of the behaviour of retarded propagators for massless
  particles, which have support only {\it on the forward light cone}
  in odd spatial dimensions, and {\it inside the forward light cone}
  in even spatial dimensions. The `leaking of the propagator' inside
  the light cone in even space dimensions has been termed a violation
  of the strong Huygen's principle (see,
  e.g. \cite{PhysRevLett.114.110505, Blasco:2015eya} for some physical
  realization of this phenomenon), since Huygen's construction of a
  propating light front is based on propagation along the light cone.
  We found that there are mathematical similarities between this
  peculiarity of the retarded propagator and the odd-even effect
  described in this paper (both are due to difference in the
  integration over angles in odd vs even dimensions), although we were
  not able to derive one as a consequence of the other \footnote{We
    thank R. Loganayagam, H. Reall and J. Samuel for illuminating
    discussions on this issue.}.
  
\item The power law approach to thermalization does not appear to have
  a holographic counterpart; e.g. quasinormal decay of black hole
  perturbations has an exponential form. In the context of an O(N)
  model in 2+1 dimensions, the above discussion would seem to suggest a
  power law decay even in the interacting model. It would be
  interesting to explore the holographic dual  in
  this context.
  
\end{itemize}

\subsection*{Acknowledgements}

We would like to thank Alexander Abanov, Sumit Das, Avinash Dhar,
Fabian Essler, Abhijit Gadde, Rajesh Gopakumar, R. Loganayagam, Juan
Maldacena, Shiraz Minwalla, Harvey Reall, Joseph Samuel and Spenta
Wadia for discussions. P.B. would like to thank the Department of
Theoretical Physics, TIFR, for the research opportunity during the
final year of her M.Sc and the Infosys Foundation for financial
support. G.M. would like to thank the International Centre for
Theoretical Sciences, Bangalore for the opportunity to present/discuss
various stages of this work during various meetings and seminars over
the past year.  This work was supported in part by the Infosys
foundation for the study of quantum structure of spacetime.

%%%%%%%%%%%%%%%%%%%%%%%%%%%%%%%%%%%%%%%%%%%%%%%%%%%%%%%%%%%%%%%%%%%%%%%%%
\appendix
%%%%%%%%
\section{Dirichlet boundary state and relation to post-quench state}\label{app:dirichlet}
A Dirichlet boundary state is defined by
\[
    \phi(x)|D\rangle =0
\]
In terms of the mode expansion (at $t=0$)
\begin{align}
    \phi(x)|D\rangle =&\int\frac{d^d k}{(2\pi)^d}e^{\iota \vec{k}\cdot\vec{x}} \phi(\vec{k})|D\rangle =\int\frac{d^d k}{(2\pi)^d} \frac{e^{\iota \vec{k}\cdot\vec{x}}}{\sqrt{2 \omega_{k}}} \left( a(\vec{k}) + a^{\dagger}(\vec{-k}) \right)|D\rangle \nonumber \\
\end{align}
Clearly this is zero if $\forall \, \vec{k}$ the following equation is satisfied
\begin{equation}
    \left[ a(\vec{k}) + a^{\dagger}(\vec{-k}) \right]|D\rangle= \left[\frac{\partial}{\partial a^\dagger(\vec{k})} + a^{\dagger}(\vec{-k}) \right]|D\rangle= 0
\end{equation}
One can check that
\begin{equation}
    |D\rangle = \exp\left[-\frac{1}{2}\int \frac{d^d k}{(2\pi)^d} a^\dagger(\vec{k})a^\dagger(\vec{-k}) \right]|0_{out}\rangle \label{dirichlet-state}
\end{equation}
easily satisfies the above equation. One can write the in ground state as (through a Bogoliubov transformation)
\[
    |0_{in}\rangle= \exp\left[\frac{1}{2}\int \frac{d^d k}{(2\pi)^d} \gamma(k) a^\dagger_{out}(\vec{k})a^\dagger_{out}(\vec{-k}) \right]|0_{out}\rangle
\]
As we shown in \cite{Mandal:2015kxi} one can write
\[
    |0_{in}\rangle= \exp\left[-\frac{1}{2}\int \frac{d^d k}{(2\pi)^d} \kappa(k) a^\dagger_{out}(\vec{k})a^\dagger_{out}(\vec{-k}) \right]|D\rangle
\]
where $\kappa(k)$ is related to $\gamma(k)$ through equation (\ref{eq:kappa-gamma}).
%%%%%%%%
\section{Recursion Relation}\label{app:recursion}
Here we derive the recursive differential operator for the for the most general gCC correlator. Starting from the 2 point function of $\phi$ fields in $d+1$ dimensions ($\omega_{out}=\sqrt{k^2+m_{out}^2}$)
\begin{align}
\langle \phi(x_1,t_1)\phi(x_2,t_2)\rangle^{(d)} _{gCC} =& \int \frac{d^d k}{(2\pi)^d} \frac{e^{i\vec{k}\cdot(\vec{x_1}-\vec{x_2})}}{2\omega_{out}} \csch(2\kappa(k)) \left[\cos \left(\omega_{out}(t_1-t_2)+ 2i\kappa(k)\right)-\cos\left(\omega_{out}(t_1+t_2)\right) \right] \nonumber \\
=& \int \frac{d^d k}{(2\pi)^d} \frac{e^{i\vec{k}\cdot(\vec{x_1}-\vec{x_2})}}{2\omega_{out}} \csch(2\kappa(k)) \left[\cos \left(\omega_{out}t_- + 2i\kappa(k)\right)-\cos\left(\omega_{out}t_+ \right) \right] 
\end{align}
Here $t_{-}=t_1-t_2$, $t_{+}=t_1+t_2$ and $\kappa(k) = \sum_{n} \kappa_{2n}|\vec{k}|^{2n-1}$ represents the most general gCC state (see eq. (\ref{eq:squeezing-function})). We now connect this to the correlator in $(d-2)+1$ dimensions. One can write
\begin{align}
&\langle \phi(x_1,t_1)\phi(x_2,t_2)\rangle^{(d)} _{gCC} = \frac{\Omega_{d-2}}{2(2\pi)^d} \int _0^{\infty} dk\frac{k^{d-1}}{\omega_{out}} \int _0^{\pi} \text{d$\theta $} e^{ikr \cos\theta} (\sin\theta)^{d-2} \left[\cos(2 i \kappa(k) + \omega_{out}t_{-}) -\cos(\omega_{out} t_+)\right] \nonumber \\
&= \frac{\Omega_{d-2}}{2(2\pi)^d} \int _0^{\infty} dk\frac{k^{d-3}}{\omega_{out}} \int _0^{\pi }\text{d$\theta $} e^{ikr \cos\theta} (\sin\theta)^{d-4} \left[\cos(2 i \kappa(k) +\omega_{out}t_{-}) -\cos(\omega_{out}t_+)\right] \times k^2 \sin^2\theta \nonumber \\
&= \frac{\Omega_{d-2}}{2(2\pi)^d} \int _0^{\infty} dk \frac{k^{d-3}}{\omega_{out}} \int _0^{\pi }\text{d$\theta $} e^{ikr \cos\theta} (\sin\theta)^{d-4} \left[\cos(2 i \kappa(k) + \omega_{out}t_{-}) -\cos(\omega_{out}t_+)\right] \times (\omega_{out}^2-m_{out}^2  -k^2 \cos^2\theta) \nonumber \\
&= \frac{\Omega_{d-2}}{2(2\pi)^d} \left(-\partial ^2 _{t_{-}} -\partial ^2 _{t_{+}} +m_{out}^2 + \partial_r^2 \right) \int_0^{\infty} dk \frac{k^{d-3}}{\omega_{out}} \int _0^{\pi }\text{d$\theta $} e^{ikr \cos\theta} (\sin\theta)^{d-4} \left[\cos(2 i \kappa(k) +\omega_{out}t_{-}) -\cos(\omega_{out}t_+)\right] \nonumber \\
&= \frac{\Omega_{d-2}}{4\pi^2 \Omega_{d-4}} \left(-\partial^2 _{t_{-}} -\partial^2 _{t_{+}} +m_{out}^2 + \partial_r^2 \right) \frac{\Omega_{d-4}}{2(2\pi)^{d-2}} \int_0^{\infty} dk\frac{k^{d-3}}{\omega_{out}} \int_0^{\pi} d\theta  e^{ikr \cos\theta} (\sin\theta)^{d-4} \nonumber \\
&\quad \times \left[\cos(2 i \kappa(k) +\omega_{out}t_{-}) -\cos(\omega_{out}t_+)\right] \nonumber \\
&= \frac{\Omega_{d-2}}{4\pi^2 \Omega_{d-4}} \left(-\partial ^2 _{t_{-}} -\partial ^2 _{t_{+}} +m_{out}^2 + \partial_r^2 \right) \langle \phi(x_1,t_1)\phi(x_2,t_2)\rangle^{(d-2)}_{gCC}
\end{align}
$\Omega_{d-1}$ is the solid angle of a $d-1$ dimensional spherical surface in $d$ spatial dimensions. In the third line we have used $k^2=\omega_{out}^2-m_{out}^2$. The main point being used here is the fact that one can obtain the extra $k^2 \sin(\theta)^2$ by acting wih the derivatives. Also notice that $t_{-}$ and $t_{+}$ act as independent variables. This derivation also holds for the ground state quench as can be verified directly from equation (\ref{eq:ground-quench}).

%%%%%%%%
\section{Details of critical quench calculations}\label{app:crit-quench} 
\subsection{2+1 dimensions}\label{app:calc-details-2+1}
The details of calculation of correlators in $d=2$ are presented here.

%%%%
\subsubsection{CC state}
The time-dependent piece of the 2-point function of $\partial_t\phi$ in arbitrary gCC state is
\begin{equation}
 \braket{\partial_t\phi(\vec{x_1,t}) \partial_t\phi(\vec{x_2},t)}_{td} = \int_{-\infty}^{\infty} \frac{d^dk}{2(2\pi)^d} e^{i\vec{k}\cdot(\vec{x_1}-\vec{x_2})}\sqrt{k^2+m_{out}^2} \cos(2t\sqrt{k^2+m_{out}^2}) \csch(2\kappa(k))
\end{equation}
which in 2+1 for critical quench is
\begin{align}
 \braket{\partial_t\phi(\vec{x_1,t}) \partial_t\phi(\vec{x_2},t)}_{td} &= \frac{1}{4\pi}\int_{0}^{\infty} dk e^{ikr} k^2 J_0(kr) \cos(2kt) \csch(2\kappa(k)) \nonumber \\
 &= \frac{1}{4\pi^2}\int_0^{\pi} d\phi \int_{0}^{\infty} dk e^{ikr} k^2 \cos(kr \sin\phi)\cos(2kt) \csch(2\kappa(k)) \nonumber \\
 &= \frac{1}{16\pi^2}\int_0^{\pi} d\phi \int_{0}^{\infty} dk e^{ikr}k^2 \csch(2\kappa(k)) \left(e^{ik\,ap}+e^{ik\,am}+e^{-ik\,ap}+e^{-ik\,am} \right)
\end{align}
where we have exponentiated the cosine in terms of $ap=2t+r\sin\phi$ and $am=2t-r\sin\phi$. For $\kappa(k)=\kappa_2 k$, these integrals are easily performed in Mathematica in terms of the Polygamma function
\[
 \int_{0}^{\infty} dk e^{ikr}k^2 \csch(2\kappa_2 k) e^{ik\,ap}= -\frac{\psi ^{(2)}\left(\frac{1}{2}-\frac{i\,ap}{4 \kappa_2}\right)}{512 \pi ^2 \kappa_2^3}
\]
After performing the $k$ integral, we expand the polygammas for large $t$, and then perform the $\phi$ integral to get
\begin{equation}
 \braket{\partial_t\phi(\vec{x_1,t}) \partial_t\phi(\vec{x_2},t)}_{td}= -\frac{1}{32 \pi  \kappa_2 t^2} -\frac{8 \kappa_2^2+3 r^2}{256 \pi \kappa_2 t^4} +\mathcal{O}(t^{-6})
\end{equation}

%%%%
\subsubsection{Ground state}
The time-dependent piece of 2-point function of $\partial_t\phi$ in the ground state is (\ref{eq:ground-quench-etc-delt})
\[
  \lan 0_{in}|\partial_t\phi(\vec{x_1},t)\partial_t\phi(\vec{x_2},t)|0_{in}\ran_{td} = -\int \frac{d^d k}{(2\pi)^d} \frac{\left(m_{out}^2-m^2\right)}{4\sqrt{k^2+m^2}} e^{i \vec{k}\cdot (\vec{x_1}-\vec{x_2})} \cos\left(2 t \sqrt{k^2+m_{out}^2}\right)
\]
which in 2+1 d for critical quench is
\begin{align}
 \lan 0_{in}|\partial_t\phi(\vec{x_1},t)\partial_t\phi(\vec{x_2},t)|0_{in}\ran_{td} &= \int_{0}^{\infty} dk \frac{k J_0(kr)}{8\pi \sqrt{k^2+m^2}} m^2 \cos(2kt) \nonumber \\
 &= \frac{m^2}{8\pi^2}\int_0^{\pi} d\phi\int_{0}^{\infty} dk \frac{k}{\sqrt{k^2+m^2}} \cos(2kt)\cos(kr \sin\phi) \nonumber \\
 &= \frac{m^2}{32\pi^2} \int_0^{\pi} d\phi\int_{0}^{\infty} dk \frac{k}{\sqrt{k^2+m^2}} \left(e^{ik\,ap}+e^{ik\,am}+e^{-ik\,ap}+e^{-ik\,am} \right)\nonumber \\
\end{align}
Naively doing the integral, we find that $\int_0^\infty \frac{m^2 k e^{ik\,a}}{32\pi^2 \sqrt{k^2+m^2}}$ diverges because as $k\rightarrow\infty$, the integrand is just $e^{ik\,a}$. We regulate this by $k\rightarrow k(1\pm i\epsilon)$, performing the k integral (alternatively Laplace transform with respect to $\epsilon$) and then taking $\epsilon\rightarrow 0$. The final result is
\[
 \lan 0_{in}|\partial_t\phi(\vec{x_1},t)\partial_t\phi(\vec{x_2},t)|0_{in}\ran_{td}= \int_0^{\pi} d\phi \frac{m^3}{32\pi} \left(-\pmb{L}_{-1}(\text{am} m)+I_1(\text{am} m)-\pmb{L}_{-1}(\text{ap} m)+I_1(\text{ap} m)\right)
\]
Now using the fact that modified Struve function has the asymptotic form
\begin{equation}
 \pmb{L}_\nu(x) - I_{-\nu}(x) \approx -\frac{1}{\pi}\sum_{j=0}^\infty \frac{(-1)^{j+1}\Gamma(j+1/2)}{\Gamma(\nu+1/2-j)(x/2)^{2j-\nu+1}} \label{eq:struve-bessel}
\end{equation}
which to leading order is
\[
 \pmb{L}_\nu(x) - I_{-\nu}(x) \approx -\frac{1}{\sqrt{\pi}\,\Gamma(\nu+1/2)}\left(\frac{2}{x}\right)^{1-\nu}
\]
gives us
\[
\lan 0_{in}|\partial_t\phi(\vec{x_1},t)\partial_t\phi(\vec{x_2},t)|0_{in}\ran_{td}= -\int_0^{\pi} d\phi \frac{m \left(r^2 \sin^2(\phi)+4 t^2\right)}{8 \pi ^2 \left(r^2 \sin ^2(\phi )-4 t^2\right)^2}= -\frac{m t}{4 \pi  \left(4 t^2-r^2\right)^{3/2}}= -\frac{m}{32\pi t^2} +\mathcal{O}(t^{-4})
\]

%%%%%%%%
\subsection{3+1 dimensions}\label{app:calc-details-3+1}
The details of calculation of correlators in $d=3$ are presented here.

%%%%
\subsubsection{CC correlator}
The integral in equation (\ref{2pt-gcc-etc}) for only $\kappa_2$ non-zero, can be evaluated directly in Mathematica. Even so we calculate it by hand as we would eventually need to do so in the gCC$_4$ case. Performing the angular integral we land up with
\begin{align}
 & \braket{f_2|\phi(x_1,t)\phi(x_2,t)|f_2} = \frac{1}{4\pi^2 r} \int_0^\infty dk \sin(k r) \left[\coth(2\kappa_2 k) - \csch(2\kappa_2 k) \cos(2 k t)\right] \nonumber \\
 & = \frac{1}{8 \pi^2 i r} \int_{-\infty}^\infty dk e^{i k r} \left[\coth(2\kappa_2 k) - \csch(2\kappa_2 k) \cos(2 k t)\right] \nonumber \\
 & = \frac{1}{8 \pi^2 i r} \int_{-\infty}^\infty dk \left[\underbrace{\frac{\cosh(2\kappa_2 k) e^{ikr}}{\sinh(2\kappa_2 k)}}_{\text{A}} - \underbrace{\frac{e^{i k (r+ t_+)}}{2\sinh(2\kappa_2 k)}}_{\text{B}}- \underbrace{\frac{e^{i k (r- t_+)}}{2\sinh(2\kappa_2 k)}}_{\text{C}}\right]
\end{align}
where we have used $t_+=2t$.
\begin{figure}[ht!]
\centering
\begin{tikzpicture}[scale = .6, every node/.style={scale=0.6}]
%axis%
\draw[->,black, thick] (-6,0) -- (6,0);
\draw[->,black] (0,-6) -- (0,6);
%C1
\draw[blue, thick,->](-5.2,0) -- (-2.5,0);
\draw[blue,thick](-2.5,0) -- (-.2,0);
\draw[blue,->, thick](.2,0) -- (2.5,0);
\draw[blue, thick](2.5,0) -- (5.2,0);
\draw [blue,thick,->] (-.3,0) arc [radius=.3, start angle=180, end angle = 90];
\draw [blue,thick] (0,.3) arc [radius=.3, start angle=90, end angle = 0];
%Contours%
%1st quarter%
\draw [->,blue,thick] (5.2,0) arc [radius=5.2, start angle=0, end angle = 45];
\draw [blue,thick] (3.67,3.67) arc [radius=5.2, start angle=45, end angle = 90];
%2nd quarter%
\draw [blue,thick] (-5.2,0) arc [radius=5.2, start angle=180, end angle = 135];
\draw [blue,thick,->] (0,5.2) arc [radius=5.2, start angle=90, end angle = 135];
%3rd quarter%
\draw [blue,->,thick] (0,-5.2) arc [radius=5.2, start angle=270, end angle = 225];
\draw [blue,thick] (-3.67,-3.67) arc [radius=5.2, start angle=225, end angle = 180];
%4rd quarter%
\draw [blue,thick] (3.67,-3.67) arc [radius=5.2, start angle=315, end angle = 270];
\draw [blue,thick,->] (5.2,0) arc [radius=5.2, start angle=360, end angle = 315];
%Poles%
\draw [fill,red] (0,0) circle [radius=0.1];
\draw [fill,red] (0,.7) circle [radius=0.1];
\draw [fill,red] (0,1.4) circle [radius=0.1];
\draw [fill,red] (0,2.1) circle [radius=0.1];
\draw [fill,red] (0,2.8) circle [radius=0.1];
\draw [fill,red] (0,3.5) circle [radius=0.1];
\draw [fill,red] (0,4.2) circle [radius=0.1];
\draw [fill,red] (0,4.9) circle [radius=0.1];
\draw [fill,red] (0,0) circle [radius=0.1];
\draw [fill,red] (0,-.7) circle [radius=0.1];
\draw [fill,red] (0,-1.4) circle [radius=0.1];
\draw [fill,red] (0,-2.1) circle [radius=0.1];
\draw [fill,red] (0,-2.8) circle [radius=0.1];
\draw [fill,red] (0,-3.5) circle [radius=0.1];
\draw [fill,red] (0,-4.2) circle [radius=0.1];
\draw [fill,red] (0,-4.9) circle [radius=0.1];
%Tags%
\node [above] at (3,.3) {$C_1$};
\node [right] at (3,3) {$C_2$};
\node [above left] at (-.3,0) {$C'_1$};
\node [right] at (3,-3) {$C'_2$};
\node[above,ultra thick] at (0,6) {$Im(k)$};
\node[right,ultra thick] at (6,0) {$Re(k)$};
\end{tikzpicture}
\caption{Singularity Structure of Integrand in eq. (\ref{2pt-gcc-etc})}
\label{fig:f4s}
\end{figure}
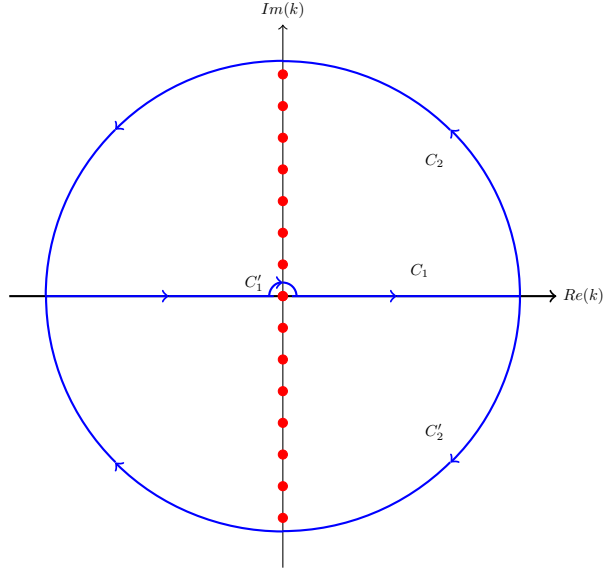
Now $\sinh(2 \kappa_2 k)$ has simple poles at $2 \kappa_2 k = i n \pi$ or $k= i n \pi/2\kappa2$ for any integer n. We are interested in the $t_+>r$ limit, so we close the contour in the upper half plane for terms A and B while in the lower half for C. The residue at these poles is easily calculated
\begin{equation*}
 \text{Res}_A(i n \pi/2\kappa2) = \frac{1}{2 \kappa_2} e^{-n \pi r/ 2 \kappa_2}
\end{equation*}
\begin{equation*}
 \text{Res}_B(i n \pi/2\kappa2) = \frac{(-1)^n}{4 \kappa_2} e^{-n \pi (t_+ +r)/ 2 \kappa_2}
\end{equation*}
\begin{equation*}
 \text{Res}_C(-i n \pi/2\kappa2) = \frac{(-1)^n}{4 \kappa_2} e^{-n \pi (t_+ -r)/ 2 \kappa_2}
\end{equation*}
We take half the contribution from the $k=0$ pole. Thus we can write the integral as
\begin{align}
 & \braket{f_2|\phi(x_1,t)\phi(x_2,t)|f_2} = \frac{2\pi i}{8 \pi^2 i r} \left[ \frac{1}{2\kappa_2}(\frac{1}{2}- \frac{1}{4}+ \frac{1}{4}) + \sum_{n=1}^\infty \left(\frac{e^{-n \pi r/ 2 \kappa_2}}{2 \kappa_2} -\frac{(-1)^n e^{-n \pi (t_+ +r)/ 2 \kappa_2}}{4 \kappa_2} + \frac{(-1)^n e^{-n \pi (t_+ -r)/ 2 \kappa_2}}{4 \kappa_2} \right)\right] \nonumber \\
 & = \frac{2\pi i}{16 \pi^2 i r \kappa_2} \left[\sum_{n=1}^\infty e^{-n \pi r/ 2\kappa_2}+\frac{1}{2} -\frac{1}{2} \left(\sum_{n=1}^\infty (-1)^n e^{-n \pi (t_+ +r)/ 2 \kappa_2} +\frac{1}{2}\right) + \frac{1}{2} \left(\sum_{n=1}^\infty(-1)^n e^{-n \pi (t_+ -r)/ 2 \kappa_2}+\frac{1}{2}\right)\right] \nonumber
\end{align}
\begin{align}
 \braket{f_2|\phi(x_1,t)\phi(x_2,t)|f_2}=& \frac{1}{32\pi \kappa_2 r}\left[ 2\coth(\frac{\pi r}{4\kappa_2}) +\tanh(\frac{\pi(2t-r)}{4\kappa_2})-\tanh(\frac{\pi(2t+r)}{4\kappa_2})\right] \nonumber \\
 =& \frac{1}{32\pi\kappa_2 r}\left[ 2\coth(\frac{\pi r}{4\kappa_2})- \tanh(\frac{\pi r}{2\kappa_2}) \left\{1- \tanh(\frac{\pi(2t-r)}{4\kappa_2})\tanh(\frac{\pi(2t+r)}{4\kappa_2}) \right\} \right]\label{eq:f2d4-2pt-answer}
\end{align}
we have already done the summations
\begin{align}
 &\sum_{n=1}^\infty e^{-n s} +\frac{1}{2}= \frac{1}{e^s-1}+\frac{1}{2} = \frac{1}{2}\coth(s/2)\nonumber \\
 &\sum_{n=1}^\infty (-1)^n e^{-n s} +\frac{1}{2} = -\frac{1}{e^s+1}+\frac{1}{2} = \frac{1}{2}\tanh(s/2) \nonumber
\end{align}
The $\frac{1}{2}$'s are coming from the $n=0$ pole at the origin. The slowest decaying transient is ($t>>\kappa_2>r$)
\begin{equation}\label{cc3d}
	 -\frac{1}{16\kappa_2^2}e^{-\pi t/\kappa_2}
\end{equation}
This is interpreted as the contribution of the pole nearest to the origin (but not the origin itself). The transients from other poles decay much faster than this. Also note that the answer for $r>2t$, for which all contours are closed in the UHP, comes out to be the same as $r<2t$.

%%%%
\subsubsection{gCC$_4$ correlator}
The singularity structure of the correlator in the gCC$_4$ state (\ref{eq:f4-state}) is similar to CC state.
\begin{align}
& \langle f_4|\phi(\vec{0},t_1) \phi(\vec{r},t_2) |f_4\rangle = \int \frac{d^3 k}{(2\pi)^3} \frac{e^{\iota \vec{k}\cdot \vec{r}}}{2|\vec{k}|} \left[\coth(2\kappa_2|\vec{k}| + 2\kappa_4 |\vec{k}|^3) - \cos(2 |\vec{k}| t)\csch(2\kappa_2|\vec{k}| + \kappa_4 |\vec{k}|^3) \right] \nonumber \\
&= \frac{1}{8 \pi^2 \iota r} \int_{\infty}^{\infty} d k \left[ \underbrace{\frac{e^{\iota k r}\cosh(2\kappa_2|\vec{k}| + \kappa_4 |\vec{k}|^3)}{\sinh(2\kappa_2|\vec{k}| + \kappa_4 |\vec{k}|^3)}}_{\text{A}} - \underbrace{\frac{e^{\iota k(r+2t)}}{2\sinh(2\kappa_2|\vec{k}| + \kappa_4 |\vec{k}|^3)}}_{B} - \underbrace{\frac{e^{\iota k(r-2t)}}{2\sinh(2\kappa_2|\vec{k}| + \kappa_4 |\vec{k}|^3)}}_{C}\right]
\end{align}
Doing this integral exactly is hard. But it is simpler, and more
instructive, to figure out the behaviour of the integral by looking at
the poles of the integrand in the complex $k$-plane.  These are
situated at the roots of $2\kappa_2 k + 2\kappa_4 k^3 = i n
\pi$. Introducing a dimensionless parameter $\bar{\kappa_4} =
\kappa_4/\kappa_2^3$, in the small $\bar{\kappa_4}$ expansion, we find
that roots of the above equation are
\begin{align}
 k_1(n) = \frac{i\pi}{2\kappa_2}\left(n+\frac{\pi^2 n^3}{4}\bar{\kappa_4} + O(\bar{\kappa_4}^2)\right) \nonumber \\
k_2(n) = \frac{i}{\kappa_2}\left(\frac{1}{\sqrt{\bar{\kappa_4}}} -\frac{\pi n}{4} - \frac{3\pi^2 n^2}{32}\sqrt{\bar{\kappa_4}}+O(\bar{\kappa_4})\right) \nonumber  \\
k_3(n) = -\frac{i}{\kappa_2}\left(\frac{1}{\sqrt{\bar{\kappa_4}}} +\frac{\pi n}{4} - \frac{3\pi^2 n^2}{32}\sqrt{\bar{\kappa_4}}+O(\bar{\kappa_4})\right)
\label{eq:rt}
\end{align}
As we saw earlier (in particular, in the previous subsection), the leading large time behaviour is given by the slowest transient. So we only need to calculate the contribution of the poles nearest to the origin i.e. from $k_1(n)$ with $n=\pm1$. Both $k_2(n)$ and $k_3(n)$ are very large due to the small $\sqrt{\bar{\kappa_4}}$ in the denominator. 

Thus forgetting about all the other poles we do the integral. The residues are
\[
 \text{Res}_A(k_1(1)) = \frac{1}{2\kappa_2}(1+\frac{3}{4\pi^2}\bar{\kappa_4} + \mathcal{O}(\bar{\kappa_4}^2)) \exp[-\frac{\pi r}{2\kappa_2}(1+\frac{\pi^2}{4}\bar{\kappa_4} + \mathcal{O}(\bar{\kappa_4}^2))]
\]
\[
 \text{Res}_B(k_1(1)) = -\frac{1}{4\kappa_2} (1+\frac{3}{4\pi^2}\bar{\kappa_4} + \mathcal{O}(\bar{\kappa_4}^2)) \exp[-\frac{\pi (2t+r)}{2\kappa_2}(1+\frac{\pi^2}{4}\bar{\kappa_4} + \mathcal{O}(\bar{\kappa_4}^2))]
\]
\[
 \text{Res}_C(k_1(-1)) = -\frac{1}{4\kappa_2} (1+\frac{3}{4\pi^2}\bar{\kappa_4} + \mathcal{O}(\bar{\kappa_4}^2)) \exp[-\frac{\pi (2t-r)}{2\kappa_2}(1+\frac{\pi^2}{4}\bar{\kappa_4} + \mathcal{O}(\bar{\kappa_4}^2))]
\]
\[
 \text{Res}_A(k_1(0)) = \frac{1}{2\kappa_2};\qquad \text{Res}_B(k_1(0)) = \frac{1}{4\kappa_2};\qquad \text{Res}_C(k_1(0)) = \frac{1}{4\kappa_2}
\]
As before we close the contour in the upper half plane for terms A and B while in the lower half for C. The result is
\begin{align}
& \langle f_4|\phi(\vec{0},t_1,) \phi(\vec{r},t_2) |f_4\rangle \nonumber \\
  =& \frac{1}{4\pi r \kappa_2} \left( \text{Res}_A(k_1(1))+ \frac{1}{2}\text{Res}_A(k_1(0)) - \text{Res}_B(k_1(1)) -\frac{1}{2}\text{Res}_B(k_1(0)) +\text{Res}_C(k_1(1)) +\frac{1}{2}\text{Res}_C(k_1(0))\right) \nonumber \\
  &= \frac{1}{16\pi r \kappa_2} + \frac{1}{16\pi r\kappa_2} \big(1 + \frac{3\pi^2}{4}\bar{\kappa_4} + \mathcal{O}(\bar{\kappa_4}^2)\big) \Bigg\{ 2\exp[-\frac{\pi}{2\kappa_2} \big(1+ \frac{\pi^2}{4}\bar{\kappa_4} + \mathcal{O}(\bar{\kappa_4}^2) \big) r] + \nonumber \\
& \exp[-\frac{\pi}{2\kappa_2}\big(1+ \frac{\pi^2}{4}\bar{\kappa_4}+ \mathcal{O}(\bar{\kappa_4}^2) \big) (2t+r)] - \exp[-\frac{\pi}{2\kappa_2} \big(1+ \frac{\pi^2}{4}\bar{\kappa_4}+ \mathcal{O}(\bar{\kappa_4}^2) \big) (2t-r) ]\Bigg\} + \mathcal{O}(\exp(-2\pi t/\kappa_2))\label{eq:f4d4-2pt-answer}
\end{align}
The exact value of $k_1(1)$ is $k_1(1)= i \gamma$ where
\[
 \gamma= \frac{-2\ 6^{2/3}+\sqrt[3]{6} \left(\sqrt{48-81 \pi ^2 \bar{\kappa}_4}+9 i \pi  \sqrt{\bar{\kappa}_4}\right)^{2/3}}{6 \kappa _2 \sqrt{\bar{\kappa}_4} \sqrt[3]{\sqrt{48-81 \pi ^2 \bar{\kappa}_4}+9 i \pi  \sqrt{\bar{\kappa}_4}}}
\]
which implies that the exponentials appearing above are of the
form
\[
e^{-\gamma r}, e^{- \gamma(2t \pm r)}
\]

%%%%
\subsubsection{Ground state correlator}
The above method of estimating the time-dependence at large
times can also be applied to the ground state as it is just a
particular squeezed state. But here we will use the special relation between correlators in $1+1$ and $3+1$ given by the action of the operation $\frac{-1}{2\pi r}\partial_r$ as already mentioned before in section \ref{sec:recursion}. Lets see this in detail. The most general gCC correlator in $1+1$ is given by
\begin{align}
\braket{f|\phi(x_1,t_1)\phi(x_2,t_2)|f} = \int \frac{dk}{(2\pi)} \frac{e^{i k r}}{2k} \csch\left(2\kappa(k)\right) \left[\cos( 2i \kappa(k) + kt_-) -\cos(kt_+)\right]
\end{align}
where $r=(x_1-x_2)$. Acting with $\frac{-1}{2\pi r}\partial_r$ on this expression we obtain
\[
 = \frac{1}{8 \pi^2 i r} \int_{-\infty}^\infty dk e^{i k r} \csch\left(2\kappa(k)\right) \left[\cos(2i\kappa(k)+ kt_-) -\cos(kt_+)\right]
\]
This is precisely the expression for the most general gCC correlator in $3+1$ after having done the angular integral. We are going to use this operation on the $1+1$ ground state correlator to obtain the ground state correlator in $3+1$. In $1+1$ we are able to do the integral exactly in terms of Meijer G-functions.
\renewcommand{\arraystretch}{1} % Default value: 1
\begin{align}
\braket{0_{in}|\phi(x_1,t)\phi(x_2,t)|0_{in}}=&-\frac{1}{16\pi}\Bigg\{-2 G_{1,3}^{2,1}\left(\frac{m^2 r^2}{4}\middle\vert
\begin{array}{c}
 \frac{3}{2} \\
 0,1,\frac{1}{2}
\end{array}
\right)+G_{1,3}^{2,1}\left(\frac{1}{4} m^2 (r+2 t)^2\middle\vert
\begin{array}{c}
 \frac{3}{2} \\
 0,1,\frac{1}{2} \\
\end{array}
\right) \nonumber \\
 &+ G_{1,3}^{2,1}\left(\frac{1}{4} m^2 (r-2 t)^2\middle\vert
\begin{array}{c}
 \frac{3}{2} \\
 0,1,\frac{1}{2} \\
\end{array}
\right)-8 K_0(m r)\Bigg\}
\end{align}
Acting with $\frac{-1}{2\pi r}\partial_r$ on this we obtain the correlator in $3+1$.
\begin{align}
\braket{0_{in}|\phi(x_1,t_1)\phi(x_2,t_2)|0_{in}}=& \frac{1}{32\pi^2 r^2 \left(r^2-4 t^2\right)}\Bigg\{-2 \left(r^2-4 t^2\right) G_{1,3}^{2,1}\left(\frac{m^2 r^2}{4}\middle\vert
\begin{array}{c}
 \frac{3}{2} \\
 0,1,\frac{1}{2} \\
\end{array}
\right) \nonumber \\
 &+ r(r+2t) G_{1,3}^{2,1}\left(\frac{1}{4} m^2 (r-2t)^2\middle\vert
\begin{array}{c}
 \frac{3}{2} \\
 0,1,\frac{1}{2} \\
\end{array}
\right) \nonumber \\
 &+ r (r-2 t) \bigg[G_{1,3}^{2,1}\left(\frac{1}{4} m^2 (r+2t)^2\middle\vert
\begin{array}{c}
 \frac{3}{2} \\
 0,1,\frac{1}{2} \\
\end{array}
\right) \nonumber \\
 &+ 2m (r+2t) \left(K_1(m (r+2 t))-K_1(2 m t-m r)+2 K_1(m r)\right)\bigg]\Bigg\}
\end{align}

\subsubsection{The Thermal Correlator}
The 2-point function in the thermal ensemble is (\ref{2pt-gge-general})
\begin{align}
 \langle \phi(\vec{x_1},t_1) \phi(\vec{x_2},t_2) \rangle_\beta =& \frac{1}{2} \int \frac{d^3 k}{(2\pi)^3} \frac{e^{\iota\vec{k}\cdot\vec{x}}}{\omega_{out}} \left[\cos(\omega_{out} t)\coth(\frac{\mu(k)}{2}) - i\sin(\omega_{out}t) \right] \nonumber \\
 =& \frac{1}{4\pi^2 r} \int_0^{\infty} dk \sin(kr) \left[\cos(kt)\coth(\beta k/2)-i\sin(kt)\right] \nonumber \\
 =& \frac{1}{8\pi^2 i r} \int_{-\infty}^{\infty} dk e^{i k r} \left[\cos(kt)\coth(\beta k/2)-i\sin(kt)\right] \nonumber \\
 =& \frac{1}{16\pi^2 i r} \int_{-\infty}^{\infty} dk \left( e^{i k(r+t)} \coth(\beta k/2) +e^{i k(r-t)} \coth(\beta k/2) - e^{i k(r+t)} +e^{i k(r-t)} \right) \nonumber
\end{align}
The last 2 terms give us Dirac-delta functions, while the first 2 terms are easily evaluated using contour integral with simple poles at the location $k=2i\pi n/\beta$. Since $t>r$, for the first term we close the contour in the UHP while for the second term we close in the LHP. The contribution of the pole at the origin cancels. Summing over all the poles we get
\begin{align}
 \langle \phi(\vec{x_1},t_1) \phi(\vec{x_2},t_2) \rangle_\beta =\frac{\coth \left(\frac{\pi(r-t)}{\beta }\right)+\coth \left(\frac{\pi(r+t)}{\beta }\right)}{8\pi \beta r} +\frac{i \delta (r-t)}{8 \pi  r}-\frac{i \delta (r+t)}{8 \pi  r}
\end{align}
The thermal autocorrelator is obtained by taking $r\rightarrow 0$
\begin{equation}
 \langle \phi(\vec{x_1},t_1) \phi(\vec{x_1},t_2) \rangle_\beta = -\frac{\csch^2\left(\frac{\pi t}{\beta}\right)}{4\beta^2}\label{thermal-auto-3d}
\end{equation}

%%%%%%%%
\subsection{4+1 dimensions}\label{app:calc-details-4+1}
Here the details of calculation of particular correlators in $d=4$ are
provided, even though a general method of calculating the
time-dependent part is presented in the text for an arbitrary gCC
state.

%%%%
\subsubsection{CC correlator}
The time dependent part of eq. (\ref{2pt-gcc-etc}) with only $\kappa_2$ non-zero is
\[
\braket{f_2|\phi(x_1,t)\phi(x_2,t)|f_2} - \langle \phi(\vec{x_1},t) \phi(\vec{x_2},t) \rangle _\beta = - \int \frac{d^d k}{(2\pi)^d} \frac{e^{i \vec{k}\cdot (\vec{x_1}-\vec{x_2})}}{2|\vec{k}|} \csch(2 \kappa_2 |\vec{k}|) \cos(2|\vec{k}|t)
\]
which in 4+1 simplifies to
\begin{align}
& -\frac{1}{8 \pi^2 r} \int _0^{\infty} dk\,k J_1(k r) \csch(2 \kappa_2 k) \cos(2 k t) \nonumber \\
&=- \frac{1}{8 \pi^3 r} \int _0^{\infty} dk\,k \int_0^\pi d\phi \cos(kr \sin\phi - \phi) \csch(2 \kappa_2 k) \cos(2 k t) \nonumber \\
&=- \frac{1}{16 \pi^3 r} \int_0^\pi d\phi \int _0^{\infty} dk\,k \csch(2 \kappa_2 k) \left(\cos(k (2t+r\sin\phi)-\phi) +\cos(k (2t-r\sin\phi)+\phi)\right)\nonumber \\
&=- \frac{1}{16 \pi^3 r} \int_0^\pi d\phi \int _0^{\infty} dk\,k \csch(2 \kappa_2 k) \left[\cos(k (2t+r\sin\phi)-\phi) +\cos(k (2t-r\sin\phi)+\phi)\right]\nonumber \\
&= -\frac{1}{256 \pi ^3 \kappa_2^2 r}\int_0^\pi d\phi \Bigg[ e^{i \phi} \left(\psi ^{(1)}\left(\frac{i (t_+ +r \sin\phi)}{4 \kappa_2}+\frac{1}{2}\right) + \psi ^{(1)}\left(\frac{1}{2}-\frac{i (t_+-r \sin\phi)}{4 \kappa_2}\right)\right) \nonumber \\
&+ e^{-i \phi}\left( \psi ^{(1)}\left(\frac{i (t_+-r \sin\phi)}{4 \kappa_2}+\frac{1}{2}\right)+\psi ^{(1)}\left(\frac{1}{2}-\frac{i (t_++r \sin\phi)}{4 \kappa_2}\right) \right) \Bigg] \nonumber \\
&\approx \int d\phi \left[\frac{\sin ^2(\phi )}{\left(\left(16 \pi ^3\right) \kappa_2\right) t_+^2}+\frac{\sin ^2(\phi ) \left(8 \kappa_2^2+r^2 (-\cos (2 \phi ))+r^2\right)}{\left(\left(32 \pi ^3\right) \kappa_2\right) t_+^4}+O\left(\frac{1}{t^6}\right) \right] \nonumber \\
&= \frac{1}{128 \pi^2 \kappa_2 t^2} + \frac{3r^2+16\kappa_2^2}{2^{11} \pi^2 \kappa_2 t^4}+ O\left(t^{-6}\right)
\end{align}
In the second line we used the integral representation $J_n(x)=\frac{1}{\pi}\int_0^\pi d\tau \cos(x\sin\tau -n\tau)$ of the Bessel funtion. Then we perform the $k$ integral to get Polygamma funcitons. In the end we do the $\phi$ integral after series expanding Polygammas around $t=\infty$ to get leading large $t$ answer.

%%%%
\subsubsection{Ground state correlator}
The time-dependent part of the equal-time ground state correlator here is
\begin{align}
 =& -\frac{m^2}{16\pi^2 r} \int_0^\infty \frac{dk}{\sqrt{k^2+m^2}} J_1(k r)\cos(k t_+) \nonumber \\
 =& -\frac{m^2}{16\pi^3 r} \int_0^\infty \frac{dk}{\sqrt{k^2+m^2}} \int_0^\pi d\phi \cos(k r\sin\phi-\phi)\cos(k t_+) \nonumber \\
 =& -\frac{m^2}{32\pi^3 r} \int_0^\infty \frac{dk}{\sqrt{k^2+m^2}} \int_0^\pi d\phi \left[\cos(k a_+ -\phi)+ \cos(k a_- +\phi) \right] \nonumber \\
 =& -\frac{m^2}{64\pi^3 r} \int_0^\infty \frac{dk}{\sqrt{k^2+m^2}} \int_0^\pi d\phi \left[e^{i(k a_+ -\phi)} +e^{-i(k a_+ -\phi)} +e^{i(k a_- +\phi)} +e^{-i(k a_- +\phi)} \right] \nonumber \\
 =& -\frac{m^2}{64\pi^3 r} \int_0^\infty \frac{dk}{\sqrt{k^2+m^2}} \int_0^\pi d\phi \left[e^{-i\phi} \left(e^{i k a_+} +e^{-i k a_-}\right) +e^{i\phi}\left(e^{-i k a_+} +e^{i k a_-}\right) \right] \nonumber \\
 =& -\frac{m^2}{64\pi^3 r} \int_0^\pi d\phi\Big[\pi \sin\phi\left(-\pmb{L}_0(m (t_++r \sin\phi))+\pmb{L}_0(m (t_+-r \sin\phi))+I_0(m (t_++r \sin\phi))-I_0(m (t_+-r \sin\phi))\right) \nonumber \\ 
 &+2 \cos (\phi ) (K_0(m (t_++r \sin\phi))+K_0(m (t_+-r \sin\phi))) \Big]
\end{align}
In the second line we have used the integral representation for $J_1(kr)$. Then after some manipulation we are able to integrate over momentum $k$. $\pmb{L}_0(x)$ is the Modified Struve funtion and $I_0(x)$ is the Bessel-I function. Now using the asymptotic form of these functions, eq.(\ref{eq:struve-bessel})
\[
 \pmb{L}_\nu(x) - I_{-\nu}(x) \approx -\frac{1}{\sqrt{\pi}\,\Gamma(\nu+1/2)}\left(\frac{2}{x}\right)^{1-\nu}
\]
and knowing that $K_0(x)\approx e^{-x}/\sqrt{x}$ is exponentially supressed at large $x$, one gets
\begin{equation}
 \int_0^\pi d\phi\frac{m \sin^2\phi}{64 \pi^3 t^2} = \frac{m}{128 \pi^2 t^2} \label{eq:ground4-2pt-answer}
\end{equation}

\subsubsection{The Thermal Correlator}
The 2-point function in the thermal ensemble is (\ref{2pt-gge-general})
\begin{align}
 \langle \phi(\vec{x_1},t_1) \phi(\vec{x_2},t_2) \rangle_\beta =& \frac{1}{2} \int \frac{d^4 k}{(2\pi)^4} \frac{e^{\iota\vec{k}\cdot\vec{x}}}{\omega_{out}} \left[\cos(\omega_{out} t)\coth(\frac{\mu(k)}{2}) - i\sin(\omega_{out}t) \right] \nonumber \\
 =& \frac{1}{8\pi^2 r} \int_0^{\infty} dk J_1(kr) k \left[\cos(kt)\coth(\beta k/2)-i\sin(kt)\right] \nonumber \\
 =& \frac{1}{8\pi^2 r} \int_0^{\infty} dk J_1(kr) k \left[\frac{2\cos(kt)}{e^{\beta k}-1} +e^{-i kt}\right] \nonumber
\end{align}
We have added and subtracted $1$ from the $\coth$ to get to the final line. Now to calculate the first term we use the same reasoning as in section \ref{sec:crit-phi-phi}. To this end we define dimensionless momentum $p=kt$ and expand the integrand at large $t$ (see eq. (\ref{eq:gcc-expansion}))
\begin{align}
 \frac{1}{8\pi^2 r} \int_0^{\infty} dk J_1(kr) k\frac{2\cos(kt)}{e^{\beta k}-1} =& \frac{1}{4\pi^2 r t^2} \int_0^{\infty} dp J_1(pr/t) p \frac{\cos(p)}{e^{\beta p/t}-1} \nonumber \\
 =& \frac{1}{16\pi^2 t^2 \beta} \left[I(1)- \frac{\beta}{2t}I(2)+ \frac{-3r^2 +2\beta^2}{24 t^2}I(3) +\mathcal{O}(t^{-3}) \right] \nonumber \\
 =& -\frac{1}{8\pi^2\beta t^2} + \frac{-3r^2 +2\beta^2}{32\pi^2\beta t^4} +\mathcal{O}(t^{-6})
\end{align}
We have already rotated the contours appropriately and used the notation $I(n)$ of eq. (\ref{eq:I(n)}). The $e^{-ikt}$ term is easily evaluted to give
\[
 \frac{1}{8\pi^2 r} \int_0^{\infty} dk J_1(kr) k e^{-ikt} = \frac{i}{8 \pi ^2 t^3 \left(1-\frac{r^2}{t^2}\right)^{3/2}}
\]
This is just the UV singularity one expects in $d=4$. Combining we have
\begin{equation}
 \langle \phi(\vec{x_1},t_1) \phi(\vec{x_2},t_2) \rangle_\beta = \frac{i}{8 \pi ^2 t^3 \left(1-\frac{r^2}{t^2}\right)^{3/2}} -\frac{1}{8\pi^2\beta t^2} + \frac{-3r^2 +2\beta^2}{32\pi^2\beta t^4} +\mathcal{O}(t^{-6})
\end{equation}
The thermal autocorrelator is obtained by taking $r\rightarrow 0$
\begin{equation}
 \langle \phi(\vec{x_1},t_1) \phi(\vec{x_2},t_2) \rangle_\beta = \frac{i}{8 \pi ^2 t^3} -\frac{1}{8\pi^2\beta t^2} + \frac{\beta}{16\pi^2 t^4} +\mathcal{O}(t^{-6}) \label{thermal-auto}
\end{equation}

\section{GGE Correlator}\label{app:gge}
For the GGE ``density matrix'' $\rho = \frac{1}{Z} \exp(-\sum_{k}\mu(k)\hat{N}(k))$, its easy to calculate the partition function
\begin{align}
 Z=& tr \exp(-\sum_{k} \mu(k) \hat{N}(k)) \nonumber \\
  =& \sum_{\{N_{k}\}} \langle \{N_{k}\}| e^{-\sum_{k} \mu(k) \hat{N}(k))}|\{N_{k}\}\rangle = \sum_{\{N_{k}\}} \langle \{N_{k}\}| \prod_{k}e^{-\mu(k) \hat{N}(k))}|\{N_{k}\}\rangle\nonumber \\
  =& \prod_{k} \sum_{N_{k}} \langle N_{k}| e^{-\mu(k) \hat{N}(k))}|N_{k}\rangle = \prod_{k} \sum_{N_{k}=1}^\infty e^{-\mu(k) \hat{N}_k} \nonumber \\
 Z=& \prod_k \left( 1-e^{-\mu(k)}\right)^{-1}
\end{align}
Starting with the GGE 2-point function
\begin{align}
\langle \phi(\vec{x_1},t_1) \phi(\vec{x_2},t_2) \rangle _\rho &= \frac{1}{Z}tr \left(e^{-\sum_{\vec{k}} \mu(\vec{k}) \Hat{N}(\vec{k})} \phi (\vec{x_1},t_1) \phi(\vec{x_2},t_2) \right) \nonumber \\
 &= \frac{1}{Z}\sum_{\{N_{\vec{k}}\}} \langle \{N_{\vec{k}}\}| e^{-\sum_{\vec{k}} \mu(\vec{k}) \Hat{N}(\vec{k})} \phi(\vec{x_1},t_1) \phi(\vec{x_2},t_2)|\{N_{\vec{k}}\} \rangle
\end{align}
In the second line we have used the occupation number basis. Using the partial Fourier transform for the field and the mode expansion
\[
\phi(\vec{x},t) = \int e^{\iota \vec{k}\cdot\vec{x}}\phi(\vec{k},t) \frac{d^d k}{(2\pi)^d}
\]
where $\phi(\vec{k},t) = a(\vec{k}) u(\vec{k},t) + a^{\dagger}(\vec{-k}) u^*(\vec{-k},t)$, gives
\begin{align}
\langle \phi(\vec{x_1},t_1)\phi(\vec{x_2},t_2) \rangle_\rho =&\frac{1}{Z}\sum_{\{N_{\vec{k}}\}} \langle \{N_{\vec{k}}\}|e^{-\sum_{\vec{k}} \mu(\vec{k}) \Hat{N}(\vec{k})} \int\int \frac{d^d k}{(2\pi)^d} \frac{d^d q}{(2\pi)^d}  e^{\iota(\vec{k}\cdot\vec{x_1}+\vec{q}\cdot\vec{x_2})} \nonumber \\
& \left( a(\vec{k})u(\vec{k},t_1) + a^{\dagger}(\vec{-k})u^*(\vec{-k},t_1) \right) \left(a(\vec{q})u(\vec{q},t_2) + a^{\dagger}(\vec{-q})u^*(\vec{-q},t_2)\right) |\{N_{\vec{k}}\}\rangle \nonumber
\end{align}
Out of the resulting four terms only two terms give non-zero values.\begin{align}
\langle \phi(\vec{x_1},t_1)\phi(\vec{x_2},t_2) \rangle_\rho =&\frac{1}{Z}\int\int \frac{d^d k}{(2\pi)^d} \frac{d^d q}{(2\pi)^d} e^{\iota(\vec{k}\cdot\vec{x_1}+\vec{q}\cdot\vec{x_2})} \sum_{\{N_{\vec{k}}\}} e^{-\sum_{\vec{k}} \mu(\vec{k}) \Hat{N}(\vec{k})} \nonumber \\
& \langle \{N_{\vec{k}}\}|(a(\vec{k})u(\vec{k},t_1)a^{\dagger}(\vec{-q})u^*(\vec{-q},t_2) + a^{\dagger}(\vec{-k})u^*(\vec{-k},t_1)a(\vec{q})u(\vec{q},t_2))|\{N_{\vec{k}}\}\rangle \nonumber
\end{align}
Using the commutation relation
\[
[a(\vec{k}),a^{\dagger}(-\vec{q})]=(2\pi)^d \delta^d(\vec{k}+\vec{q})
\]
and the form of the number operator
\[
a^{\dagger}(-\vec{q}) a(\vec{k})= N_{\vec{k}}(2\pi)^d\delta^d(\vec{k}+\vec{q})
\]
Therefore
\begin{align}
\langle \phi(\vec{x_1},t_1) \phi(\vec{x_2},t_2) \rangle_\rho =&\frac{1}{Z}\int\int \frac{d^d k}{(2\pi)^d} \frac{d^d q}{(2\pi)^d}  e^{\iota(\vec{k}\cdot\vec{x_1}+\vec{q}\cdot\vec{x_2})} (2\pi)^d\delta^d(\vec{k}+\vec{q}) \sum_{\{N_{\vec{k}}\}} e^{-\sum_{\vec{k}} \mu(\vec{k}) \Hat{N}(\vec{k})}\nonumber \\
& \sum_{\{N_{\vec{k}}\}} \left(\langle \{N_{\vec{k}}\}| (N_{\vec{k}}+1)|\{N_{\vec{k}}\}\rangle u(\vec{k},t_1)u^*(\vec{-q},t_2) +\langle \{N_{\vec{k}}\}| N_{\vec{q}} |\{N_{\vec{k}}\}\rangle u^*(\vec{-k},t_1)u(\vec{q},t_2) \right) \nonumber
\end{align}
Doing the q integral for the first term and k integral for the second and then writing it in terms of a single dummy variable:
\begin{align}
\langle \phi(\vec{x_1},t_1) \phi(\vec{x_2},t_2) \rangle_\rho &=\frac{1}{Z}\int \frac{d^d k}{(2\pi)^d}\sum_{\{N_{\vec{k}}\}} e^{-\sum_{\vec{k}} \mu(\vec{k}) \Hat{N}(\vec{k})} \Big[\langle \{N_{\vec{k}}\}| (N_{k}+1) |\{N_{\vec{k}}\}\rangle \nonumber \\
& u(\vec{k},t_1)u^*(\vec{k},t_2) e^{\iota\vec{k}\cdot(\vec{x_1}-\vec{x_2})} + \langle \{N_{\vec{k}}\}|N_{\vec{k}} |\{N_{\vec{k}}\}\rangle u^*(\vec{k},t_1)u(\vec{k},t_2) e^{-\iota\vec{k}\cdot(\vec{x_1}-\vec{x_2})} \Big] \nonumber
\end{align}
Directly using $ \langle N_{\vec{k}} \rangle = \frac{1}{Z} \sum_{\{N_{\vec{k}}\}} \langle \{N_{\vec{k}}\}|e^{-\sum_{\vec{k}} \mu(\vec{k}) \Hat{N}(\vec{k})} N(k)|\{N_{\vec{k}}\}\rangle=\bar{N}(k) = (e^{\mu(k)}-1)^{-1}$, we get
\[
\langle \phi(\vec{x_1},t_1) \phi(\vec{x_2},t_2) \rangle_\rho = \frac{1}{2} \int \frac{d^d k}{(2\pi)^d} e^{\iota\vec{k}\cdot(\vec{x_1}-\vec{x_2})} \left[\frac{u(\vec{k},t_1)u^*(\vec{k},t_2)} {1- e^{-\mu(k)}} +\frac{u^*(\vec{k},t_1)u(\vec{k},t_2)}{e^{\mu(k)}-1} \right]
\]
Defining $\vec{x}=\vec{x_1}-\vec{x_2}$, $t=t_1-t_2$ and 
\[
G_{\pm}=\frac{1}{\omega_{out}(\pm e^{\pm \mu(k)} \mp 1)}
\]
\[
\langle \phi(\vec{x_1},t_1) \phi(\vec{x_2},t_2) \rangle_\mu = \frac{1}{2} \int \frac{d^d k}{(2\pi)^d} e^{\iota\vec{k}\cdot\vec{x}} \Big(G_- e^{-\iota \omega_{out}t} +G_+ e^{\iota \omega_{out}t} \Big)
\]
We can further simplify
\begin{align}
 & \frac{e^{-\iota \omega_{out}t}}{(1- e^{-\mu(k)})} +\frac{e^{\iota \omega_{out}t}}{(e^{\mu(k)}-1)} = \frac{e^{-\iota \omega_{out}t} e^{\mu(k)/2}+e^{\iota \omega_{out}t} e^{-\mu(k)/2}} {e^{\mu(k)/2}- e^{-\mu(k)/2}} \nonumber \\
 &= \frac{(\cos(\omega_{out}t)- i\sin(\omega_{out}t)) e^{\mu(k)/2}+ (\cos(\omega_{out}t)+ i\sin(\omega_{out}t)) e^{-\mu(k)/2}} {e^{\mu(k)/2}- e^{-\mu(k)/2}} \nonumber \\
 &= \cos(\omega_{out} t)\coth(\frac{\mu(k)}{2}) - i\sin(\omega_{out}t) \nonumber
\end{align}
So
\begin{equation}
 \langle \phi(\vec{x_1},t_1) \phi(\vec{x_2},t_2) \rangle_\mu = \frac{1}{2} \int \frac{d^d k}{(2\pi)^d} \frac{e^{\iota\vec{k}\cdot\vec{x}}}{\omega_{out}} \left[\cos(\omega_{out} t)\coth(\frac{\mu(k)}{2}) - i\sin(\omega_{out}t) \right] \label{2pt-gge-general}
\end{equation}
We would be particularly interested in the ETC when $t_1=t_2 \Rightarrow t=0$ then
\begin{equation}
\langle \phi(\vec{x_1},t) \phi(\vec{x_2},t) \rangle_\mu = \frac{1}{2} \int \frac{d^d k}{(2\pi)^d} \frac{e^{\iota\vec{k}\cdot\vec{x}}}{\omega_{out}} \coth(\frac{\mu(k)}{2}) \label{2pt-equilibrium}
\end{equation}

%%%%%%%%%%%%%%%%%%%%%%%%%%%%%%%
%%%%%%%%% BIBLIOGRAPHY %%%%%%%%%%
%%%%%%%%%%%%%%%%%%%%%%%%%%%%%%%
%\enlargethispage{10pt}
%{\normalsize 
\bibliographystyle{JHEP} 
\bibliography{arbit-dim} 

\providecommand{\href}[2]{#2}\begingroup\raggedright\begin{thebibliography}{10}

\bibitem{Mandal:2015kxi}
G.~{Mandal}, S.~{Paranjape}, and N.~{Sorokhaibam}, {\it {Thermalization in 2D
  critical quench and UV/IR mixing}},
  \href{http://arxiv.org/abs/1512.02187}{{\tt arXiv:1512.02187}}.

\bibitem{Rigol:2007}
M.~Rigol, V.~Dunjko, V.~Yurovsky, and M.~Olshanii, {\it Relaxation in a
  completely integrable many-body quantum system: An \textit{Ab Initio} study
  of the dynamics of the highly excited states of 1d lattice hard-core bosons},
   {\em Phys. Rev. Lett.} {\bf 98} (Feb, 2007) 050405,
  [\href{http://arxiv.org/abs/0604476}{{\tt 0604476}}].

\bibitem{Rigol:2007a}
M.~{Rigol}, V.~{Dunjko}, and M.~{Olshanii}, {\it {Thermalization and its
  mechanism for generic isolated quantum systems}},  {\em Nature} {\bf 452}
  (Apr., 2008) 854--858, [\href{http://arxiv.org/abs/0708.1324}{{\tt
  arXiv:0708.1324}}].

\bibitem{Barthel:2008GGE}
T.~{Barthel} and U.~{Schollw{\"o}ck}, {\it {Dephasing and the Steady State in
  Quantum Many-Particle Systems}},  {\em Physical Review Letters} {\bf 100}
  (Mar., 2008) 100601, [\href{http://arxiv.org/abs/0711.4896}{{\tt
  arXiv:0711.4896}}].

\bibitem{Cramer:2008GGE}
M.~{Cramer}, C.~M. {Dawson}, J.~{Eisert}, and T.~J. {Osborne}, {\it {Exact
  Relaxation in a Class of Nonequilibrium Quantum Lattice Systems}},  {\em
  Physical Review Letters} {\bf 100} (Jan., 2008) 030602,
  [\href{http://arxiv.org/abs/cond-mat/0703314}{{\tt cond-mat/0703314}}].

\bibitem{Mussardo:2009GGE}
D.~{Fioretto} and G.~{Mussardo}, {\it {Quantum quenches in integrable field
  theories}},  {\em New Journal of Physics} {\bf 12} (May, 2010) 055015,
  [\href{http://arxiv.org/abs/0911.3345}{{\tt arXiv:0911.3345}}].

\bibitem{Iucci:2010GGE}
A.~{Iucci} and M.~A. {Cazalilla}, {\it {Quantum quench dynamics of the
  Luttinger model}},  {\em Physical} {\bf 80} (2009) 063619,
  [\href{http://arxiv.org/abs/1003.5170}{{\tt arXiv:1003.5170}}].

\bibitem{Calabrese:2012GGE}
P.~{Calabrese}, F.~H.~L. {Essler}, and M.~{Fagotti}, {\it {Quantum quench in
  the transverse field Ising chain: I. Time evolution of order parameter
  correlators}},  {\em Journal of Statistical Mechanics: Theory and Experiment}
  {\bf 7} (July, 2012) 16, [\href{http://arxiv.org/abs/1204.3911}{{\tt
  arXiv:1204.3911}}].

\bibitem{Calabrese:2011GGE}
P.~{Calabrese}, F.~H.~L. {Essler}, and M.~{Fagotti}, {\it {Quantum Quench in
  the Transverse-Field Ising Chain}},  {\em Physical Review Letters} {\bf 106}
  (June, 2011) 227203, [\href{http://arxiv.org/abs/1104.0154}{{\tt
  arXiv:1104.0154}}].

\bibitem{Calabrese:2012GGE-II}
P.~{Calabrese}, F.~H.~L. {Essler}, and M.~{Fagotti}, {\it {Quantum quenches in
  the transverse field Ising chain: II. Stationary state properties}},  {\em
  Journal of Statistical Mechanics: Theory and Experiment} {\bf 7} (July, 2012)
  22, [\href{http://arxiv.org/abs/1205.2211}{{\tt arXiv:1205.2211}}].

\bibitem{Essler:2014GGE}
B.~{Bertini}, D.~{Schuricht}, and F.~H.~L. {Essler}, {\it {Quantum quench in
  the sine-Gordon model}},  {\em Journal of Statistical Mechanics: Theory and
  Experiment} {\bf 10} (Oct., 2014) 35,
  [\href{http://arxiv.org/abs/1405.4813}{{\tt arXiv:1405.4813}}].

\bibitem{Essler:2014qza}
F.~H.~L. Essler, G.~Mussardo, and M.~Panfil, {\it {Generalized Gibbs Ensembles
  for Quantum Field Theories}},  {\em Phys. Rev.} {\bf A91} (2015), no.~5
  051602, [\href{http://arxiv.org/abs/1411.5352}{{\tt arXiv:1411.5352}}].

\bibitem{gogolin2015equilibration}
C.~Gogolin and J.~Eisert, {\it Equilibration, thermalisation, and the emergence
  of statistical mechanics in closed quantum systems},  tech. rep.,
  arXiv:1503.07538, Mar., 2015.

\bibitem{scully1997quantum}
M.~Scully and M.~Zubairy, {\em Quantum Optics}.
\newblock Cambridge University Press, 1997.

\bibitem{squeezed-2011}
E.~Tiesinga and P.~R. Johnson, {\it Collapse and revival dynamics of
  number-squeezed superfluids of ultracold atoms in optical lattices},  {\em
  Phys. Rev. A} {\bf 83} (Jun, 2011) 063609.

\bibitem{Mandal:2015jla}
G.~Mandal, R.~Sinha, and N.~Sorokhaibam, {\it {Thermalization with chemical
  potentials, and higher spin black holes}},  {\em JHEP} {\bf 08} (2015) 013,
  [\href{http://arxiv.org/abs/1501.04580}{{\tt arXiv:1501.04580}}].

\bibitem{Cardy:2015xaa}
J.~Cardy, {\it {Quantum Quenches to a Critical Point in One Dimension: some
  further results}},  \href{http://arxiv.org/abs/1507.07266}{{\tt
  arXiv:1507.07266}}.

\bibitem{birrell1984quantum}
N.~Birrell and P.~Davies, {\em Quantum Fields in Curved Space}.
\newblock Cambridge Monographs on Mathematical Physics. Cambridge University
  Press, 1984.

\bibitem{Bhattacharyya:2016nhn}
S.~Bhattacharyya, A.~K. Mandal, M.~Mandlik, U.~Mehta, S.~Minwalla, U.~Sharma,
  and S.~Thakur, {\it {Currents and Radiation from the large $D$ Black Hole
  Membrane}},  {\em JHEP} {\bf 05} (2017) 098,
  [\href{http://arxiv.org/abs/1611.09310}{{\tt arXiv:1611.09310}}].

\bibitem{Sotiriadis:2010si}
S.~Sotiriadis and J.~Cardy, {\it {Quantum quench in interacting field theory: A
  Self-consistent approximation}},  {\em Phys. Rev.} {\bf B81} (2010) 134305,
  [\href{http://arxiv.org/abs/1002.0167}{{\tt arXiv:1002.0167}}].

\bibitem{AK-GM:2019}
A.~Kaushal and G.~Mandal, {\it Work in progress}, .

\bibitem{PhysRevLett.114.110505}
R.~H. Jonsson, E.~Mart\'{\i}n-Mart\'{\i}nez, and A.~Kempf, {\it Information
  transmission without energy exchange},  {\em Phys. Rev. Lett.} {\bf 114}
  (Mar, 2015) 110505.

\bibitem{Blasco:2015eya}
A.~Blasco, L.~J. Garay, M.~Martin-Benito, and E.~Martin-Martinez, {\it
  {Violation of the Strong Huygen’s Principle and Timelike Signals from the
  Early Universe}},  {\em Phys. Rev. Lett.} {\bf 114} (2015), no.~14 141103,
  [\href{http://arxiv.org/abs/1501.01650}{{\tt arXiv:1501.01650}}].

\end{thebibliography}\endgroup
%}

\end{document}